\documentclass[11pt]{article}

\usepackage{amssymb}
\usepackage{makeidx}
\usepackage[totalwidth=460truept,totalheight=600truept]{geometry}
\usepackage{latexsym,amssymb,amsmath,graphicx,accents,eucal,slashed,subfigure}
\usepackage{dsfont}
\usepackage[T1]{fontenc}
\usepackage[hidelinks]{hyperref}

\newtheorem{theorem}{Theorem}

\global\arraycolsep=1truept

\numberwithin{equation}{section}

\begin{document}

\bigskip \phantom{C}

\vskip1truecm

\begin{center}
{\huge \textbf{Adler-Bardeen Theorem}}

\vskip .5truecm

{\huge \textbf{And Cancellation Of\ Gauge Anomalies}}

\vskip .5truecm

{\huge \textbf{To All Orders}}

\vskip .5truecm

{\huge \textbf{In Nonrenormalizable Theories}}

\vskip 1truecm

\textsl{Damiano Anselmi}

\vskip .2truecm

\textit{Dipartimento di Fisica ``Enrico Fermi'', Universit\`{a} di Pisa, }

\textit{and INFN, Sezione di Pisa,}

\textit{Largo B. Pontecorvo 3, I-56127 Pisa, Italy}

\vskip .2truecm

damiano.anselmi@df.unipi.it

\vskip 1truecm

\textbf{Abstract}
\end{center}

\medskip

We prove the Adler-Bardeen theorem in a large class of general gauge
theories, including nonrenormalizable ones. We assume that the gauge
symmetries are general covariance, local Lorentz symmetry and Abelian and
non-Abelian Yang-Mills symmetries, and that the local functionals of
vanishing ghost numbers satisfy a variant of the Kluberg-Stern--Zuber
conjecture. We show that if the gauge anomalies are trivial at one loop, for
every truncation of the theory there exists a subtraction scheme where\ they
manifestly vanish to all orders, within the truncation. Outside the
truncation the cancellation of gauge anomalies can be enforced by
fine-tuning local counterterms. The framework of the proof is worked out by
combining a recently formulated chiral dimensional regularization with a
gauge invariant higher-derivative regularization. If the higher-derivative
regularizing terms are placed well beyond the truncation, and the energy
scale $\Lambda $ associated with them is kept fixed, the theory is
super-renormalizable and has the property that, once the gauge anomalies are
canceled at one loop, they manifestly vanish from two loops onwards by
simple power counting. When the $\Lambda $ divergences are subtracted away
and $\Lambda $ is sent to infinity, the anomaly cancellation survives in a
manifest form within the truncation and in a nonmanifest form outside. The
standard model coupled to quantum gravity satisfies all the assumptions, so
it is free of gauge anomalies to all orders.

\vskip 1truecm

\vfill\eject

\section{Introduction}

\label{s1}

\label{RGren}\setcounter{equation}{0}

The Adler-Bardeen theorem \cite{adlerbardeen,review} is crucial to prove the
consistency of a wide class of perturbative quantum field theories. Its main
consequence is that the cancellation of gauge anomalies at one loop ensures
the cancellation of gauge anomalies to all orders. Thanks to this result, a
finite number of conditions is sufficient to determine when a potentially
anomalous theory is actually anomaly free. The cancellation conditions can
be worked out rather easily, because they just involve simplified
divergences of one-loop diagrams. If a similar theorem did not hold, a
chiral gauge theory, such as the standard model, would have to satisfy
infinitely many independent cancellation conditions, to be consistent. The
solutions would be very few, or contain infinitely many fields.

So far, the Adler-Bardeen theorem has been proved in Abelian and non-Abelian
power counting renormalizable gauge theories, including the standard model,
but not in more general classes of theories. In this paper we overcome this
limitation by working out a more powerful proof that applies to a large
class of nonrenormalizable theories and allows us to infer that the standard
model coupled to quantum gravity, which is known to be free of gauge
anomalies at one loop \cite{peskin}, is also free of gauge anomalies to all
orders, and so are most of its extensions.

In general, we must show that when the gauge anomalies are trivial at one
loop, there exists a subtraction scheme where they vanish to all orders.
Once we know that the scheme exists, we can build it order by order by
fine-tuning finite local counterterms. A more powerful result is to provide
the right scheme from the beginning, that is to say define a framework where
all potentially anomalous contributions cancel out at one loop and are
automatically zero from two loops onwards. We call a statement identifying
such a scheme \textit{manifest} Adler-Bardeen theorem. In perturbatively
unitary renormalizable theories the manifest Adler-Bardeen theorem has been
proved recently \cite{ABrenoYMLR}. For reasons that we explain in the paper,
in nonrenormalizable theories we are not able to determine the subtraction
scheme where anomaly cancellation is manifest from two loops onwards. We
have to content ourselves with a weaker, yet powerful enough, result, which
we call \textit{almost manifest} Adler-Bardeen theorem: given an appropriate
truncation $T$ of the theory, we find a subtraction scheme where the gauge
anomalies manifestly vanish from two loops onwards within the truncation.

The most common regularization techniques are not very convenient to work
out general proofs of the Adler-Bardeen theorem, because they give us no
clue about the right subtraction scheme. In ref. \cite{ABrenoYMLR} a better
regularization technique was built by merging the dimensional regularization
with a suitable gauge invariant higher-derivative (HD) regularization \cite%
{higherder} and used to prove the manifest Adler-Bardeen theorem in
four-dimensional renormalizable perturbatively unitary gauge theories.
Unfortunately, several difficulties of the dimensional regularization make
it hard to generalize that proof to nonrenormalizable theories. To overcome
those problems, in ref. \cite{chiraldimreg} a \textit{chiral} \textit{%
dimensional} (CD) regularization technique was defined. 
Nevertheless, the CD\ technique alone does not identify the subtraction
scheme where gauge anomalies manifestly cancel and must still be merged with
a suitable gauge invariant HD regularization. The resulting technique,
called chiral-dimensional/higher-derivative (CDHD) regularization, is the
right one to generalize the proof of the Adler-Bardeen theorem to
nonrenormalizable theories. It has two rregularizing parameters: $%
\varepsilon =d-D$, where $d$ is the physical spacetime dimension and $D$ is
the continued dimension, and the energy scale $\Lambda $ associated with the
higher-derivative terms. The limit $\varepsilon \rightarrow 0$ must be
studied before the limit $\Lambda \rightarrow \infty $.

The CDHD\ technique is organized so that the higher-derivative regularizing
terms fall well beyond the truncation. When $\Lambda $ is kept fixed, a
peculiar super-renormalizable higher-derivative theory is obtained, which we
call \textit{HD theory}. The HD theory satisfies the manifest Adler-Bardeen
theorem by simple power counting arguments. The limit $\Lambda \rightarrow
\infty $ on the HD theory defines the \textit{final theory}, which is the
one we are interested in. We show that we can renormalize the $\Lambda $
divergences so as to preserve the cancellation of gauge anomalies to all
orders within the truncation.

The proof we provide holds under certain assumptions. First, we assume that
the gauge symmetries are general covariance, local Lorentz symmetry and
Abelian and non-Abelian Yang-Mills symmetries. At this stage, we cannot
include local supersymmetry. Second, we assume that the local functionals of
vanishing ghost numbers satisfy a variant of the Kluberg-Stern--Zuber
conjecture \cite{kluberg}. The standard model coupled to quantum gravity
does not satisfy the ordinary Kluberg-Stern--Zuber conjecture, but satisfies
the variant that we assume in this paper. The other key assumption is of
course that the one-loop gauge anomalies $\mathcal{A}^{(1)}$ are trivial. In
our approach the functional $\mathcal{A}^{(1)}$ is extremely simple, since
it can only depend on the gauge fields, their ghosts and some matter fields.
We call $\mathcal{A}^{(1)}$ trivial if there exists a local functional $\chi 
$ of the fields such that $\mathcal{A}^{(1)}=(S_{d},\chi )$, where $S_{d}$
is the $d$-dimensional tree-level action and $(X,Y)$ are the
Batalin-Vilkovisky (BV) antiparentheses \cite{bata}, recalled in formula (%
\ref{usa}). Other mild technical assumptions needed for the proof (all of
which are satisfied by most common theories of fields of spins $\leqslant 2$%
) are described along the way.

Here are the main statements that we consider in this paper. The most
general Adler-Bardeen theorem for the cancellation of gauge anomalies states
that

\begin{theorem}
If the gauge anomalies are trivial at one loop, the subtraction scheme can
be fine-tuned so that they vanish to all orders.\label{bardo}
\end{theorem}

In renormalizable theories we actually have a stronger result, the \textit{%
manifest} Adler-Bardeen theorem \cite{ABrenoYMLR}, stating that

\begin{theorem}
If the gauge anomalies are trivial at one loop, there exists a subtraction
scheme where they cancel at one loop and manifestly vanish from two loops
onwards.\label{bardo2}
\end{theorem}

In nonrenormalizable theories, instead, we can prove a result that is
stronger than \ref{bardo}, but weaker than \ref{bardo2}, the almost manifest
Adler-Bardeen theorem, which states that

\begin{theorem}
If the gauge anomalies are trivial at one loop, for every appropriate
truncation of the theory there exists a subtraction scheme where they cancel
at one loop and manifestly vanish from two loops onwards within the
truncation.\label{bardo3}
\end{theorem}

The proper way to truncate a nonrenormalizable theory is specified in the
next section. We stress again that in nonrenormalizable theories we are not
able to prove statement \ref{bardo2}, namely find the right subtraction
scheme independently of the truncation. We can just find a good subtraction
scheme for every truncation. This result is still satisfactory, because
theorem \ref{bardo3} implies theorem \ref{bardo}. Indeed, let $s_{T}$ denote
the subtraction scheme associated with the truncation $T$ by the proof of
theorem \ref{bardo3}. There, the gauge anomalies $\mathcal{A}$ vanish within
the truncation. Let $\mathcal{A}_{>T}$ denote a finite class of
contributions to the gauge anomalies that lie outside the truncation $T$, in
the scheme $s_{T}$. Clearly, the contributions of class $\mathcal{A}_{>T}$
are fully contained in some truncation $T^{\prime }>T$. There, however, they
must vanish. Since two schemes differ by finite local counterterms, there
must exist finite local counterterms that cancel the contributions of class $%
\mathcal{A}_{>T}$ in the scheme $s_{T}$. In conclusion, the scheme $s_{T}$
satisfies theorem \ref{bardo3} within the truncation, and theorem \ref{bardo}
outside.

It is worthwhile to compare our approach with other approaches to the
Adler-Bardeen theorem that can be found in the literature. The original
proof given by Adler and Bardeen \cite{adlerbardeen} was designed to work in
QED. Most generalizations to renormalizable non-Abelian gauge theories used\
arguments based on the renormalization group \cite{zee,collins,tonin,sorella}%
. Those arguments work well unless the first coefficients of the beta
functions satisfy peculiar conditions \cite{sorella} (for example, they
should not vanish). If the theory is nonrenormalizable, we can build
infinitely many dimensionless couplings, and can hardly exclude that the
first coefficients of their beta functions satisfy peculiar conditions.
Algebraic/geometric derivations \cite{witten} based on the Wess-Zumino
consistency conditions \cite{wesszumino} and the quantization of the
Wess-Zumino-Witten action also do not seem suitable to be generalized to
nonrenormalizable theories. Another method to prove the Adler-Bardeen
theorem in renormalizable theories is obtained by extending the coupling
constants to spacetime-dependent fields \cite{kraus}. A tentative
regularization-independent approach in nonrenormalizable theories can be
found in ref. \cite{barnich}.

We stress that the proof provided in this paper is the first proof that the
standard model coupled to quantum gravity is free of gauge anomalies to all
orders. Our arguments and results also apply to the study of
higher-dimensional composite fields in renormalizable and nonrenormalizable
theories.

In this paper, the powers of $\hbar $ are merely used as tools to denote the
appropriate orders of the loop expansion. They are not written explicitly
unless necessary. It is understood that the functionals depend analytically
on the parameters that are treated perturbatively.

The paper is organized as follows. In section \ref{s2} we provide the
setting of the proof. We specify the truncation, recall the properties of
the CD regularization technique, and explain how it can be combined with a
suitable higher-derivative regularization to build the CDHD regularized
theory. In section \ref{s3} we study the properties of the HD theory. In
particular, we show that it is super-renormalizable and study the structures
of its counterterms and potential anomalies. In section \ref{s4} we work out
the renormalization of the HD theory. In section \ref{s5} we study its
one-loop anomalies. In section \ref{s6} we prove that the HD theory
satisfies the manifest Adler-Bardeen theorem. In section \ref{s7} we
subtract the $\Lambda $ divergences and prove that the final theory
satisfies the almost manifest Adler-Bardeen theorem, as well as theorem \ref%
{bardo}. In section \ref{s9} we show that the standard model coupled to
quantum gravity, as well as most of its extensions, belongs to the class of
nonrenormalizable theories to which our results apply. Section \ref{s10}
contains our conclusions.

\section{General setting}

\label{s2}

\setcounter{equation}{0}

In this section we give the general setup of the proof and specify most of
the assumptions we need. First we recall the properties of the CD
regularization and explain how it is merged with the HD regularization to
build the CDHD regularization. Then we explain how to truncate the theory.
Instead of working directly with the standard model coupled to quantum
gravity, we formulate a general approach and give specific examples along
the way.

Throughout the paper, $d$ denotes the physical spacetime dimension, and $%
D=d-\varepsilon $ is the continued complex dimension introduced by the
dimensional regularization (see subsection \ref{s22} for details). We work
in $d>2$. We use the symbol $\phi $ to collect the \textquotedblleft
physical fields\textquotedblright , that is to say the Yang-Mills gauge
fields $A_{\bar{\mu}}^{a}$, the matter fields, and (if gravity is dynamical)
the metric tensor $g_{\bar{\mu}\bar{\nu}}$ or the vielbein $e_{\bar{\mu}}^{%
\bar{a}}$. The indices $a,b,\ldots $, refer to the Yang-Mills gauge group,
while $\bar{a},\bar{b},\ldots $, refer to the Lorentz group. The indices $%
\bar{\mu},\bar{\nu},\ldots $, refer to the physical $d$-dimensional
spacetime $\mathbb{R}^{d}$, as opposed to the continued spacetime $\mathbb{R}%
^{D}$.

We denote the classical action by $S_{c}(\phi )$. In the case of the
standard model coupled to quantum gravity, we take $S_{c}=S_{c\text{SMG}%
}+\Delta S_{c}$, where 
\begin{equation}
S_{c\text{SMG}}=\int \sqrt{|g|}\left[ -\frac{1}{2\kappa ^{2}}(R+2\Lambda _{%
\text{c}})-\frac{1}{4}F_{\bar{\mu}\bar{\nu}}^{a}F^{a\bar{\mu}\bar{\nu}}+%
\mathcal{L}_{m}\right]  \label{basi}
\end{equation}%
and $\Delta S_{c}$ collects the invariants generated as counterterms by
renormalization, multiplied by independent parameters. Here, $R$ is the
Ricci curvature, $F_{\bar{\mu}\bar{\nu}}^{a}$ are the Yang-Mills field
strengths, $\mathcal{L}_{m}$ is the matter Lagrangian coupled to the metric
tensor or vielbein, $g$ is the determinant of the metric tensor $g_{\bar{\mu}%
\bar{\nu}}$, $\Lambda _{\text{c}}$ is the cosmological constant, and $\kappa
^{2}=8\pi G$, where $G$ is Newton's constant.

We use the Batalin-Vilkovisky formalism \cite{bata}, because it is very
efficient to keep track of gauge invariance throughout the renormalization
algorithm. An enlarged set of fields $\Phi ^{\alpha }=\{\phi ,C,\bar{C},B\}$
is introduced, to collect the physical fields $\phi $, the Fadeev-Popov
ghosts $C$, the antighosts $\bar{C}$, and the Lagrange multipliers $B$ for
the gauge fixing. Next, external sources $K_{\alpha }=\{K_{\phi },K_{C},K_{%
\bar{C}},K_{B}\}$ are coupled to the $\Phi ^{\alpha }$ symmetry
transformations $R^{\alpha }(\Phi )$ in a way specified below.

If $X$ and $Y$ are functionals of $\Phi $ and $K$, their \textit{%
antiparentheses} are defined as 
\begin{equation}
(X,Y)\equiv \int \left( \frac{\delta _{r}X}{\delta \Phi ^{\alpha }}\frac{%
\delta _{l}Y}{\delta K_{\alpha }}-\frac{\delta _{r}X}{\delta K_{\alpha }}%
\frac{\delta _{l}Y}{\delta \Phi ^{\alpha }}\right) ,  \label{usa}
\end{equation}%
where the integral is over spacetime points associated with repeated indices
and the subscripts $l$ and $r$ in $\delta _{l}$ and $\delta _{r}$ denote the
left and right functional derivatives, respectively. The \textit{master
equation} is the condition $(S,S)=0$ and must be solved in $D$ dimensions
with the \textquotedblleft boundary condition\textquotedblright\ $S=S_{c}$
at $C=\bar{C}=B=K=0$. At the practical level, we first solve the equation $%
(S,S)=0$ in $d$ dimensions, and then interpret its solution $S$ as a $D$%
-dimensional action, according to the rules of the CD\ regularization (see
subsection \ref{s22}). We denote the non-gauge-fixed solution of the master
equation by $\bar{S}_{d}(\Phi ,K)$. The subscript $d$ reminds us that,
although $\bar{S}_{d}$ solves $(\bar{S}_{d},\bar{S}_{d})=0$ in $D$
dimensions, it is just the $d$-dimensional action interpreted from the $D$%
-dimensional point of view. In particular, it may not be well regularized as
a $D$-dimensional action. Once we regularize it, we may not be able to
preserve the master equation exactly in $D\neq d$. The violations of the
master equation at $D\neq d$ are the origins of potential anomalies.

(I)\ We assume that the gauge symmetries are general covariance, local
Lorentz symmetry and Abelian and non-Abelian Yang-Mills symmetries. In
particular, the gauge algebra is irreducible and closes off shell. We use
the second order formalism for gravity and choose the fields $\Phi $ and the
sources $K$ so that the non-gauge-fixed solution $\bar{S}_{d}(\Phi ,K)$ of
the master equation reads 
\begin{equation}
\bar{S}_{d}(\Phi ,K)=S_{c}(\phi )+S_{K}(\Phi ,K),\qquad S_{K}(\Phi ,K)=-\int
R^{\alpha }(\Phi )K_{\alpha },  \label{sk}
\end{equation}%
where the functional $S_{K}$ (with left-handed fermions $\psi _{L}$ and
scalars $\varphi $, for definiteness) reads 
\begin{eqnarray}
S_{K} &=&\int (C^{\bar{\rho}}\partial _{\bar{\rho}}A_{\bar{\mu}}^{a}+A_{\bar{%
\rho}}^{a}\partial _{\bar{\mu}}C^{\bar{\rho}}-\partial _{\bar{\mu}%
}C^{a}-gf^{abc}A_{\bar{\mu}}^{b}C^{c})K_{A}^{\bar{\mu}a}+\int \left( C^{\bar{%
\rho}}\partial _{\bar{\rho}}C^{a}+\frac{g}{2}f^{abc}C^{b}C^{c}\right)
K_{C}^{a}  \notag \\
&&+\int (C^{\bar{\rho}}\partial _{\bar{\rho}}e_{\bar{\mu}}^{\bar{a}}+e_{\bar{%
\rho}}^{\bar{a}}\partial _{\bar{\mu}}C^{\bar{\rho}}+C^{\bar{a}\bar{b}}e_{%
\bar{\mu}\bar{b}})K_{\bar{a}}^{\bar{\mu}}+\int C^{\bar{\rho}}(\partial _{%
\bar{\rho}}C^{\bar{\mu}})K_{\bar{\mu}}^{C}+\int (C^{\bar{a}\bar{c}}\eta _{%
\bar{c}\bar{d}}C^{\bar{d}\bar{b}}+C^{\bar{\rho}}\partial _{\bar{\rho}}C^{%
\bar{a}\bar{b}})K_{\bar{a}\bar{b}}^{C}  \notag \\
&&\!\!\!\!\!\!{+\int \left( C^{\bar{\rho}}\partial _{\bar{\rho}}\bar{\psi}%
_{L}-\frac{i}{4}\bar{\psi}_{L}\sigma ^{\bar{a}\bar{b}}C_{\bar{a}\bar{b}}+g%
\bar{\psi}_{L}T^{a}C^{a}\right) K_{\psi }+\int K_{\bar{\psi}}\left( C^{\bar{%
\rho}}\partial _{\bar{\rho}}\psi _{L}-\frac{i}{4}\sigma ^{\bar{a}\bar{b}}C_{%
\bar{a}\bar{b}}\psi _{L}+gT^{a}C^{a}\psi _{L}\right) }  \notag \\
&&+\int \left( C^{\bar{\rho}}(\partial _{\bar{\rho}}\varphi )+g\mathcal{T}%
^{a}C^{a}\varphi \right) K_{\varphi }-\int B^{a}K_{\bar{C}}^{a}-\int B_{\bar{%
\mu}}K_{\bar{C}}^{\bar{\mu}}-\int B_{\bar{a}\bar{b}}K_{\bar{C}}^{\bar{a}\bar{%
b}}.  \label{skexpl}
\end{eqnarray}%
Here, $T^{a}$ and $\mathcal{T}^{a}$ are the anti-Hermitian matrices
associated with the fermion and scalar representations, respectively. The
ghosts of Yang-Mills symmetry are $C^{a}$, those of local Lorentz symmetry
are $C^{\bar{a}\bar{b}}$, and those of diffeomorphisms are $C^{\bar{\mu}}$.
The pairs $\bar{C}^{a}$-$B^{a}$, $\bar{C}_{\bar{a}\bar{b}}$-$B^{\bar{a}\bar{b%
}}$, and $\bar{C}_{\bar{\mu}}$-$B_{\bar{\mu}}$ collect the antighosts and
the Lagrange multipliers of Yang-Mills symmetry, local Lorentz symmetry, and
diffeomorphisms, respectively. The functional $S_{K}$ satisfies $%
(S_{K},S_{K})=0$ in arbitrary $D$ dimensions.

We can gauge fix the theory with the help of a gauge fermion $\Psi (\Phi )$,
which is a local functional of ghost number $-1$ that depends only on the
fields $\Phi $ and contains the gauge-fixing functions $G(\phi )$. For
example, $G(\phi )=\partial ^{\bar{\mu}}A_{\bar{\mu}}$ for the Lorenz gauge
in Yang-Mills theories. The typical form of $\Psi (\Phi )$ is 
\begin{equation}
\Psi (\Phi )=\int \sqrt{|g|}\bar{C}\left( G(\phi ,\xi )+\frac{1}{2}P(\phi
,\xi ^{\prime },\partial )B\right) ,  \label{phi}
\end{equation}%
where $\xi $, $\xi ^{\prime }$ are gauge-fixing parameters and $P$ is an
operator that may contain derivatives acting on $B$. Typically, if the gauge
fields $\phi _{g}$ have dominant kinetic terms (which are the quadratic
terms that have the largest numbers of derivatives) of the form%
\begin{equation}
\sim \frac{1}{2}\int \phi _{g}\partial ^{N_{\phi _{g}}}\phi _{g}
\label{domo}
\end{equation}%
inside $S_{c}$, we choose $G$ and $P$ such that%
\begin{equation}
G(\phi ,\xi )\sim \partial ^{N_{\phi _{g}}-1+a}\phi _{g}+\text{nonlinear
terms,\qquad }P(\phi ,\xi ^{\prime },\partial )\sim \xi ^{\prime }\partial
^{N_{\phi _{g}}-2+b}+\mathcal{O}(\phi ),  \label{choice}
\end{equation}%
up to terms with fewer derivatives, where $a=b=0$ for diffeomorphisms and
Yang-Mills symmetries, while $a=1$, $b=2$ for local Lorentz symmetry. See
formula (\ref{psi1}) for more details. In the case of three-dimensional
Chern-Simons theories ($N_{\phi _{g}}=1$) we take $a=1$ and $P=0$.

The gauge-fixed action $S_{d}$ is obtained by adding $(S_{K},\Psi )$ to $%
\bar{S}_{d}$: 
\begin{equation}
S_{d}(\Phi ,K)=\bar{S}_{d}+(S_{K},\Psi )=S_{c}+(S_{K},\Psi )+S_{K}.
\label{sid}
\end{equation}%
Alternatively, $S_{d}$ is obtained from $\bar{S}_{d}$ by applying the
canonical transformation generated by 
\begin{equation}
F(\Phi ,K^{\prime })=\int \Phi ^{\alpha }K_{\alpha }^{\prime }+\Psi (\Phi ).
\label{cang}
\end{equation}%
We still have $(S_{d},S_{d})=0$ in $D$ dimensions, but we stress again that
in general the action $S_{d}$ may not be well regularized.

Let $\{\mathcal{G}_{i}(\phi )\}$ denote a basis of local gauge invariant
functionals of the physical fields $\phi $, i.e. local functionals such that 
$(S_{K},\mathcal{G}_{i})=0$. Expand the classical action as 
\begin{equation}
S_{c}(\phi )=\sum_{i}\lambda _{i}\mathcal{G}_{i}(\phi ),  \label{scf}
\end{equation}%
where $\lambda _{i}$ are independent constants. We call such constants
\textquotedblleft physical parameters\textquotedblright , since they
include, or are related to, the gauge coupling constants, the masses, etc.
If the theory is power counting renormalizable, $S_{c}(\phi )$ is restricted
accordingly, and contains just a finite number of independent parameters $%
\lambda _{i}$. If the theory is nonrenormalizable, (\ref{scf}) must include
all the invariants $\mathcal{G}_{i}$ required by renormalization, which are
typically infinitely many.

In several cases, the set $\{\mathcal{G}_{i}(\phi )\}$ is restricted to the
invariants that are inequivalent, where two functionals are considered
equivalent if they differ by terms proportional to the $S_{c}$ field
equations. The reason why such a restriction is meaningful is that the
counterterms proportional to the field equations can be subtracted away by
means of canonical transformations of the BV type, instead of $\lambda _{i}$
redefinitions. However, for some arguments of this paper it is convenient to
include the terms proportional to the $S_{c}$ field equations inside the set 
$\{\mathcal{G}_{i}(\phi )\}$, which we assume from now on. We can remove
them at the end, by means of a convergent canonical transformation and the
procedure of ref. \cite{ABward}. There, it is shown that, after the
transformation, it is always possible to re-renormalize the theory and
re-fine-tune its finite local counterterms so as to preserve the
cancellation of gauge anomalies. The renormalized $\Gamma $ functional of
the transformed theory is related to the renormalized $\Gamma $ functional
of the starting theory by a (convergent, nonlocal) canonical transformation.
See \cite{ABward} for more details.

We say that an action $\mathcal{S}$ satisfies the Kluberg-Stern--Zuber
assumption \cite{kluberg}, if every local functional $X$ of ghost number
zero that solves the equation $(\mathcal{S},X)=0$ has the form 
\begin{equation}
\qquad X=\sum_{i}a_{i}\mathcal{G}_{i}+(\mathcal{S},Y),  \label{coho}
\end{equation}%
where $a_{i}$ are constants depending on the parameters of the theory, and $%
Y $ is a local functional of ghost number $-1$. The Kluberg-Stern--Zuber
assumption is very useful to study the counterterms. It is satisfied, for
example, when the Yang-Mills gauge group is semisimple and the action $%
\mathcal{S}$ meets other mild requirements \cite{coho2}. Unfortunately, the
standard model coupled to quantum gravity does not satisfy it, unless its
accidental symmetries are completely broken. This forces us to search for a
more general version of the assumption.

The accidental symmetries are the continuous global symmetries unrelated to
the gauge transformations. Some of them are anomalous, others are
nonanomalous. If the gauge group has $U(1)$ factors, let $G_{\text{nas}}$
denote the group of nonanomalous accidental symmetries. If the gauge group
has no $U(1)$ factors, we take $G_{\text{nas}}$ equal to the identity. We
denote the local gauge invariant functionals of $\phi $ that break the group 
$G_{\text{nas}}$ by $\mathcal{\check{G}}_{i}(\phi )$. We exclude the
invariants $\mathcal{\check{G}}_{i}$ from the set $\{\mathcal{G}_{i}(\phi
)\} $ and the actions $S_{c}$, $S_{d}$, but include them in more general
actions $\check{S}_{c}$ and $\check{S}_{d}=\check{S}_{c}+(S_{K},\Psi )+S_{K}$%
, multiplied by independent parameters $\check{\lambda}_{i}$. The invariants
that explicitly break the anomalous accidental symmetries are instead
included in the set $\{\mathcal{G}_{i}(\phi )\}$.

It is consistent to switch the invariants $\mathcal{\check{G}}_{i}$ off,
since, when they are absent, renormalization is unable to generate them back
as counterterms. However, for some arguments of the proof it is necessary to
temporarily switch them on. For this reason, we need to work with both
actions $S_{d}$ and $\check{S}_{d}$.

The action $\mathcal{S}$ of (\ref{coho}) is assumed to be invariant under
the group $G_{\text{nas}}$. We say that an action $\mathcal{\check{S}}$ that
breaks $G_{\text{nas}}$ satisfies the extended Kluberg-Stern--Zuber
assumption if every local functional $X$ of ghost number zero that solves
the equation $(\mathcal{\check{S}},X)=0$ has the form 
\begin{equation}
\qquad X=\sum_{i}a_{i}\mathcal{G}_{i}+\sum_{i}b_{i}\mathcal{\check{G}}_{i}+(%
\mathcal{\check{S}},Y),  \label{coho2}
\end{equation}%
where $b_{i}$ are other constants and $Y$ is local. We say that the action $%
S_{d}$ is \textit{cohomologically complete} if its extension $\check{S}_{d}$
satisfies the extended Kluberg-Stern--Zuber assumption. In section \ref{s9}
we prove that the standard model coupled to quantum gravity is
cohomologically complete.

The variant of the Kluberg-Stern--Zuber assumption that we need for the
proof of the Adler-Bardeen theorem is formulated in subsection \ref{key}. In
section \ref{s9} we show that it is satisfied by the standard model coupled
to quantum gravity, as well as most of its extensions. We also prove that
the standard model coupled to quantum gravity satisfies a \textquotedblleft
physical\textquotedblright\ variant of the Kluberg-Stern--Zuber assumption.

It is straightforward to show that the results of this paper, which we
derive for theories with unbroken $G_{\text{nas}}$, also hold when $G_{\text{%
nas}}$ is completely, or partially, broken. In the end, it is our choice to
decide which symmetries of $G_{\text{nas}}$ should be preserved and which
ones should be broken. It should also be noted that it may not be easy to
establish which accidental symmetries are anonalous and which ones are
nonanomalous \textit{a priori}. We have arranged our statements to make them
work in any case, under this respect. In the safest case, we can extend the
action $S_{d}$ till $G_{\text{nas}}=\mathds{1}$ and $S_{d}=\check{S}_{d}$.

\subsection{Chiral dimensional regularization}

\label{s22}

If we want to identify the subtraction scheme where the anomaly cancellation
is (almost) manifest, we must provide a regularization and a set of specific
prescriptions to handle the counterterms and the potentially anomalous
contributions in convenient ways. The best regularization technique is
obtained by merging the chiral dimensional regularization recently
introduced in ref. \cite{chiraldimreg} with a suitable gauge invariant
higher-derivative regularization.

Going through the derivation of ref. \cite{ABrenoYMLR}, where the manifest
Adler-Bardeen theorem was proved in perturbatively unitary, power counting
renormalizable four-dimensional gauge theories, it is easy to spot several
crucial arguments that do not generalize to wider classes of models in a
straightforward way. The main obstacles are due to the dimensional
regularization as it is normally understood \cite{dimreg}. Besides the
nuisances associated with the definition of $\gamma _{5}$, the dimensionally
continued Dirac algebra is responsible for other serious difficulties. For
example, it allows us to build infinitely many inequivalent evanescent terms
of the same dimensions, and the Fierz identities involve infinite sums.
Moreover, it generates ambiguities that plague the classification of
counterterms and make it difficult to extract the divergent parts from the
antiparentheses of functionals. The CD regularization overcomes these
problems. In this subsection we recall how it works.

As usual, we split the $D$-dimensional spacetime manifold $\mathbb{R}^{D}$
into the product $\mathbb{R}^{d}\times \mathbb{R}^{-\varepsilon }$ of the
physical $d$-dimensional spacetime $\mathbb{R}^{d}$ times a residual $%
(-\varepsilon )$-dimensional evanescent space $\mathbb{R}^{-\varepsilon }$,
where $\varepsilon $ is a complex number. Spacetime indices $\mu ,\nu
,\ldots $, of vectors and tensors are split into bar indices $\bar{\mu},\bar{%
\nu},\ldots $, which take the values $0,1,\cdots ,d-1$, and formal hat
indices $\hat{\mu},\hat{\nu},\ldots $, which denote the $\mathbb{R}%
^{-\varepsilon }$ components. For example, the momenta $p^{\mu }$ are split
into the pairs $p^{\bar{\mu}}$, $p^{\hat{\mu}}$, also denoted by $\bar{p}%
^{\mu }$, $\hat{p}^{\mu }$, and the coordinates $x^{\mu }$ are split into $%
\bar{x}^{\mu }$, $\hat{x}^{\mu }$. The formal flat-space metric $\eta _{\mu
\nu }$ is split into the physical $d\times d$ flat-space metric $\eta _{\bar{%
\mu}\bar{\nu}}=$diag$(1,-1,\cdots ,-1)$ and the formal evanescent metric $%
\eta _{\hat{\mu}\hat{\nu}}=-\delta _{\hat{\mu}\hat{\nu}}$ (the off-diagonal
components $\eta _{\bar{\mu}\hat{\nu}}$ being equal to zero). When we
contract evanescent components, we use the metric $\eta _{\hat{\mu}\hat{\nu}%
} $, so for example $\hat{p}^{2}=p^{\hat{\mu}}\eta _{\hat{\mu}\hat{\nu}}p^{%
\hat{\nu}}$.

The fields $\Phi (x)$ have the same components they have in $d$ dimensions,
and each of them is a function of $\bar{x}$ and $\hat{x}$. For example,
spinors $\psi ^{\alpha }$ have $2^{[d/2]_{\text{int}}}$ components, where $%
[d/2]_{\text{int}}$ is the integral part of $d/2$, vectors have $d$
components $A_{\bar{\mu}}$, symmetric tensors with two indices have $%
d(d+1)/2 $ components, and so on. In particular, the metric tensor $g_{\mu
\nu }$ is made of the diagonal blocks $g_{\bar{\mu}\bar{\nu}}$ and $\eta _{%
\hat{\mu}\hat{\nu}}$, while the off-diagonal components $g_{\bar{\mu}\hat{\nu%
}}$ vanish.

The $\gamma $ matrices are the usual, $d$-dimensional ones, and satisfy the
Dirac algebra $\{\gamma ^{\bar{a}},\gamma ^{\bar{b}}\}=2\eta ^{\bar{a}\bar{b}%
}$. If $d=2k$ is even, the $d$-dimensional generalization of $\gamma _{5}$
is 
\begin{equation*}
\tilde{\gamma}=-i^{k+1}\gamma ^{0}\gamma ^{1}\cdots \gamma ^{2k-1},
\end{equation*}%
which satisfies $\tilde{\gamma}^{\dagger }=\tilde{\gamma}$, $\tilde{\gamma}%
^{2}=1$. Left and right projectors $P_{L}=(1-\tilde{\gamma})/2$, $P_{R}=(1+%
\tilde{\gamma})/2$ are defined as usual. The tensor $\varepsilon ^{\bar{a}%
_{1}\cdots \bar{a}_{d}}$ and the charge-conjugation matrix $\mathcal{C}$
also coincide with the usual ones. Full $SO(1,D-1)$ invariance is lost in
most expressions, replaced by $SO(1,d-1)\times SO(-\varepsilon )$ invariance.

We endow the fields with well-behaved propagators by adding suitable
higher-derivative evanescent kinetic terms to the action. We multiply them
by inverse powers of some mass $M$. For example, the regularized action of
(left-handed) chiral fermions in curved space reads 
\begin{equation*}
\int e\bar{\psi}_{L}ie_{\bar{a}}^{\bar{\mu}}\gamma ^{\bar{a}}\mathcal{D}_{%
\bar{\mu}}\psi _{L}+S_{\text{ev}\psi },
\end{equation*}%
where $\mathcal{D}_{\bar{\mu}}$ denotes the covariant derivative and 
\begin{equation}
S_{\text{ev}\psi }=\frac{i}{2M}\int e\left( \varsigma _{\psi }\psi _{L}^{T}%
\mathcal{\tilde{C}}\hat{\partial}^{2}\psi _{L}-\varsigma _{\psi }^{\ast }%
\bar{\psi}_{L}\mathcal{\tilde{C}}\hat{\partial}^{2}\bar{\psi}_{L}^{T}\right)
,  \label{w2}
\end{equation}%
while $e$ is the determinant of the vielbein $e_{\bar{\mu}}^{\bar{a}}$, $%
\varsigma _{\psi }$ are constants, and $\mathcal{\tilde{C}}$ coincides with
the matrix $\mathcal{C}$ of charge conjugation if $d=4$ mod 8; otherwise $%
\mathcal{\tilde{C}}=-i\gamma ^{0}\gamma ^{2}$ (in $d>2$).

In the case of Yang-Mills gauge fields in curved space, we choose the gauge
fermion%
\begin{equation*}
\Psi =\int \sqrt{|g|}\bar{C}^{a}\left( g^{\bar{\mu}\bar{\nu}}\partial _{\bar{%
\mu}}A_{\bar{\nu}}^{a}+\frac{\xi ^{\prime }}{2}B^{a}\right) .
\end{equation*}%
The regularized gauge-fixed action reads 
\begin{equation}
-\frac{1}{4}\int \sqrt{|g|}F_{\bar{\mu}\bar{\nu}}^{a}F^{\bar{\mu}\bar{\nu}%
a}+\int \sqrt{|g|}B^{a}\left( g^{\bar{\mu}\bar{\nu}}\partial _{\bar{\mu}}A_{%
\bar{\nu}}^{a}+\frac{\xi ^{\prime }}{2}B^{a}\right) -\int \sqrt{|g|}\bar{C}%
^{a}g^{\bar{\mu}\bar{\nu}}\partial _{\bar{\mu}}\mathcal{D}_{\bar{\nu}%
}C^{a}+S_{\text{ev}A}+S_{\text{ev}C},  \notag
\end{equation}%
where 
\begin{eqnarray}
S_{\text{ev}A} &=&\frac{1}{2}\int \sqrt{|g|}g^{\bar{\mu}\bar{\nu}}\left[ 
\frac{\varsigma _{A}}{M^{2}}(\hat{\partial}^{2}A_{\bar{\mu}}^{a})(\hat{%
\partial}^{2}A_{\bar{\nu}}^{a})-\frac{\eta _{A}}{M}g^{\bar{\mu}\bar{\nu}}(%
\hat{\partial}_{\hat{\rho}}A_{\bar{\mu}}^{a})(\hat{\partial}^{\hat{\rho}}A_{%
\bar{\nu}}^{a})\right] ,  \notag \\
S_{\text{ev}C} &=&-\int \sqrt{|g|}\left[ \frac{\varsigma _{C}}{M^{2}}(\hat{%
\partial}^{2}\bar{C}^{a})^{2}(\hat{\partial}^{2}C^{a})-\frac{\eta _{C}}{M}%
\int \sqrt{|g|}(\hat{\partial}_{\hat{\rho}}\bar{C}^{a})(\hat{\partial}^{\hat{%
\rho}}C^{a})\right] ,  \label{w3}
\end{eqnarray}%
while $\varsigma _{A}$, $\varsigma _{C}$, $\eta _{A}$, and $\eta _{C}$ are
constants. Quantum gravity can be dealt with in a similar fashion, both in
the metric tensor formalism and in the vielbein formalism \cite{chiraldimreg}%
.

Thanks to the higher-derivative evanescent kinetic terms introduced by the
CD regularization, the propagators of all the fields have denominators that
are equal to products of polynomials 
\begin{equation}
D(\bar{p},\hat{p},m,\varsigma ,\eta )=\bar{p}^{2}-m^{2}-\varsigma \frac{(%
\hat{p}^{2})^{2}}{M^{2}}+\eta \frac{\hat{p}^{2}}{M}+i0,  \label{denni}
\end{equation}%
where $\varsigma $ is a nonvanishing constant of order one and $\eta $ is
another constant. The propagators fall off in all directions $\bar{p}$, $%
\hat{p}$ for large momenta $p$. However, they decrease more rapidly or more
slowly depending on whether the evanescent or physical components, $\hat{p}$
or $\bar{p}$, of the momenta become large. The structure (\ref{denni})
suggests that $\bar{p}$ and $\hat{p}^{2}$ should be regarded as equally
important in the ultraviolet limit. The key point of the CD regularization
is to define \textquotedblleft weights\textquotedblright so that $\bar{p}$
and $\hat{p}^{2}$ are equally weighted, and use the weights to replace the
dimensions in units of mass that are normally used for power counting. Doing
so, we arrive at a \textit{weighted} power counting\textit{\ }\cite{halat},
which gives us an efficient control over the locality of counterterms when
the denominators of propagators are products of polynomials of the form (\ref%
{denni}).

Weights are defined in $D=d$, since the corrections of order $\varepsilon $
are not important for the weighted power counting. We conventionally take $%
\bar{p}$ to have weight 1, so the evanescent components $\hat{p}$ of momenta
have weight 1/2. Call the kinetic terms with the largest number of
derivatives $\bar{\partial}$ \textit{dominant} kinetic terms. Once they are
diagonalized, we write the dominant kinetic terms of the fields $\Phi $ as 
\begin{equation}
\frac{1}{2}\int \Phi \bar{\partial}^{N_{\Phi }}\Phi ,\qquad \text{or}\qquad
\int \bar{\Phi}\bar{\partial}^{N_{\Phi }}\Phi ,  \label{dom}
\end{equation}%
depending on the case. Clearly, the weight of $\Phi $ is equal to $%
(d-N_{\Phi })/2$ and coincides with its dimension in units of mass. Weights
can be unambiguously assigned to the parameters of the theory and the
sources $K$, by demanding that the action and the scale $M$ be weightless.

The $\Phi $ propagators are rational functions of the momenta, of the form 
\begin{equation}
\frac{P_{2w-N_{\Phi }}^{\prime }(\bar{p},\hat{p})}{P_{2w}(\bar{p},\hat{p})},
\label{propag}
\end{equation}%
where $P_{2w-N_{\Phi }}^{\prime }$ and $P_{2w}$ are $SO(-\varepsilon )$%
-scalar polynomials of weighted degrees $2w-N_{\Phi }$ and $2w$,
respectively, such that ($a$) $P_{2w}$ is a scalar under $SO(1,d-1)$, ($b$)
the parameters contained in $P_{2w}$ admit a nontrivial range of values
where $P_{2w}$ is positive definite in the Euclidean framework, and ($c$)
the monomials $(\bar{p}^{2})^{w}$ and $(\hat{p}^{2})^{2w}$ of $P_{2w}(\bar{p}%
,\hat{p})$ are multiplied by nonvanishing coefficients. The
\textquotedblleft weighted degree\textquotedblright\ of a $SO(-\varepsilon )$%
-scalar polynomial $Q(\bar{p},\hat{p})$ is its ordinary degree once $Q$ is
rewritten as a polynomial $\tilde{Q}(\bar{p},\hat{p}^{2})$ of $\bar{p}$ and $%
\hat{p}^{2}$.

The theories that contain only parameters of non-negative weights (and are
such that the propagators fall off with the correct behaviors in the
ultraviolet limit) are renormalizable by weighted power counting. The
theories that contain some parameters of strictly negative weights are
nonrenormalizable. In all cases, the propagators (\ref{propag}) must contain
only parameters of non-negative weights.

Weighted power counting also ensures that the scale $M$ does not propagate
into the physical sector of the theory. Precisely, $M$ is an arbitrary,
renormalization-group invariant parameter that belongs to the evanescent
sector of the theory from the beginning to the end, so there is no need to
take the limit $M\rightarrow \infty $ at any stage.

In ref. \cite{chiraldimreg} we showed that it is possible to find
appropriate higher-derivative evanescent kinetic terms for all most common
fields, such as scalars, fermions, Yang-Mills fields, gravity in the metric
formalism, gravity in the vielbein formalism, Chern-Simons fields, and so
on, and arrange the regularized action so that the requirements listed above
are fulfilled. The total action is the one that contains all monomials
compatible with weighted power counting, as well as the nonanomalous
symmetries of the theory, multiplied by the maximum number of independent
coefficients.

Some aspects of the CD\ regularization are reminiscent of Siegel's
dimensional reduction \cite{siegel1}, which is a popular modified
dimensional regularization taylored for supersymmetric theories. Among other
things, both techniques make use of the ordinary $d$-dimensional Dirac
algebra. However, in Siegel's approach it is necessary to think that $D$ is
\textquotedblleft smaller\textquotedblright\ than $d$. Then, it is possible
to define a $D$-dimensional gauge covariant derivative and build gauge
invariant schemes for gauge theories. Using the CD technique, on the other
hand, only the $d$-dimensional gauge covariant derivative is consistent.
Moreover, in Siegel's framework ordinary vectors and tensors are decomposed
into multiplets made of vectors/tensors and extra components that behave
like scalars (called $\varepsilon $-scalars). The latter are absent in the
CD\ regularization. Another aspect in common is the important role played by
the evanescent couplings, although they have different features in the two
cases. The dimensional reduction, in its original formulation, has
inconsistencies \cite{siegel2}, and the evanescent terms can be used to
overcome some of those, in both supersymmetric and nonsupersymmetric
theories \cite{jones}.

The CD technique has several advantages, which we now recall. In the
ordinary, as well as chiral dimensional regularization we can distinguish
divergent, nonevanescent and evanescent terms, depending on how they behave
in the limit\ $D\rightarrow d$. The nonevanescent\ terms are those that have
a regular limit for $D\rightarrow d$ and coincide with the value of that
limit. The evanescent terms are those that vanish when $D\rightarrow d$.
They\ can be of two types: \textit{formal} or \textit{analytic}. The
analytically evanescent terms are those that factorize at least one $%
\varepsilon $, such as $\varepsilon F_{\bar{\mu}\bar{\nu}}F^{\bar{\mu}\bar{%
\nu}}$, $\varepsilon \bar{\psi}_{L}ie_{\bar{a}}^{\bar{\mu}}\gamma ^{\bar{a}}%
\mathcal{D}_{\bar{\mu}}\psi _{L}$. The formally evanescent terms are those
that do not factorize powers of $\varepsilon $, such as $\psi _{L}^{T}\hat{%
\partial}^{2}\psi _{L}$. The divergences are poles in $\varepsilon $ and can
multiply either nonevanescent terms or formally evanescent terms. The former
are called \textit{nonevanescent divergences}. The latter are called \textit{%
evanescent divergences}, or \textit{divergent evanescences}, an example
being $\psi _{L}^{T}\hat{\partial}^{2}\psi _{L}/\varepsilon $. The divergent
evanescences must be subtracted away like any other divergences, because the
locality of counterterms is much clearer that way.

Using the ordinary dimensional regularization, the classification of
divergent evanescences in the nonrenormalizable sector presents several
problems \cite{ABrenoYMLR}. Consider the fermionic bilinears $\bar{\psi}%
_{1}\gamma ^{\rho _{1}\cdots \rho _{k}}\psi _{2}$, where $\gamma ^{\rho
_{1}\cdots \rho _{k}}$ denotes the completely antisymmetric product of $%
\gamma ^{\rho _{1}},\cdots ,\gamma ^{\rho _{k}}$. The independent bilinears
of this type are infinitely many, because they do not vanish for $k>d$.
Infinitely many Lagrangian terms of the same dimensions can be built with
them, such as the four fermion vertices $(\bar{\psi}_{1}\gamma ^{\rho
_{1}\cdots \rho _{k}}\psi _{2})(\bar{\psi}_{3}\gamma _{\rho _{1}\cdots \rho
_{k}}\psi _{4})$. The Fierz identities contain infinite sums and can be used
to relate certain divergent evanescences to finite terms, which makes the
classification of both ambiguous. No such problems are present using the CD
regularization, because the $\gamma $ matrices are just the ordinary $d$%
-dimensional ones.

Second,\ the CD\ technique simplifies the extraction of divergent parts out
of the antiparentheses of functionals, which is a key step in all
renormalization algorithms. We have to take some precautions to ensure that
this operation can safely cross the antiparentheses, so that, for example, $%
(S,X)_{\text{div}}=(S,X_{\text{div}})$. The first thing to do to achieve
this goal is define the tree-level action $S$ so that it does not contain
analytically evanescent terms, but only nonevanescent and formally
evanescent terms, multiplied by $\varepsilon $-independent coefficients. In
this way, $S$ does not contain dangerous factors of $\varepsilon $, which
could simplify the divergences of $X$ inside $(S,X)$. Moreover, the
antiparentheses cannot generate factors of $\varepsilon $. Indeed, since the 
$\gamma $ matrices are $d$ dimensional, and the fields $\Phi $ and the
sources $K$ only have $d$-dimensional components, the formally evanescent
quantities that we have are just $\eta ^{\hat{\mu}\hat{\nu}}$ and the
evanescent components $\hat{p}$ and $\hat{x}$ of momenta and coordinates.
These objects can generate factors of $\varepsilon $ only by means of the
contractions $\eta ^{\hat{\mu}\hat{\nu}}\eta _{\hat{\mu}\hat{\nu}%
}=-\varepsilon $, $\partial _{\hat{\mu}}x^{\hat{\mu}}=-\varepsilon $, $\hat{%
\partial}^{2}\hat{x}^{2}=-2\varepsilon $, etc. However, the functional
derivatives $\delta /\delta \Phi ^{\alpha }$ and $\delta /\delta K_{\alpha }$
due to the antiparentheses cannot generate $\eta ^{\hat{\mu}\hat{\nu}}\eta _{%
\hat{\mu}\hat{\nu}}$, because fields and sources have no evanescent
components. At the same time, the antiparentheses just multiply correlation
functions in momentum space, which are $SO(-\varepsilon )$-scalar, so they
cannot generate factors of $\varepsilon $, poles in $\varepsilon $ or
expressions such as $\partial _{\hat{\mu}}x^{\hat{\mu}}=-\varepsilon $, $%
\hat{\partial}^{2}\hat{x}^{2}=-2\varepsilon $, and cannot convert formal
evanescences into analytic ones. Ultimately, we can freely cross the sign of
antiparentheses, when we extract the divergent parts of local functionals
using the CD regularization.

Third, the CD regularization is compatible with invariance under rigid
diffeomorphisms, which are the $GL(d,\mathbb{R})$ coordinate transformations 
\begin{equation}
x^{\bar{\mu}\hspace{0.01in}\prime }=M_{\bar{\nu}}^{\bar{\mu}}x^{\bar{\nu}%
},\qquad x^{\hat{\mu}\hspace{0.01in}\prime }=x^{\hat{\mu}},  \label{genco}
\end{equation}%
where $M_{\bar{\nu}}^{\bar{\mu}}$ is an arbitrary invertible real constant
matrix. We can choose the tree-level action $S$ to be completely invariant
under this symmetry, even in the gauge-fixing and regularization sectors. To
fulfill this requirement, we write the fields $\Phi $ and the derivatives $%
\bar{\partial}$ using lower spacetime indices $\bar{\mu},\bar{\nu},\ldots $,
and the sources $K$ using upper spacetime indices. Then, we contract those
indices by means of the metric tensor $g_{\bar{\mu}\bar{\nu}}$, its inverse $%
g^{\bar{\mu}\bar{\nu}}$, or the Kronecker tensor $\delta _{\bar{\nu}}^{\bar{%
\mu}}$. Finally, we multiply by an appropriate power of $\sqrt{|g|}$, to
obtain a scalar density of weight 1, and integrate over spacetime. The
derivatives $\hat{\partial}$ must be contracted by means of $\eta ^{\hat{\mu}%
\hat{\nu}}$, to ensure $SO(-\varepsilon )$ invariance.

We formulate the theory without introducing \textquotedblleft second
metrics\textquotedblright\ $h_{\mu \nu }$, i.e. additional metrics besides
the metric tensor $g_{\bar{\mu}\bar{\nu}}$ and the background metric $g_{B%
\bar{\mu}\bar{\nu}}$ around which we expand $g_{\bar{\mu}\bar{\nu}}$
perturbatively. Since field translations leave the functional integral
invariant, the correlation functions are independent of $g_{B\bar{\mu}\bar{%
\nu}}$, so we do not consider $g_{B\bar{\mu}\bar{\nu}}$ a second metric.
However, the correlation functions may depend on true second metrics $h_{\mu
\nu }$, which may enter the classical action through the gauge fixing or the
regularization. Several common gauge-fixing functions $G(\phi )$, such as $%
\eta ^{\rho \nu }\partial _{\rho }g_{\mu \nu }$, do introduce a second
metric, which is often the flat-space metric $\eta _{\mu \nu }$.

When two independent metrics $g_{\mu \nu }$ and $h_{\mu \nu }$ are present,
the classifications of counterterms and contributions to anomalies are
plagued with unnecessary complications. For example, the divergent parts can
contain arbitrary dimensionless functions of $g_{\mu \nu }h^{\mu \nu }$, $%
g_{\mu \nu }h^{\nu \rho }g_{\rho \sigma }h^{\sigma \mu }$, and similar
contractions. If the theory contains a unique metric (and a unique
vielbein), these arbitrary functions do not appear.

In the approach of this paper, invariance under rigid diffeomorphisms is not
completely preserved. If the action $S$ is invariant, the $\Gamma $
functional is also invariant, as well as its divergent parts. However,
sometimes we need to express certain divergent terms $\Delta \Gamma _{\text{%
div}}$ or potentially anomalous terms $\mathcal{A}_{\text{pot}}$ in the form 
$(S,\chi )$, where $\chi \left( \Phi ,K\right) $ is a local functional. Even
when $\Delta \Gamma _{\text{div}}$ and $\mathcal{A}_{\text{pot}}$ are
invariant under rigid diffeomorphisms, $\chi $ may be noninvariant. The
divergent terms $\Delta \Gamma _{\text{div}}=(S,\chi )$ are iteratively
subtracted by means of canonical transformations generated by 
\begin{equation*}
F(\Phi ,K^{\prime })=\int \Phi ^{\alpha }K_{\alpha }^{\prime }-\chi (\Phi
,K^{\prime }).
\end{equation*}%
Instead, the potentially anomalous terms $\mathcal{A}_{\text{pot}}=(S,\chi )$
are subtracted by redefining the action $S$ as $S-\chi /2$. In these ways,
the violation of invariance under rigid diffeomorphisms can propagate into
the renormalized action $S_{R}$. When no second metrics are present, such a
violation is parametrized by multiplicative functions of the determinant $g$
of the metric tensor, which are relatively easy to handle.

To simplify various arguments, we assume that the gauge fermion $\Psi (\Phi
) $ is independent of the matter fields. For example, a good gauge fermion
for Yang-Mills symmetries, local Lorentz symmetry, and diffeomorphisms in
perturbatively unitary theories [where $N_{\phi _{g}}=2$ in formulas (\ref%
{domo}) and (\ref{choice})] is \cite{chiraldimreg} 
\begin{eqnarray}
\Psi (\Phi ) &=&\int \sqrt{|g|}\bar{C}^{a}\left( g^{\bar{\mu}\bar{\nu}%
}\partial _{\bar{\mu}}A_{\bar{\nu}}^{a}+\frac{\xi ^{\prime }}{2}B^{a}\right)
+\int e\bar{C}_{\bar{a}\bar{b}}\left( \frac{1}{\kappa }e^{\bar{\rho}\bar{a}%
}g^{\bar{\mu}\bar{\nu}}\partial _{\bar{\mu}}\partial _{\bar{\nu}}e_{\bar{\rho%
}}^{\bar{b}}+\frac{\xi _{L}}{2}B^{\bar{a}\bar{b}}+\frac{\xi _{L}^{\prime }}{2%
}g^{\bar{\mu}\bar{\nu}}\partial _{\bar{\mu}}\partial _{\bar{\nu}}B^{\bar{a}%
\bar{b}}\right)  \notag \\
&&{-\int }\sqrt{|g|}\bar{C}_{\bar{\mu}}\left( \frac{1}{\kappa }\partial _{%
\bar{\nu}}g^{\bar{\mu}\bar{\nu}}+\frac{\xi _{G}}{\kappa }g^{\bar{\mu}\bar{\nu%
}}g_{\bar{\rho}\bar{\sigma}}\partial _{\bar{\nu}}g^{\bar{\rho}\bar{\sigma}}-%
\frac{\xi _{G}^{\prime }}{2}g^{\bar{\mu}\bar{\nu}}B_{\bar{\nu}}\right) ,
\label{psi1}
\end{eqnarray}%
where the constants $\xi ^{\prime }$, $\xi _{L}$, $\xi _{L}^{\prime }$, $\xi
_{G}$, and $\xi _{G}^{\prime }$ are gauge-fixing parameters. We have
arranged $\Psi (\Phi )$ so that it is invariant under rigid diffeomorphisms.
The factors $1/\kappa $ are inserted to be consistent with the $\kappa $
structure (\ref{para}), explained in the next subsection, which becomes
manifest once we expand the vielbein around flat space and make the other
replacements of formula (\ref{repla}). The gauge fixing of local Lorentz
symmetry contained in (\ref{psi1}) takes inspiration from the less common
gauge condition $\partial ^{\mu }\omega _{\mu }^{ab}=0$, rather than the
more common condition of symmetric vielbein, because the latter is not
compatible with the requirement of having a unique metric. In
higher-derivative theories we choose a gauge fermion with a similar
structure, the only difference being that the gauge conditions $G(\phi ,\xi
) $ and the operators $P(\phi ,\xi ^{\prime },\partial )$ of formula (\ref%
{phi}) also include higher-derivative terms, to fulfill the conditions (\ref%
{choice}).

Finally, the CD technique preserves the good properties of the dimensional
regularization. The most important ones are that ($a$) the
Batalin-Vilkovisky master equation is simply $(S,S)=0$ in $D=d$ (a
correction appears on the right-hand side in most nondimensional
regularizations), and ($b$) the local perturbative changes of field
variables have Jacobian determinants identically equal to one. Property ($b$%
) follows from the fact that the integrals of polynomials $P(p)$ of the
momenta in $\mathrm{d}^{D}p$ vanish.

Summarizing, when the gauge algebra closes off shell, the CD regularized
action has the form 
\begin{equation}
S(\Phi ,K)=S_{c}(\phi )+(S_{K},\Psi )+S_{K}+S_{\text{ev}}=S_{d}+S_{\text{ev}%
}=\bar{S}_{d}+(S_{K},\Psi )+S_{\text{ev}}\text{,}  \label{sfk}
\end{equation}%
where $S_{c}(\phi )$ is given by (\ref{scf}) and the evanescent part $S_{%
\text{ev}}$ collects the evanescent terms required by the CD\
regularization, such as $S_{\text{ev}\psi }$, $S_{\text{ev}A}$, and $S_{%
\text{ev}C}$ of (\ref{w2}) and (\ref{w3}). For the reasons explained above,
we assume that $S_{d}$ is nonevanescent and $S_{\text{ev}}$ is formally
evanescent, so $S$ does not contain any analytically evanescent terms.
Moreover, the action (\ref{sfk}) does not contain second metrics and is
invariant under $SO(-\varepsilon )$ and the other global nonanomalous
symmetries of the theory. We do not require that $S_{\text{ev}}$ be
invariant under rigid diffeomorphisms, but just that it be built with a
unique metric tensor or vielbein. We denote the parameters contained in $S_{%
\text{ev}}$ by $\varsigma _{I}$ and $\eta _{I}$, where $\varsigma _{I}$
multiply the dominant evanescent kinetic terms, and $\eta _{I}$ multiply the
other terms, as shown by formulas (\ref{w3}) and (\ref{denni}). For
convenience, we assume that $S_{\text{ev}}$ depends linearly on $\varsigma $
and $\eta $, and vanishes for $\varsigma =\eta =0$. We extend $S_{\text{ev}}$
till it includes all the evanescent terms allowed by weighted power
counting, constructed with the fields $\Phi $, the sources $K_{\phi }$ and $%
K_{C}$, and their derivatives, multiplied by the maximum number of
independent parameters $\varsigma $ and $\eta $. This will allow us to
renormalize the divergent evanescences by means of $\varsigma $ and $\eta $
redefinitions. It is consistent to choose $S_{\text{ev}}$ independent of the
sources $K_{\bar{C}}$ and $K_{B}$. Indeed, if we do so, the action $S$ does
not contain $K_{B}$ and depends on $K_{\bar{C}}$ only through the last three
terms of (\ref{skexpl}). Then, $K_{\bar{C}}$ and $K_{B}$ cannot contribute
to nontrivial diagrams, so the counterterms are also independent of them.

In total, we have physical parameters $\lambda $, contained in $S_{c}$,
gauge-fixing parameters $\xi $, contained in $\Psi $, and regularizing
parameters $\varsigma $ and $\eta $, contained in $S_{\text{ev}}$. The
action (\ref{sfk}) is also written as $S(\Phi ,K,\lambda ,\xi ,\varsigma
,\eta )$.

Clearly, the CD regularized action $S=S_{d}+S_{\text{ev}}$ satisfies the 
\textit{deformed master equation} 
\begin{equation}
(S,S)=\mathcal{\hat{O}}(\varepsilon ),  \label{masterd}
\end{equation}%
where \textquotedblleft $\mathcal{\hat{O}}(\varepsilon )$\textquotedblright\
denotes formally evanescent local terms. The right-hand side is the source
of potential anomalies.

Given a regularized classical action $S(\Phi ,K)$, the regularized
generating functionals $Z$ and $W$ are given by 
\begin{equation}
Z(J,K)=\int [\mathrm{d}\Phi ]\exp \left( iS(\Phi ,K)+i\int \Phi ^{\alpha
}J_{\alpha }\right) =\exp iW(J,K).  \label{zg}
\end{equation}%
The Legendre transform $\Gamma (\Phi ,K)=W(J,K)-\int \Phi ^{\alpha
}J_{\alpha }$ of $W(J,K)$ with respect to $J$ is the generating functional
of one-particle irreducible diagrams. The anomaly functional is 
\begin{equation}
\mathcal{A}=(\Gamma ,\Gamma )=\langle (S,S)\rangle _{S},  \label{anom}
\end{equation}%
where $\langle \cdots \rangle _{S}$ denotes the average defined by the
action $S$ at arbitrary sources $J$ and $K$. A quick way to prove the last
equality of (\ref{anom}) is to make the change of field variables $\Phi
^{\alpha }\rightarrow \Phi ^{\alpha }+\bar{\theta}(S,\Phi ^{\alpha })$
inside $Z(J,K)$, where $\bar{\theta}$ is a constant anticommuting parameter.
For details, see for example the appendixes of \cite{back,ABrenoYMLR}.

\subsection{Truncation}

\label{s21}

When we quantize a nonrenormalizable theory, or study composite fields of
high dimensions in any kind of theory, it may be convenient to truncate the
tree-level action $S_{d}$ in some way. For the arguments of this paper, the
truncation is necessary to define a suitable higher-derivative
regularization. Indeed, to make the HD\ theory super-renormalizable at fixed 
$\Lambda $, the higher-derivative regularizing terms must be placed well
beyond the truncation.

Denote the gauge coupling of minimum dimension with $\kappa $. If there are
more than one gauge coupling of minimum dimension we call one of them $%
\kappa $ and write any other as $r\kappa $, where the dimensionless ratio $r$
is treated as a parameter of order one. The other gauge couplings $g$ are
written as $g=r_{+}\kappa $, where the ratios $r_{+}$ have positive
dimensions and are also of order one. We parametrize the non-gauge-fixed
solution $\bar{S}_{d}(\Phi ,K,\kappa ,\zeta )$ of the master equation as 
\begin{equation*}
\bar{S}_{d}(\Phi ,K,\kappa ,\zeta )=\frac{1}{\kappa ^{2}}\bar{S}_{d}^{\prime
}(\kappa \Phi ,\kappa K,\zeta ),
\end{equation*}%
where $\zeta $ are any other parameters besides $\kappa $, including $r$ and 
$r_{+}$, and $\bar{S}_{d}^{\prime }$ is analytic in $\zeta $. We assume that
each field $\Phi $ has a dominant kinetic term (\ref{dom}) normalized to one
or multiplied by a dimensionless parameter of order one.

The gauge fixing must be parametrized similarly. We choose a gauge fermion $%
\Psi $ of the form 
\begin{equation*}
\Psi (\Phi ,\kappa ,\xi )=\frac{1}{\kappa ^{2}}\Psi ^{\prime }(\kappa \Phi
,\xi ),
\end{equation*}%
where $\xi $ are the gauge-fixing parameters and $\Psi ^{\prime }$ depends
analytically on $\xi $. We know that if the gauge algebra closes off shell,
we can choose an $\bar{S}_{d}$ that is linear in $K$, as in formula (\ref{sk}%
). Then, the gauge-fixed solution $S_{d}=\bar{S}_{d}+(\bar{S}_{d},\Psi )$ of
the master equation has the structure 
\begin{equation}
S_{d}(\Phi ,K,\kappa ,\zeta ,\xi )=\frac{1}{\kappa ^{2}}S_{d}^{\prime
}(\kappa \Phi ,\kappa K,\zeta ,\xi ).  \label{para}
\end{equation}%
We parametrize the evanescent sector $S_{\text{ev}}$ in the same way and
define the parameters $\varsigma ,\eta $ so that 
\begin{equation}
S_{\text{ev}}(\Phi ,K,\kappa ,\varsigma ,\eta )=\frac{1}{\kappa ^{2}}S_{%
\text{ev}}^{\prime }(\kappa \Phi ,\kappa K,\varsigma ,\eta ).  \label{para2}
\end{equation}

In the end, the total action $S$, and all the tree-level functionals we work
with, have the $\kappa $ structure 
\begin{equation}
X_{\text{tree}}(\Phi ,K,\kappa )=\frac{1}{\kappa ^{2}}X_{\text{tree}%
}^{\prime }(\kappa \Phi ,\kappa K).  \label{liable}
\end{equation}%
Then, it is easy to prove that every loop carries an additional factor $%
\kappa ^{2}$. Therefore, the renormalized action, the $\Gamma $ functional,
and the renormalized $\Gamma $ functional have the $\kappa $ structure 
\begin{equation}
X(\Phi ,K,\kappa )=\sum_{L\geqslant 0}\kappa ^{2(L-1)}X_{L}^{\prime }(\kappa
\Phi ,\kappa K),  \label{liable2}
\end{equation}%
where $X_{L}$ collects the $L$-loop contributions.

The $\kappa $ structures (\ref{liable}) and (\ref{liable2})\ are preserved
by the antiparentheses: if two functionals $X(\Phi ,K,\kappa )$ and $Y(\Phi
,K,\kappa )$ satisfy (\ref{liable}), or (\ref{liable2}), then the functional 
$(X,Y)$ satisfies (\ref{liable}), or (\ref{liable2}), respectively.

In perturbatively unitary theories, the propagating fields have standard
dimensions in units of mass (because $N_{\Phi }=2$ and $N_{\Phi }=1$ for
bosons and fermions, respectively). When the theory is not perturbatively
unitary, such as higher-derivative quantum gravity \cite{stelle}, fields of
negative or vanishing dimensions may be present. This is not a problem, as
long as the tree-level action has the structure \ (\ref{para}) and the other
assumptions we make are fulfilled.

In the presence of gravity, the square root $\kappa _{N}$ of Newton's
constant is equal to $\kappa $ times a ratio of non-negative dimension. The $%
\kappa $ structure of the action becomes explicit when we expand around a
background metric or vielbein. We also need to rescale the ghosts and the
sources associated with diffeomorphisms and local Lorentz symmetry. For
simplicity, we expand around flat space, although flat space may not be a
solution of the classical field equations, because the renormalization of
the theory and its anomalies do not depend on the background we choose. In
that case, we can make the $\kappa $ structures (\ref{para}), (\ref{liable}%
), and (\ref{liable2}) explicit by means of the canonical transformation 
\begin{eqnarray}
e_{\bar{\mu}}^{\bar{a}} &\rightarrow &\delta _{\bar{\mu}}^{\bar{a}}+\kappa
_{N}\phi _{\bar{\mu}}^{\bar{a}},\qquad C^{\bar{\rho}}\rightarrow \kappa
_{N}C^{\bar{\rho}},\qquad C^{\bar{a}\bar{b}}\rightarrow \kappa _{N}C^{\bar{a}%
\bar{b}},  \notag \\
K_{\bar{a}}^{\bar{\mu}} &\rightarrow &\frac{1}{\kappa _{N}}K_{\bar{a}}^{\bar{%
\mu}},\qquad K_{\bar{\mu}}^{C}\rightarrow \frac{1}{\kappa _{N}}K_{\bar{\mu}%
}^{C},\qquad K_{\bar{a}\bar{b}}^{C}\rightarrow \frac{1}{\kappa _{N}}K_{\bar{a%
}\bar{b}}^{C}.  \label{repla}
\end{eqnarray}%
Check this fact in formulas (\ref{skexpl}) and (\ref{psi1}). Whenever we
speak of $\kappa $ structures we understand the replacements (\ref{repla}),
although we do not make them explicit all the time.

Now we define the truncation. We organize the set of parameters $\zeta ,\xi
,\varsigma ,\eta $ into two subsets $\bar{s}$ and $s_{-}$. The subset $\bar{s%
}$ contains the parameters of positive dimensions, as well as those of
vanishing dimensions that are not treated perturbatively. Examples are the
parameters that appear in the propagators. The parameters $r$ and $r_{+}$
(but not $\kappa $) are also included in the set $\bar{s}$, because they are
considered of order one. The set $\bar{s}$ also includes the parameters that
cure infrared problems when super-renormalizable interactions are present.
Examples are the masses, the cosmological constant $\Lambda _{\text{c}}$ of
formula (\ref{basi}) and the Chern-Simons coupling in three dimensions. If $%
\kappa $ has a negative dimension (such as the square root of Newton's
constant in Einstein gravity), the set $\bar{s}$ also includes the
parameters $\zeta ,\xi $ that multiply the power counting renormalizable
vertices. An example is the constant $\lambda _{4}^{\prime }=\lambda
_{4}/\kappa ^{2}$ that appears when the four-scalar vertex $\lambda
_{4}\varphi ^{4}$ is written as $\lambda _{4}^{\prime }(\kappa \varphi
)^{4}/\kappa ^{2}$ in the four-dimensional $\varphi ^{4}$-theory coupled to
Einstein gravity.\ If $[\kappa ]=0$, the parameters such as $\lambda
_{4}^{\prime }$ can be assumed to be of order one and also included in $\bar{%
s}$. We express each parameter contained in $\bar{s}$ as a dimensionless
constant of order one times $m^{\Delta }$, where $\Delta $ is a non-negative
number and $m$ is a generic mass scale.

The subset $s_{-}$ contains the parameters $\zeta ,\xi ,\varsigma ,\eta $ of
negative dimensions. We write them as dimensionless constants of order one
times $\Lambda _{-}^{-\Delta _{-}}$, where $\Lambda _{-}$ is some energy
scale and $\Delta _{-}$ is a positive number. The subset $s_{-}$ includes
the coefficients of the quadratic terms $\sim \Phi \partial ^{N_{\Phi
}^{\prime }}\Phi $ with $N_{\Phi }^{\prime }>N_{\Phi }$, which have to be
treated perturbatively, since the dominant quadratic terms we perturb around
are (\ref{dom}). Observe that $\kappa $ is not included in the set $s_{-}$,
even if it may have a negative dimension.

The Feynman diagrams are multiplied by various factors, but their core
integrals depend only on the parameters of the subset $\bar{s}$ and the
external momenta. Therefore, if we assume that $m$ and the overall energy $E$
are of the same order, each field $\Phi $ of dimension $d_{\Phi }$
contributes to the amplitudes as a power $\sim E^{d_{\Phi }}\sim m^{d_{\Phi
}}$.

We assume that there exists a range of energies $E$ such that 
\begin{equation}
m\sim E\ll \Lambda _{-},  \label{range}
\end{equation}%
and that $\kappa $ is small enough; that is to say, 
\begin{equation}
\kappa \Lambda _{-}^{-[\kappa ]}\ll 1,\qquad \kappa E^{-[\kappa ]}\ll 1.
\label{range2}
\end{equation}%
If $[\kappa ]<0$, the first of these conditions, combined with (\ref{range}%
), implies the second one. If $[\kappa ]>0$, the second condition implies
the first one. If $[\kappa ]=0$, the two conditions obviously coincide.

It is easy to show that the conditions (\ref{range}) and (\ref{range2}) are
sufficient to have a well-defined perturbative expansion. Consider the
contributions to the action $S$ and the logarithmic divergences. Factorizing
the parameters in front of a generic local Lagrangian term $V(\partial ,\Phi
,K)$, we find the structure%
\begin{equation*}
\frac{\kappa ^{a}m^{c}}{\Lambda _{-}^{b}}\left( 1+\cdots +\frac{\kappa
^{a^{\prime }}m^{c^{\prime }}}{\Lambda _{-}^{b^{\prime }}}+\cdots \right)
\int V(\partial ,\Phi ,K)
\end{equation*}%
where the first factor is the tree-level coefficient and the ratio inside
the parentheses is a generic contribution coming from the divergent parts of
Feynman diagrams. We have $a\geqslant -1$ \footnote{%
According to the $\kappa $ structure (\ref{liable}), the terms with $a=-1$
are linear in the fields $\Phi $ or the sources $K$. Such terms may be
present when we expand around a configuration that is not a minimum of the
action (for example when we expand the metric tensor around flat space in
the presence of a cosmological term). All other terms have $a\geqslant 0$.}, 
$b\geqslant 0$, $c\geqslant 0$, $a[\kappa ]+c=b$, $a^{\prime }>0$, and $%
a^{\prime }[\kappa ]+c^{\prime }=b^{\prime }$. The tree-level vertices have
either $b=0$ or $c=0$. Then, $b^{\prime }\geqslant 0$ or $c^{\prime
}\geqslant 0$, respectively, so we can write 
\begin{equation}
\frac{\kappa ^{a^{\prime }}m^{c^{\prime }}}{\Lambda _{-}^{b^{\prime }}}%
=\left( \kappa m^{-[\kappa ]}\right) ^{a^{\prime }}\left( \frac{m}{\Lambda
_{-}}\right) ^{b^{\prime }}\ll 1,\qquad \text{or\qquad }\frac{\kappa
^{a^{\prime }}m^{c^{\prime }}}{\Lambda _{-}^{b^{\prime }}}=\left( \kappa
\Lambda _{-}^{-[\kappa ]}\right) ^{a^{\prime }}\left( \frac{m}{\Lambda _{-}}%
\right) ^{c^{\prime }}\ll 1,  \label{cases}
\end{equation}%
which shows that the expansion does work.

Next, consider the finite contributions to the $\Gamma $ functional. They
have the form%
\begin{equation}
\sim \frac{\kappa ^{a}E^{b-a[\kappa ]}}{\Lambda _{-}^{b}}=\left( \kappa
E^{-[\kappa ]}\right) ^{a}\left( \frac{E}{\Lambda _{-}}\right) ^{b},
\label{forma}
\end{equation}%
where $a\geqslant -1$ and $b\geqslant 0$. The power of $E$ can be arbitrary
and comes from the fields $\Phi $, the sources $K$, the powers of $m\sim E$,
and the evaluations of the core integrals of the Feynman diagrams. Clearly,
formula (\ref{forma}) shows that the expansion works. It also ensures that a
finite number of diagrams can contribute for each $a$ and $b$. Indeed, by
formula (\ref{liable2}) $a$ bounds the number of loops. Moreover, we can use
only a finite number of vertices, because the power of $\kappa $ bounds the
numbers of $\Phi $ and $K$ legs, while the power of $1/\Lambda _{-}$ bounds
the number of derivatives.

It should be noticed that assumptions (\ref{range}) and (\ref{range2}) are
merely tools to organize the perturbative expansion and the proof of the
Adler-Bardeen theorem. They ensure that we can reach all types of
contributions (vertices, diagrams, counterterms, potential anomalies, etc.),
working with finitely many of them at a time. They are not crucial for the
validity of the proof itself. What we mean is that the proof of the theorem
also holds when assumptions (\ref{range}) and (\ref{range2}) are not valid,
and the perturbative expansion is organized in a different way.

Now we define the truncation $T$ of the theory. We divide it into two
prescriptions, (T1) and (T2), which play different roles.

(T1) We switch off the $o(1/\Lambda _{-}^{T})$ terms of the action $%
S=S_{d}+S_{\text{ev}}$. All the terms of $S_{c}$ and $S_{\text{ev}}$ that
are not $o(1/\Lambda _{-}^{T})$ and satisfy the other assumptions of this
paper are kept and multiplied by the maximum number of independent
parameters.

In subsection \ref{s23} we explain that this prescription is also sufficient
to truncate the action $S_{\Lambda }=S+S_{\mathrm{HD}}$ of the HD theory,
because the higher-derivative terms $S_{\mathrm{HD}}$ can be chosen to be $%
\Lambda _{-}$ independent. We can also take a $\Lambda _{-}$-independent
gauge fermion $\Psi $. The actions determined by the truncation T1 are
denoted by $S_{cT}$, $\bar{S}_{dT}$, $S_{dT}$, $S_{T}$, $S_{\Lambda T}$, and
so on.

Note that the prescription T1 just switches off portions of $S$, but leaves
arbitrary powers of $1/\Lambda _{-}$ in the radiative corrections. This is
sufficient to renormalize the HD theory, at $\Lambda $ fixed, and prove that
it satisfies the manifest Adler-Bardeen theorem.

(T2) For $[\kappa ]<0$, define $\sigma =-[\kappa ]$ and 
\begin{equation}
\bar{\ell}=\left[ \frac{T}{2\sigma }\right] _{\text{int}},  \label{lmax}
\end{equation}%
$[\ldots ]_{\text{int}}$ denoting the integral part. For $[\kappa ]\geqslant
0$, define $\sigma =0$, $\bar{\ell}=\infty $ . We define the truncation T2
as the truncation that keeps the $\ell $-loop contributions up to $%
o(1/\Lambda _{-}^{T-2\ell \sigma })$, for $0\leqslant \ell \leqslant \bar{%
\ell}$, and neglects the rest.

The truncation T2 is useful for the second part of the proof, when we study
the limit $\Lambda \rightarrow \infty $ on the HD theory, renormalize the $%
\Lambda $ divergences and prove that the final theory satisfies the almost
manifest Adler-Bardeen theorem. Indeed, these results are all proved within
the truncation T2. This fact illustrates the meaning of the almost manifest
Adler-Bardeen theorem, i.e. statement \ref{bardo3} of the introduction.

Both prescriptions T1 and T2 are gauge invariant at $\varepsilon =0$, since
the gauge symmetries do not involve $\Lambda _{-}$. In power-counting
renormalizable theories with $[\kappa ]=0$ we have $T=0$.

If $[\kappa ]<0$, the quantity $\sigma $ is strictly positive, so the
prescription T2 reduces the powers of $1/\Lambda _{-}$ when the number of
loops increases. The area that is covered by the truncation forms a triangle
in the plane with axes $T$ and $L$. In particular, the truncation only
contains a finite number of loops, up to and including $\bar{\ell}$.

Note that we do not truncate the powers of $\kappa $. If we did, we would
explicitly break the gauge invariant terms into gauge noninvariant pieces.
For various arguments of the proof, it is convenient to define a truncation
that is gauge invariant at $\varepsilon =0$. Nevertheless, at the practical
level, a sort of truncation on the powers of $\kappa $ is implicitly
contained in the conditions (\ref{range2}), because they imply that the
contributions carrying sufficiently large powers of $\kappa $ are smaller
than certain contributions neglected by the truncation. We keep the higher
powers of $\kappa $ anyway, because we want to concentrate on the potential
anomalies that may break gauge invariance dynamically, so it is not wise to
break gauge invariance artificially at the same time.

The reason why we adopt the prescription T2, when we renormalize the final
theory, can be understood as follows. Consider an invariant $\mathcal{G}%
(\kappa \phi )$, equal to the integral of a local function of dimension $d_{%
\mathcal{G}}$. By power counting and formula (\ref{liable2}), at $L$ loops $%
\mathcal{G}$ may appear as a counterterm with the structure%
\begin{equation}
\frac{(\kappa ^{2})^{L}m^{p}\Lambda ^{q}}{\kappa ^{2}\Lambda _{-}^{\Delta
+2L[\kappa ]}}(\ln \Lambda )^{q^{\prime }}\mathcal{G}(\kappa \phi ),
\label{conto}
\end{equation}%
times a product of dimensionless constants, where $\Delta =p+q+d_{\mathcal{G}%
}-d-2[\kappa ]$ and $q,q^{\prime }\geqslant 0$. If the counterterm (\ref%
{conto}) is contained within the truncation, prescription T2 tells us that%
\begin{equation}
\Delta +2L[\kappa ]\leqslant T-2L\sigma .  \label{ina}
\end{equation}%
Then we also have the inequality $\Delta \leqslant T$. This ensures that the
truncated classical action $S_{cT}$, which obeys T1, also contains the
invariant $\mathcal{G}$. There, it appears with one of the structures%
\begin{equation}
\frac{\zeta }{\kappa ^{2}\Lambda _{-}^{\Delta -p-q}}\mathcal{G}(\kappa \phi
),\qquad \frac{\zeta m^{p+q-\Delta }}{\kappa ^{2}}\mathcal{G}(\kappa \phi ),
\label{custom}
\end{equation}%
depending on whether $\Delta >p+q$ or $\Delta \leqslant p+q$, where $\zeta $
is a dimensionless constant. In the end, a divergence of the form (\ref%
{conto}) can be subtracted by redefining $\zeta $. If we replaced (\ref{ina}%
) by a different prescription, i.e. $\Delta +2L[\kappa ]\leqslant T$, we
could be unable to subtract the counterterms (\ref{conto}) by redefining the
parameters of $S_{cT}$, for $[\kappa ]<0$.

The same argument applies to the counterterms that depend on both $\kappa
\Phi $ and $\kappa K$ and fall within the truncation. In particular, thanks
to the prescriptions T1 and T2, the counterterms that are formally
evanescent can be subtracted by redefining the parameters $\varsigma $ and $%
\eta $ of $S_{\text{ev}T}$. The counterterms that fall within the truncation
but do not belong to either this class or the class (\ref{conto}) will be
subtracted by means of canonical transformations.

For example, in pure quantum gravity ($[\kappa ]=-1$) we have the
counterterms%
\begin{equation}
\int \sqrt{|g|}R^{2},\qquad \int \sqrt{|g|}R_{\bar{\mu}\bar{\nu}}R^{\bar{\mu}%
\bar{\nu}},  \label{squa}
\end{equation}%
at one loop, which are $\Lambda _{-}$ independent and have $\Delta =2$. The
minimal truncation containing them is the one that neglects $o(1/\Lambda
_{-}^{0})$ at one loop, which means $T+2[\kappa ]=0$, i.e. $T=2$. At the
tree level, the same terms appear as%
\begin{equation}
\frac{\zeta _{1}}{\kappa ^{2}\Lambda _{-}^{2}}\int \sqrt{|g|}R^{2},\qquad 
\frac{\zeta _{2}}{\kappa ^{2}\Lambda _{-}^{2}}\int \sqrt{|g|}R_{\bar{\mu}%
\bar{\nu}}R^{\bar{\mu}\bar{\nu}},  \label{squa0}
\end{equation}%
where $\zeta _{1,2}$ are dimensionless constants. Thus, if we truncated the
powers of $\Lambda _{-}$ by neglecting $o(1/\Lambda _{-}^{0})$ at the tree
level, the truncated classical action $S_{cT}$ would not contain the terms (%
\ref{squa0}), and we would not be able to subtract the divergences (\ref%
{squa}) by redefining appropriate parameters.

Now we discuss the truncated actions. We have $\bar{S}_{dT}=S_{cT}+S_{K}$, $%
S_{dT}=\bar{S}_{dT}+(S_{K},\Psi )$, where, as anticipated before, we assume
that $\Psi $ is $\Lambda _{-}$ independent. Since the truncation does not
conflict with the gauge symmetries, $S_{dT}$ and $\bar{S}_{dT}$ satisfy the
master equations $(S_{dT},S_{dT})=(\bar{S}_{dT},\bar{S}_{dT})=0$. Observe
that, by prescription T1, $S_{dT}$ does not contain any invariants $\mathcal{%
G}_{i}$ that fall beyond the truncation. We stress that, at the tree level,
it is not enough to neglect those invariants: we must really switch them
off. Indeed, if they were present, we would be unable to properly HD
regularize the truncated theory. On the other hand, all the invariants $%
\mathcal{G}_{i}$ that are multiplied by powers $1/\Lambda _{-}^{t}$ with $%
t\leqslant T$ and satisfy the other assumptions of this paper [check, in
particular, (II-$i$)-(II-$iv$) right below] must be contained in $S_{dT}$,
multiplied by independent parameters, since we want to renormalize the
divergences proportional to $\mathcal{G}_{i}$ that fall within the
truncation by redefining those parameters. The evanescent part $S_{\text{ev}%
} $ of the action $S$ is truncated according to the same rules. In
particular, the $o(1/\Lambda _{-}^{T})$ monomials of $S_{\text{ev}T}$ must
also be switched off and all the monomials of $S_{\text{ev}}$ that are not $%
o(1/\Lambda _{-}^{T})$ must be contained in $S_{\text{ev}T}$, multiplied by
independent parameters.

In the end, the truncated version of the action $S$ is 
\begin{equation}
S_{T}(\Phi ,K)=S_{cT}(\phi )+(S_{K},\Psi )+S_{K}+S_{\text{ev}T}=S_{dT}+S_{%
\text{ev}T}  \label{st}
\end{equation}%
and satisfies the master equation up to evanescent terms: $(S_{T},S_{T})=%
\mathcal{\hat{O}}(\varepsilon )$.

In general, the number of terms contained in the truncation\ may be
infinite, because there can be fields $\Phi $ with $[\kappa \Phi ]=0$, or,
as far as we know now, even fields with $[\kappa \Phi ]<0$. Now we make some
assumptions that give us relative control on the power counting.

(II) We assume that

($i$) $[\kappa \Phi ]\geqslant 0$ for every $\Phi $;

($ii$) there exists at least one field with $N_{\Phi }\geqslant 1$;

($iii$) every field $\Phi $ with $[\kappa \Phi ]=0$ has $N_{\Phi }\geqslant
2 $;

($iv$) the fields with $N_{\Phi }=0$ are just the Lagrange multipliers $B$
for the gauge fixing.

\noindent The integers $N_{\Phi }$ are those defined by formula (\ref{dom}).

Clearly, the standard model coupled to quantum gravity, as well as most of
its extensions, satisfies these assumptions, with the gauge fermion (\ref%
{psi1}). Assumption (II-$i$) excludes, for example, four-dimensional
higher-derivative Yang-Mills theory coupled to Einstein gravity, because in
that case $[A]\leqslant 0$ and $[\kappa ]=-1$. Assumption (II-$ii$) just
excludes nonpropagating theories.

Assumptions (II-$ii$) and (II-$iii$) allow us to prove that the sources $%
K_{\Phi }$ satisfy $[\kappa K_{\Phi }]\geqslant N_{\Phi }/2$. Indeed, we
know that 
\begin{equation}
\lbrack \Phi ]=\frac{d-N_{\Phi }}{2},\qquad \lbrack K_{\Phi }]=\frac{%
d+N_{\Phi }}{2}-1,  \label{fk}
\end{equation}%
because $[R^{\alpha }]=[\Phi ^{\alpha }]+1$, while the form of $S_{K}$
ensures that $[\Phi ^{\alpha }]+[K^{\alpha }]=d-1$. Now, if there exists a
field $\bar{\Phi}$ with $[\kappa \bar{\Phi}]=0$, we have $d=2[\bar{\Phi}]+N_{%
\bar{\Phi}}=-2[\kappa ]+N_{\bar{\Phi}}\geqslant 2-2[\kappa ]$, which implies 
$[\kappa ]\geqslant 1-(d/2)$ and $[\kappa K_{\Phi }]\geqslant N_{\Phi }/2$
for every $\Phi $. If all fields satisfy $[\kappa \Phi ]>0$, we have $%
d>N_{\Phi }-2[\kappa ]$, which implies $[\kappa ]>(N_{\Phi }-d)/2$ for every 
$\Phi $. Since there must be at least a $\Phi $ with $N_{\Phi }\geqslant 1$,
we conclude that $[\kappa ]>(1-d)/2$ and $[\kappa K_{\Phi }]>(N_{\Phi }-1)/2$
for every $\Phi $. If $g$ denotes the gauge coupling associated with the
gauge field $\phi _{g}$ [which is the fluctuation $\phi _{\bar{\mu}}^{\bar{a}%
}$ of formula (\ref{repla}) in the case of gravity], and $s_{g}$ denotes the
spin of $\phi _{g}$, we have $[g\phi _{g}]=2-s_{g}$, which is integer or
semi-integer. Since $[\Phi ]$ and $[K_{\Phi }]$ are also integer or
semi-integer, so is $[g]$, as well as $[\kappa ]$, $[\kappa \Phi ]$ and $%
[\kappa K_{\Phi }]$. Then, the inequality $[\kappa K_{\Phi }]>(N_{\Phi
}-1)/2 $ gives $[\kappa K_{\Phi }]\geqslant N_{\Phi }/2$.

We have already remarked that the sources $K_{B}$ and $K_{\bar{C}}$ do not
contribute to nontrivial one-particle irreducible diagrams. Thus, assumption
(II-$iv$) ensures that all sources that contribute to nontrivial diagrams
satisfy the stronger inequality $[\kappa K_{\Phi }]\geqslant 1/2$.

It is easy to check that the relations $[\kappa _{N}\phi _{\bar{\mu}}^{\bar{a%
}}]=0$, $[gA_{\bar{\mu}}]=1$, $[\kappa \phi _{\bar{\mu}}^{\bar{a}}]\geqslant
0$, $[\kappa A_{\bar{\mu}}]\geqslant 0$ and formula (\ref{fk}) imply $%
[g]\geqslant \lbrack \kappa _{N}]$ and $N_{A}\leqslant N_{\phi }\leqslant
N_{A}+2$, where $N_{\phi }$ and $N_{A}$ are the numbers of $\bar{\partial}$
derivatives of the dominant kinetic terms (\ref{dom}) of the graviton field $%
\phi _{\bar{\mu}}^{\bar{a}}$ and the Yang-Mills gauge fields $A_{\bar{\mu}}$%
, respectively. Thus, in the presence of gravity the square root $\kappa
_{N} $ of Newton's constant is always a gauge coupling of minimum dimension,
and we can take $\kappa =\kappa _{N}$.

Note that the remarks made after formula (\ref{fk}) ensure that the powers
of $1/\Lambda _{-}$ appearing in the action are also integer or semi-integer.

\subsection{Key assumptions}

\label{key}

Now we formulate the key assumptions that allow us to characterize the
counterterms and ensure the triviality of the one-loop gauge anomalies. The
action obtained from $S_{d}$ by switching off all parameters $\zeta $ that
belong to the subset $s_{-}$ is called \textit{basic action} and is denoted
by $S_{d\text{b}}$. The basic action can also be formally obtained from $%
S_{dT}$ by taking the limit $\Lambda _{-}\rightarrow \infty $.

For example, in the case of the standard model coupled to quantum gravity,
the basic action $S_{d\text{b}}$ is equal to $S_{c\text{SMG}}+(S_{K},\Psi
)+S_{K}$, where $S_{c\text{SMG}}$ is the low-energy\ classical action of
formula (\ref{basi}), if $\mathcal{L}_{m}$ is extended appropriately. Note
that the matter Lagrangian $\mathcal{L}_{m}$ of $S_{c\text{SMG}}$ is at most
linear in $D_{\bar{\mu}}\psi $, and at most quadratic in $D_{\bar{\mu}}H$,
where $\psi $ are the fermions and $H$ is the Higgs field. The scalar mass
terms, the Yukawa couplings, and the vertices $(H^{\dag }H)^{2}$ and $%
R(H^{\dag }H)$ have the structures%
\begin{equation}
\frac{m^{2}}{\kappa ^{2}}\int \sqrt{|g|}(\kappa \varphi )^{2},\quad \frac{m}{%
\kappa ^{2}}\int \sqrt{|g|}(\kappa \varphi )(\kappa \bar{\psi})(\kappa \psi
),\quad \frac{m^{2}}{\kappa ^{2}}\int \sqrt{|g|}(\kappa \varphi )^{4},\quad 
\frac{\zeta }{\kappa ^{2}}\int \sqrt{|g|}R(\kappa \varphi ^{\dag })(\kappa
\varphi ),  \label{lh2}
\end{equation}%
where $\zeta $ is dimensionless. Therefore, they survive the limit $\Lambda
_{-}\rightarrow \infty $ and are contained in $S_{d\text{b}}$. For the same
reason, arbitrary powers of $\kappa \varphi $ are contained in $\mathcal{L}%
_{m}$. The basic action $\check{S}_{d\text{b}}$ associated with the extended
theory $\check{S}_{dT}$ contains the vertices $(LH)^{2}$ and the four
fermion vertices that break baryon number conservation. Indeed, although
those vertices are power counting nonrenormalizable, they also survive the
limit $\Lambda _{-}\rightarrow \infty $, because their structures are 
\begin{equation}
\frac{m}{\kappa ^{2}}\int \sqrt{|g|}(\kappa \varphi )^{2}(\kappa \bar{\psi}%
)(\kappa \psi ),\qquad \frac{\lambda }{\kappa ^{2}}\int \sqrt{|g|}(\kappa 
\bar{\psi})^{2}(\kappa \psi )^{2},  \label{4f}
\end{equation}%
where $\lambda $ is dimensionless.

If the nonanomalous accidental symmetries are unbroken, the standard model
coupled to quantum gravity does not satisfy the Kluberg-Stern--Zuber
assumption (\ref{coho}). Nevertheless, we can formulate a less restrictive
assumption that is sufficient to give us control over the counterterms.
Precisely, we assume that

(III) the basic action $S_{d\text{b}}$ is cohomologically complete [that is
to say, $\check{S}_{d\text{b}}$ satisfies the extended Kluberg-Stern--Zuber
assumption (\ref{coho2})] and the group $G_{\text{nas}}$ is compact.

Moreover, we assume that

(IV) the basic action $S_{d\text{b}}$ has trivial one-loop gauge anomalies $%
\mathcal{A}_{\hspace{0.01in}\text{b}}^{(1)}$; i.e. there exists a local
functional $\mathcal{X}(\Phi ,K)$ such that $\mathcal{A}_{\hspace{0.01in}%
\text{b}}^{(1)}=(S_{d\text{b}},\mathcal{X})$.

To subtract the potential anomalies of the higher-derivative theory, which
is defined at $\Lambda $ fixed, in a way that preserves its structure and
nice properties, we actually need a stronger assumption, that is to say,

(V) a local functional $\mathcal{F}(\Phi )$ of ghost number one that is
trivial in the $S_{d\text{b}}$ cohomology is also trivial in the $S_{K}$
cohomology; i.e. if there exists a local functional $\mathcal{X}(\Phi ,K)$
such that $\mathcal{F}=(S_{d\text{b}},\mathcal{X})$, then there also exists
a local functional $\chi (\Phi )$ such that $\mathcal{F}=(S_{K},\chi )$.

In section \ref{s9} we show that the standard model coupled to quantum
gravity satisfies all the assumptions of our proof, so it is free of gauge
anomalies to all orders.

When assumptions (IV) and (V) do not hold, or only one of them holds, we may
replace them with the assumption that

(IV$^{\prime }$) the one-loop anomalies of the higher-derivative\ theory
defined in subsection \ref{s23} are trivial in the $S_{K}$ cohomology; i.e.
there exists a local functional $\chi (\Phi )$ such that they can be written
as $(S_{K},\chi )$.

Indeed, assumptions (IV)\ and (V)\ are just needed to prove (IV$^{\prime }$)
[see the arguments of section \ref{s5} from formula (\ref{algol}) to formula
(\ref{anomcanc})]. In some practical situations it may be easier to prove (IV%
$^{\prime }$) rather than (III) and (IV).

\subsection{CDHD regularization}

\label{s23}

To find the subtraction scheme where the Adler-Bardeen theorem is almost
manifest, we must merge the CD regularization with a suitable gauge
invariant higher-derivative regularization. The resulting technique is
called chiral-dimensional/higher-derivative regularization. It resembles the
dimensional/higher-derivative (DHD) regularization of ref. \cite{ABrenoYMLR}
in various respects, but there are a few crucial differences. First, the
usual dimensional regularization is replaced by the CD regularization to
overcome the difficulties mentioned in subsection \ref{s22}. Second, the DHD
regularization is good for renormalizable theories, while we also want to
apply the CDHD technique to nonrenormalizable theories. To this purpose, the
HD regularizing terms must be adapted to the truncation. For several
arguments of our derivations, we actually need to place them well beyond the
truncation, and we must show that it is always possible to arrange them to
meet our needs. As in ref. \cite{ABrenoYMLR}, the HD regularization must
preserve gauge invariance in $d$ dimensions, to ensure that it is as
transparent as possible to potential anomalies.

In this section we build the HD and CDHD regularizations. In general terms,
they can be defined independently of the truncation, so we first work with
the untruncated theory. Nevertheless, we cannot satisfy all the requirements
we need in this paper, until we introduce the truncation. We do that at a
second stage and emphasize why the truncation is so crucial for our purposes.

We introduce higher-derivative local functionals $S_{\mathrm{HD}}^{I}$,
where $I$ is an index labeling them, a higher-derivative gauge fermion $\Psi
_{\mathrm{HD}}$, and higher-derivative formally evanescent terms $S_{\text{ev%
}\Lambda }$. We use them to define a regularized action $S_{\Lambda } $
whose propagators fall off as rapidly as we want, when the momenta $p$
become large.

We take the functionals $S_{\mathrm{HD}}^{I}$ to be gauge invariant in $d$
dimensions, i.e. satisfy $(S_{K},S_{\mathrm{HD}}^{I})=0$ and are of the form 
$S_{\mathrm{HD}}^{I}(\kappa \phi ,r,r_{+})$. In particular, they just depend
on the physical fields $\phi $. We normalize each $S_{\mathrm{HD}}^{I}$ so
that its quadratic terms (if any) have the form $\sim \kappa ^{2}\phi
\partial ^{\bar{N}_{I}+N_{\phi }}\phi $, where $\bar{N}_{I}$ are
non-negative integers and $N_{\phi }$ are the integers of formula (\ref{dom}%
). The invariants $S_{\mathrm{HD}}^{I}$ are extended from $d$ to $D$
dimensions by preserving the identity $(S_{K},S_{\mathrm{HD}}^{I})=0$,
according to the rules of the CD regularization \cite{chiraldimreg}.

Specifically, for the standard model coupled to quantum gravity, examples of
the functionals $S_{\mathrm{HD}}^{I}$ are the integrals of $\sqrt{|g|}$
times 
\begin{eqnarray}
&&g^{\bar{\mu}\bar{\nu}}(\kappa \mathcal{D}_{\bar{\mu}}\bar{\varphi})(%
\mathcal{D}^{2})^{\bar{N}_{\varphi }/2}(\kappa \mathcal{D}_{\bar{\nu}%
}\varphi ),\qquad (\kappa \bar{\psi})(\gamma ^{\bar{\mu}}\mathcal{D}_{\bar{%
\mu}})^{\bar{N}_{\psi }+1}(\kappa \psi ),\qquad (\kappa F_{\mu \nu })(%
\mathcal{D}^{2})^{\bar{N}_{A}/2}(\kappa F^{\mu \nu }),  \notag \\
&&\qquad \qquad \qquad R_{\mu \nu }(\mathcal{D}^{2})^{(\bar{N}%
_{G}-2)/2}R^{\mu \nu },\qquad R(\mathcal{D}^{2})^{(\bar{N}_{G}-2)/2}R,
\label{regulat}
\end{eqnarray}%
where $\mathcal{D}_{\bar{\mu}}$ denotes the covariant derivative, $\mathcal{D%
}^{2}=g^{\bar{\mu}\bar{\nu}}\mathcal{D}_{\bar{\mu}}\mathcal{D}_{\bar{\nu}}$,
and the integers $\bar{N}_{\varphi }$, $\bar{N}_{\psi },$ $\bar{N}_{A}$, $%
\bar{N}_{G}$ are large enough (see below). The same invariants work for any
Einstein--Yang-Mills theory, as well as any higher-derivative theories of
quantum gravity, Yang-Mills gauge fields, scalars, and fermions.

The classical action $S_{c}(\phi )$ is extended to 
\begin{equation}
S_{c\Lambda }(\phi )=S_{c}(\phi )+\frac{1}{\kappa ^{2}}\sum_{I}\frac{1}{%
\Lambda ^{2\bar{N}_{I}}}S_{\mathrm{HD}}^{I}(\kappa \phi ,r,r_{+}),
\label{scl}
\end{equation}%
where $\Lambda $ is the energy scale associated with the HD regularization.
The new non-gauge-fixed action then reads 
\begin{equation}
\bar{S}_{d\Lambda }(\Phi ,K)=S_{c\Lambda }(\phi )+S_{K}=S_{c\Lambda }(\phi
)-\int R^{\alpha }(\Phi )K_{\alpha }  \label{link}
\end{equation}%
and solves $(\bar{S}_{d\Lambda },\bar{S}_{d\Lambda })=0$ in arbitrary $D$
dimensions.

Divide the set $\phi $ of the physical fields into two subsets, called $\phi
_{g}^{\prime }$ and $\phi _{m}$. The set $\phi _{m}$ contains the matter
fields $\phi $ that have $[\kappa \phi ]>0$. The set $\phi _{g}^{\prime }$
contains the gauge fields $\phi _{g}$, plus the matter fields $\phi $ that
have $[\kappa \phi ]=0$. We decompose $\Phi $ as $\{\Phi _{g}^{\prime },\phi
_{m}\}$, where $\Phi _{g}^{\prime }$ contains the fields $\phi _{g}^{\prime
} $, the ghosts $C$, the antighosts $\bar{C}$ and the Lagrange multipliers $%
B $. Similarly, we decompose the sources $K$ as $\{K_{g}^{\prime },K_{m}\}$.
The transformations $R_{g}(\Phi )$ of the fields $\Phi _{g}^{\prime }$ are
independent of $\phi _{m}$, and the transformations $R_{m}(\Phi )$ of the
fields $\phi _{m}$ are linear in the fields $\phi _{m}$ themselves and
vanish at $\phi _{m}=0$.

In the case of the standard model coupled to quantum gravity, the set $\phi
_{g}^{\prime }$ contains the bosons, while the set $\phi _{m}$ contains the
fermions.

If we organize the HD regularization properly, we can show that the
counterterms and the local contributions to potential anomalies at finite $%
\Lambda $ are independent of the matter fields $\phi _{m}$. The
transformations $R^{\alpha }(\Phi ,g)$ do not depend on other parameters
besides the gauge couplings $g$, so, after the replacements (\ref{repla}),
we can write 
\begin{equation}
S_{K}(\Phi ,K,\kappa )=-\int R^{\alpha }(\Phi ,g)K_{\alpha }=-\frac{1}{%
\kappa ^{2}}\int R^{\prime \alpha }(\kappa \Phi ,r,r_{+})(\kappa K_{\alpha
}).  \label{sko}
\end{equation}

We organize the invariants $S_{\mathrm{HD}}^{I}$ into invariants $S_{g%
\mathrm{HD}}^{I}$ that are $\phi _{m}$-independent and invariants $S_{m%
\mathrm{HD}}^{I}$ that are quadratic in the fields $\phi _{m}$. We ignore
any $\phi _{m}$-dependent invariants $S_{\mathrm{HD}}^{I}$ that are not
quadratic in $\phi _{m}$ because they are not necessary for our purposes.
The examples (\ref{regulat}) fulfill this requirement.

We require that the modified gauge fermion $\Psi _{\mathrm{HD}}$ be
invariant under rigid diffeomorphisms and independent of the matter fields.
Moreover, we organize it so that each term contains an even power $2k$ of $%
1/\Lambda $, and at least $k$ derivatives $\bar{\partial}$ act on the
antighosts $\bar{C}$ and $k$ derivatives $\bar{\partial}$ act on the
Lagrange multipliers $B$, whenever $\bar{C}$ and/or $B$\ are present. The
prototype of this kind of gauge fermion is 
\begin{equation}
\Psi _{\mathrm{HD}}(\Phi )=\sum_{i}\int \sqrt{|g|}\bar{C}_{i}\left(
Q_{i}(\square )G_{i}(\phi ,\xi )+\frac{1}{2}Q_{i}^{\prime }(\square
)B_{i}\right) ,  \label{psihd}
\end{equation}%
where $i$ is a generic label to distinguish different types of
contributions, and $Q_{i}$ and $Q_{i}^{\prime }$ are operators acting as
follows: 
\begin{eqnarray}
\int \sqrt{|g|}\bar{C}_{i}Q_{i}(\square )G_{i}(\phi ) &\equiv &\int \sqrt{|g|%
}\sum_{k=0}^{N_{i}}\frac{c_{ik}}{\Lambda ^{2k}}(\partial _{\bar{\rho}%
_{1}}\cdots \partial _{\bar{\rho}_{2k}}\bar{C}_{I})g^{\bar{\rho}_{1}\bar{\rho%
}_{2}}\cdots g^{\bar{\rho}_{2k-1}\bar{\rho}_{2k}}G^{I}(\phi ,\xi ),  \notag
\\
\int \sqrt{|g|}\bar{C}_{i}Q_{i}^{\prime }(\square )B_{i} &\equiv &\int \sqrt{%
|g|}\sum_{k=0}^{N_{i}^{\prime }}\frac{c_{ik}^{\prime }}{\Lambda ^{2k}}%
(\partial _{\bar{\rho}_{1}}\cdots \partial _{\bar{\rho}_{k}}\bar{C}_{I})g^{%
\bar{\rho}_{1}\bar{\sigma}_{1}}\cdots g^{\bar{\rho}_{k}\bar{\sigma}%
_{k}}g^{IJ}P(\phi ,\xi ^{\prime },\partial )(\partial _{\bar{\sigma}%
_{1}}\cdots \partial _{\bar{\sigma}_{k}}B_{J}).  \notag \\
&&  \label{antigh}
\end{eqnarray}%
The functions $G^{I}(\phi ,\xi )$ and the operators $P(\phi ,\xi ^{\prime
},\partial )$ can be read by comparing $\Psi _{\mathrm{HD}}$ with the gauge
fermion $\Psi $ of $S_{d}$ in the limit $\Lambda =\infty $, while $N_{i}$, $%
N_{i}^{\prime }$ are integer numbers and $c_{ik}$, $c_{ik}^{\prime }$ are
constants. In the case of diffeomorphisms, $\bar{C}_{I}=\bar{C}_{\bar{\mu}}$%
, $G^{I}=G^{\bar{\mu}}$, $B_{J}=B_{\bar{\nu}}$, and ${g^{IJ}=g^{\bar{\mu}%
\bar{\nu}}}$. In the case of Yang-Mills symmetries, $\bar{C}_{I}=\bar{C}^{a}$%
, $G^{I}=G^{a}$, $B_{J}=B^{b}$, and ${g^{IJ}=\delta }^{ab}$. In the case of
local Lorentz symmetry, $\bar{C}_{I}=\bar{C}_{\bar{a}\bar{b}}$, $G^{I}=G^{%
\bar{a}\bar{b}}$, $B_{J}=B_{\bar{c}\bar{d}}$, and ${g^{IJ}}=(\delta ^{\bar{a}%
\bar{c}}\delta ^{\bar{b}\bar{d}}-\delta ^{\bar{a}\bar{d}}\delta ^{\bar{b}%
\bar{c}})/2$. Thanks to the structure (\ref{psihd}), we will be able to
prove that the antighosts and the Lagrange multipliers cannot contribute to
the counterterms and the potential anomalies at finite $\Lambda $.

Specifically, in the case of perturbatively unitary theories, such as the
standard model coupled to quantum gravity, we extend (\ref{psi1}) to 
\begin{eqnarray}
\Psi _{\mathrm{HD}}(\Phi ) &=&\int \sqrt{|g|}\bar{C}^{a}\left( Q_{1}(\square
)g^{\bar{\mu}\bar{\nu}}\partial _{\bar{\mu}}A_{\bar{\nu}}^{a}+\frac{1}{2}%
Q_{1}^{\prime }(\square )B^{a}\right)  \notag \\
&&+\int \sqrt{|g|}\bar{C}_{\bar{a}\bar{b}}\left( \frac{1}{\kappa }%
Q_{2}(\square )e^{\bar{\rho}\bar{a}}g^{\bar{\mu}\bar{\nu}}\partial _{\bar{\mu%
}}\partial _{\bar{\nu}}e_{\bar{\rho}}^{\bar{b}}+\frac{1}{2}Q_{2}^{\prime
}(\square )B^{\bar{a}\bar{b}}\right)  \label{phiHD} \\
&&{-\int }\sqrt{|g|}\bar{C}_{\bar{\mu}}\left( \frac{1}{\kappa }Q_{3}(\square
)\partial _{\bar{\nu}}g^{\bar{\mu}\bar{\nu}}+\frac{1}{\kappa }Q_{4}(\square
)g^{\bar{\mu}\bar{\nu}}g_{\bar{\rho}\bar{\sigma}}\partial _{\bar{\nu}}g^{%
\bar{\rho}\bar{\sigma}}-\frac{Q_{3}^{\prime }(\square )}{2}g^{\bar{\mu}\bar{%
\nu}}B_{\bar{\nu}}\right) {.}  \notag
\end{eqnarray}

The gauge-fixed action is then 
\begin{equation}
S_{d\Lambda }(\Phi ,K)=\bar{S}_{d\Lambda }+(S_{K},\Psi _{\mathrm{HD}})
\label{sdlam}
\end{equation}%
and satisfies $(S_{d\Lambda },S_{d\Lambda })=0$ in arbitrary $D$. It is
obvious that the higher-derivative terms can make the propagators of all
fields fall off as rapidly as we want, when the physical components $\bar{p}$
of the momenta $p$ become large.

Finally, the HD\ regularized action 
\begin{equation}
S_{\Lambda }=S_{d\Lambda }+S_{\text{ev}\Lambda }=S_{c}(\phi )+\frac{1}{%
\kappa ^{2}}\sum_{I}\frac{1}{\Lambda ^{2\bar{N}_{I}}}S_{\mathrm{HD}%
}^{I}(\kappa \phi ,r,r_{+})+(S_{K},\Psi _{\mathrm{HD}})+S_{K}+S_{\text{ev}%
\Lambda }  \label{slambda}
\end{equation}%
is obtained by adding suitable formally evanescent terms $S_{\text{ev}%
\Lambda }$ compatible with weighted power counting and the nonanomalous
global symmetries of the theory. We also require that $S_{\text{ev}\Lambda }$
be built with a unique metric tensor or vielbein. The scale $\Lambda $ has
weight 1, equal to its dimension. The important terms of $S_{\text{ev}%
\Lambda }-S_{\text{ev}}$ are the kinetic ones, which must complete the
regularized propagators, according to weighted power counting (more details
on this are given in the next subsection). We can choose the other
contributions to $S_{\text{ev}\Lambda }-S_{\text{ev}}$ at our discretion, or
suppress them. The kinetic terms of $S_{\text{ev}\Lambda }$ can be
constructed, for example, by inserting higher derivatives $\bar{\partial}%
/\Lambda $ and $\hat{\partial}^{2}/(M\Lambda )$ into the evanescent terms of 
$S_{\text{ev}}$, such as (\ref{w2}) and (\ref{w3}). We assume that the
difference $S_{\text{ev}\Lambda }-S_{\text{ev}}$ is $K$ independent, since $%
K $-dependent higher-derivative terms are unnecessary for our purposes. We
also assume that $S_{\text{ev}\Lambda }-S_{\text{ev}}$ is a sum of terms
that are either independent of the fields $\phi _{m}$ or quadratic in $\phi
_{m}$, and that the $\phi _{m}$-dependent terms are independent of $\bar{C}$
and $B$. Finally, we assume that each term of $S_{\text{ev}\Lambda }-S_{%
\text{ev}}$ contains an even power $2k$ of $1/\Lambda $, and at least $k$
derivative operators $\bar{\partial}\sim $ $\hat{\partial}^{2}/M$ act on the
antighosts $\bar{C}$ and $k$ derivative operators $\bar{\partial}\sim $ $%
\hat{\partial}^{2}/M$\ act on the Lagrange multipliers $B$, whenever $\bar{C}
$ and/or $B$\ are present.

The action (\ref{slambda}) clearly satisfies 
\begin{equation}
(S_{\Lambda },S_{\Lambda })=\mathcal{\hat{O}}(\varepsilon ).  \label{sleva}
\end{equation}

The HD sector $S_{\mathrm{HD}}\equiv S_{\Lambda }-S$ is also $K$
independent. It must have the $\kappa $ structure (\ref{liable}) and be
organized so that all the propagators have the structure (\ref{propag}). The
parameters on which $S_{\mathrm{HD}}$ depends, besides $\kappa ,r,r_{+}$ and 
$\Lambda $, must have non-negative dimensions. We include them in a set $%
\lambda _{+}$, together with $r,r_{+}$, and write 
\begin{equation}
S_{\mathrm{HD}}=S_{\mathrm{HD}}(\Phi ,\kappa ,\Lambda ,\lambda _{+})=\frac{1%
}{\kappa ^{2}}S_{\mathrm{HD}}^{\prime }(\kappa \Phi ,\Lambda ,\lambda _{+}).
\label{shd}
\end{equation}%
Note that each contribution to $S_{\mathrm{HD}}$ is either independent of
the fields $\phi _{m}$, or quadratic in them. Formula (\ref{shd}) is also
implicitly assuming that $S_{\mathrm{HD}}$ is $\Lambda _{-}$ independent.
Then, it coincides with its own truncation. More conditions on the
higher-derivative sector $S_{\mathrm{HD}}$ are given in the next section.

Now we come to the truncation. The prescription T1 of subsection \ref{s21}
tells us that the truncated action $S_{\Lambda T}$ is obtained by switching
off the $o(1/\Lambda _{-}^{T})$ terms of $S_{\Lambda }$. Since $S_{\mathrm{HD%
}}$ is $\Lambda _{-}$ independent, we just get the sum of $S_{T}$ and $S_{%
\mathrm{HD}}$: 
\begin{equation}
S_{\Lambda T}=S_{T}+S_{\mathrm{HD}}=S_{cT}(\phi )+\frac{1}{\kappa ^{2}}%
\sum_{I}\frac{1}{\Lambda ^{2\bar{N}_{I}}}S_{\mathrm{HD}}^{I}(\kappa \phi
,r,r_{+})+(S_{K},\Psi _{\mathrm{HD}})+S_{K}+S_{\text{ev}\Lambda T}.
\label{tHD}
\end{equation}%
Again, the action $S_{\Lambda T}$ satisfies the master equation up to
formally evanescent terms, which means 
\begin{equation}
(S_{\Lambda T},S_{\Lambda T})=\mathcal{\hat{O}}(\varepsilon ).
\label{sleva2}
\end{equation}

At finite $\Lambda $, the theory defined by the action $S_{\Lambda T}$,
regularized and renormalized by means the CD\ technique, is called
(truncated)\ \textquotedblleft higher-derivative theory\textquotedblright ,
or HD theory. The theory defined by the same action $S_{\Lambda T}$, but
regularized and renormalized by means of the CDHD technique, is called
(truncated) \textit{final} theory. The HD\ theory is renormalized by
studying the limit $\varepsilon \rightarrow 0$ and removing the divergences
and potential anomalies at $\Lambda $ fixed. Once that is done, the final
theory is reached by studying the limit $\Lambda \rightarrow \infty $ on the
HD\ theory, removing the $\Lambda $ divergences and proving that the
cancellation of anomalies survives these operations.

At this point, we have two regulators and two types of divergences: the
poles in $\varepsilon $ and the $\Lambda $ divergences. The latter are
products $\Lambda ^{k}\ln ^{k^{\prime }}\Lambda $, with $k,k^{\prime
}\geqslant 0$, $k+k^{\prime }>0$, times local monomials of the fields, the
sources and their derivatives. From the point of view of the CD
regularization, those monomials may be nonevanescent or formally evanescent,
and their coefficients must be evaluated in the analytic limit $\varepsilon
\rightarrow 0$. To complete the CDHD regularization, we must specify how the
regularization parameters $\varepsilon $ and $\Lambda $ are removed. If the
HD sector of the regularization is organized in a suitable way, which we
specify in the next section, the HD theory is super-renormalizable and only
a few one-loop diagrams diverge. After studying the poles in $\varepsilon $
and the one-loop potential anomalies, at $\Lambda $ fixed, we prove that it
is possible to remove both. We also show that these operations are
sufficient to remove both divergences and anomalies to all orders, in the
HD\ theory. Then we study the limit $\Lambda \rightarrow \infty $ and show
that we can remove the divergences and potential anomalies appearing in that
limit, preserving gauge invariance. We call the set of such operations the%
\textit{\ CDHD limit}.

For more clarity, we describe how the CDHD limit works with the help of a
set of symbolic expressions. When we study the HD theory, we expand around $%
\varepsilon =0$ at $\Lambda $ fixed. Then we find poles, finite terms, and
evanescent terms of the form 
\begin{equation*}
\frac{1}{\varepsilon },\qquad \frac{\hat{\delta}}{\varepsilon },\qquad
\varepsilon ^{0},\qquad \hat{\delta}\varepsilon ^{0},\qquad \varepsilon
,\qquad \hat{\delta}\varepsilon ,
\end{equation*}%
where $1/\varepsilon $ denotes any divergent expression, $\hat{\delta}$ is
any formally evanescent expression, $\varepsilon ^{0}$ is any expression
that is convergent and nonevanescent in the analytic limit $\varepsilon
\rightarrow 0$, and $\varepsilon $ denotes any analytic evanescence. Next,
we subtract the divergent parts, that is to say, the first two terms of the
list. The coefficients of the surviving terms, which are 
\begin{equation}
\varepsilon ^{0},\qquad \hat{\delta}\varepsilon ^{0},\qquad \varepsilon
,\qquad \hat{\delta}\varepsilon ,  \label{survi}
\end{equation}%
are then expanded around $\Lambda =\infty $, which gives the structures 
\begin{eqnarray}
&&\varepsilon ^{0}\Lambda ,\qquad \hat{\delta}\varepsilon ^{0}\Lambda
,\qquad \varepsilon ^{0}\Lambda ^{0},\qquad \hat{\delta}\varepsilon
^{0}\Lambda ^{0},\qquad \frac{\varepsilon ^{0}}{\Lambda },\qquad \frac{\hat{%
\delta}\varepsilon ^{0}}{\Lambda },  \notag \\
&&\varepsilon \Lambda ,\qquad \hat{\delta}\varepsilon \Lambda ,\qquad
\varepsilon \Lambda ^{0},\qquad \hat{\delta}\varepsilon \Lambda ^{0},\qquad 
\frac{\varepsilon }{\Lambda },\qquad \frac{\hat{\delta}\varepsilon }{\Lambda 
},  \label{assu}
\end{eqnarray}%
where $\Lambda $ denotes any expression that diverges when $\Lambda
\rightarrow \infty $ (i.e. it is multiplied by a coefficient that behaves
like $\Lambda ^{k}\ln ^{k^{\prime }}\Lambda $, with $k,k^{\prime }\geqslant
0 $, $k+k^{\prime }>0$), $\Lambda ^{0}$ is any expression that is
convergent, but not evanescent, in the same limit, while $1/\Lambda $ is any
expression that vanishes in the limit. The first two terms of the list (\ref%
{assu})\ are the $\Lambda $ divergences of the CDHD limit and must be
subtracted. For convenience, we include the terms $\hat{\delta}\varepsilon
^{0}\Lambda $ (which are local) in this subtraction, although they are going
to be dropped at a later stage. We cannot include the terms $\varepsilon
\Lambda $, instead, because they are not local. After these new
subtractions, we remain with 
\begin{equation}
\varepsilon ^{0}\Lambda ^{0},\qquad \hat{\delta}\varepsilon ^{0}\Lambda
^{0},\qquad \frac{\varepsilon ^{0}}{\Lambda },\qquad \frac{\hat{\delta}%
\varepsilon ^{0}}{\Lambda },\qquad \varepsilon \Lambda ,\qquad \hat{\delta}%
\varepsilon \Lambda ,\qquad \varepsilon \Lambda ^{0},\qquad \hat{\delta}%
\varepsilon \Lambda ^{0},\qquad \frac{\varepsilon }{\Lambda },\qquad \frac{%
\hat{\delta}\varepsilon }{\Lambda }.  \label{survo}
\end{equation}%
Finally, the CDHD limit is taken by dropping all the contributions of this
list but the $\varepsilon ^{0}\Lambda ^{0}$ terms. Note that the terms
proportional to $\varepsilon $ vanish in the CDHD limit, even if they are
divergent in $\Lambda $, because the limit $\varepsilon \rightarrow 0$ is
taken before the limit $\Lambda \rightarrow \infty $.

\section{Properties of the HD theory}

\label{s3}

\setcounter{equation}{0}

In this section we organize the higher-derivative regularization and study
its properties. We want to show that for every truncation T1 of subsection %
\ref{s21} we can arrange the higher-derivative sector $S_{\mathrm{HD}%
}=S_{\Lambda T}-S_{T}$ so that it satisfies a number of conditions that will
be useful to prove the Adler-Bardeen theorem. So far, for example, we have
not specified the numbers of higher derivatives that we need. We anticipate
that, besides being sufficiently many, they should not conflict with the
truncated action $S_{T}$, that is to say, they should all be placed well
beyond the truncation. The tree-level truncation T1 will be enough to give
us complete control on the radiative corrections of the HD theory, to all
orders in $\hbar $ and for arbitrarily large powers of $1/\Lambda _{-}$. We
do not apply the truncation T2 till section \ref{s7}, where we study the
limit $\Lambda \rightarrow \infty $ and the final theory.

The numbers of higher derivatives are governed by the $\Lambda $ exponents $%
\bar{N}_{I}$ appearing in formula (\ref{scl}), analogous exponents $\hat{N}%
_{I}$ appearing inside $S_{\text{ev}\Lambda }$, and the exponents $N_{i}$, $%
N_{i}^{\prime }$ of $\Psi _{\mathrm{HD}}$, appearing in (\ref{antigh}). The $%
\Phi $ kinetic terms of $S_{\mathrm{HD}}$ that are dominant in the large
momentum limits $\bar{p}\rightarrow \infty $ and $\hat{p}\rightarrow \infty $
have the form 
\begin{equation}
\bar{c}_{\Phi }\int \Phi \left( \frac{\bar{\partial}^{2}}{\Lambda ^{2}}%
\right) ^{\bar{N}_{\Phi }}\bar{\partial}^{N_{\Phi }}\Phi +\hat{c}_{\Phi
}\int \Phi \left( \frac{\hat{\partial}^{2}}{M\Lambda }\right) ^{2\hat{N}%
_{\Phi }}\left( \frac{\hat{\partial}^{2}}{M}\right) ^{N_{\Phi }}\Phi ,
\label{domHD}
\end{equation}%
where $\bar{c}_{\Phi }$ and $\hat{c}_{\Phi }$ are weightless constants, $2%
\bar{N}_{\Phi }$ is the maximum number of higher derivatives $\bar{\partial}$
and $4\hat{N}_{\Phi }$ the maximum number of higher derivatives $\hat{%
\partial}$. Weighted power counting requires $\bar{N}_{\Phi }=\hat{N}_{\Phi
} $. For reasons that will be clear below, we need to take the same $\bar{N}%
_{\phi _{g}^{\prime }}=\hat{N}_{\phi _{g}^{\prime }}\equiv N_{+}$ for all
fields $\phi _{g}^{\prime }$, and the same $\bar{N}_{\phi _{m}}=\hat{N}%
_{\phi _{m}}\equiv N_{-}$ for all fields $\phi _{m}$. Then we set $%
N_{i}=N_{i}^{\prime }=N_{+}$ in (\ref{antigh}). We switch off all terms of $%
S_{\mathrm{HD}}$ that are multiplied by more than $2N_{+}$ powers of $%
1/\Lambda $, and all $\phi _{m}$-dependent $S_{\mathrm{HD}}$ terms that are
multiplied by more than $2N_{-}$ powers of $1/\Lambda $. We also need to
take $N_{+}$, $N_{-}$, and $N_{+}-N_{-}>0$ sufficiently large. The first
task of this section is to determine the bounds on these numbers and show
that it is always possible to choose them so that they satisfy the
requirements we need.

Define tilde fields and sources as 
\begin{equation}
\tilde{\Phi}_{g}^{\prime }=\frac{\Phi _{g}^{\prime }}{\Lambda ^{N_{+}}}%
,\qquad \tilde{\phi}_{m}=\frac{\phi _{m}}{\Lambda ^{N_{-}}},\qquad \tilde{K}%
_{g}^{\prime }=\Lambda ^{N_{+}}K_{g}^{\prime },\qquad \tilde{K}_{m}=\Lambda
^{N_{-}}K_{m},  \label{canoti}
\end{equation}%
and tilde parameters $\tilde{\kappa}=\kappa \Lambda ^{N_{+}}$, $\tilde{r}=r$%
, and $\tilde{r}_{+}=r_{+}$. We have 
\begin{equation}
\tilde{\kappa}\tilde{\Phi}_{g}^{\prime }=\kappa \Phi _{g}^{\prime },\qquad 
\tilde{\kappa}\tilde{K}_{g}^{\prime }=\Lambda ^{2N_{+}}\kappa K_{g}^{\prime
},\qquad \tilde{\kappa}\tilde{\phi}_{m}=\kappa \phi _{m}\Lambda
^{N_{+}-N_{-}},\qquad \tilde{\kappa}\tilde{K}_{m}=\Lambda
^{N_{+}+N_{-}}\kappa K_{m}.  \label{tildi}
\end{equation}%
Observe that (\ref{canoti}) is a canonical transformation. After the
redefinitions, the dominant kinetic terms (\ref{domHD}) of $S_{\mathrm{HD}}$
are $\Lambda $ independent. Those of the fields $\phi _{g}^{\prime }$ and $%
\phi _{m}$ are 
\begin{equation*}
\int \tilde{\phi}_{g}^{\prime }\left[ \bar{c}_{g}\bar{\partial}%
^{2N_{+}+N_{\phi _{g}^{\prime }}}+\hat{c}_{g}\left( \frac{\hat{\partial}^{2}%
}{M}\right) ^{2N_{+}+N_{\phi _{g}^{\prime }}}\right] \tilde{\phi}%
_{g}^{\prime }+\int \tilde{\phi}_{m}\left[ \bar{c}_{m}\bar{\partial}%
^{2N_{-}+N_{\phi _{m}}}+\hat{c}_{m}\left( \frac{\hat{\partial}^{2}}{M}%
\right) ^{2N_{-}+N_{\phi _{m}}}\right] \tilde{\phi}_{m}.
\end{equation*}%
Those of the ghosts $C$, the antighosts $\bar{C}$, and the Lagrange
multipliers $B$ follow from the choices of $G(\phi ,\xi )$ and $P(\phi ,\xi
^{\prime },\partial )$ in (\ref{phi}).

Recall that $S_{\mathrm{HD}}$ has the structure (\ref{shd}), $\Psi _{\mathrm{%
HD}}$ is independent of the matter fields, and each contribution to $S_{%
\mathrm{HD}}$ is either quadratic in the matter fields $\phi _{m}$ or
independent of them. Then, we can write%
\begin{equation}
S_{\mathrm{HD}}=\frac{1}{\tilde{\kappa}^{2}}S_{\mathrm{HD}}^{\prime \prime }(%
\tilde{\kappa}\tilde{\Phi}_{g}^{\prime },\tilde{\kappa}\tilde{\phi}_{m},%
\tilde{\lambda}_{+}),  \label{sghj}
\end{equation}%
where $S_{\mathrm{HD}}^{\prime \prime }$ is $\Lambda $ independent in the
tilde parametrization and $\tilde{\lambda}_{+}$ are parameters of
non-negative dimensions, equal to products $\lambda _{+}\Lambda ^{k}$, with $%
k\geqslant 0$. To simplify some arguments, we switch off all the parameters $%
\lambda _{+}$ such that $\tilde{\lambda}_{+}=\lambda _{+}\Lambda ^{k}$ with $%
k>0$, because they are not necessary to make the higher-derivative
regularization work. Thus, from now on we assume that the parameters $%
\lambda _{+}$ have non-negative weights and satisfy $\lambda _{+}=\tilde{%
\lambda}_{+}$. Examples are the ratios $r=\tilde{r}$, $r_{+}=\tilde{r}_{+}$
between the gauge couplings $g$ and $\kappa $.

As far as the truncated action $S_{\Lambda T}$ is concerned, we have 
\begin{equation}
S_{\Lambda T}(\Phi ,K)=\frac{\Lambda ^{2N_{+}}}{\tilde{\kappa}^{2}}%
S_{T}^{\prime }(\tilde{\kappa}\tilde{\Phi}_{g}^{\prime },\Lambda
^{N_{-}-N_{+}}\tilde{\kappa}\tilde{\phi}_{m},\Lambda ^{-2N_{+}}\tilde{\kappa}%
\tilde{K}_{g}^{\prime },\Lambda ^{-N_{+}-N_{-}}\tilde{\kappa}\tilde{K}_{m})+%
\frac{1}{\tilde{\kappa}^{2}}S_{\mathrm{HD}}^{\prime \prime }(\tilde{\kappa}%
\tilde{\Phi}_{g}^{\prime },\tilde{\kappa}\tilde{\phi}_{m},\tilde{\lambda}%
_{+}),  \label{sltil}
\end{equation}%
where $S_{T}^{\prime }=S_{dT}^{\prime }+S_{\text{ev}T}^{\prime }$ and $%
S_{dT}^{\prime }$ and $S_{\text{ev}T}^{\prime }$ are defined by applying the
truncation T1 to formulas (\ref{para}) and (\ref{para2}).

If $N_{+}$ is large enough, the dimension $[\tilde{\kappa}]$ of $\tilde{%
\kappa}$ is strictly positive, which is a necessary condition to have
super-renormalizability. Actually, for later use we assume that $[\tilde{%
\kappa}]$ is greater than some given $t>0$, that is to say, 
\begin{equation}
N_{+}>t-[\kappa ].  \label{req1}
\end{equation}

The right-hand side of (\ref{sghj}) contains only parameters of non-negative
dimensions in units of mass, apart from the overall factor $1/\tilde{\kappa}%
^{2}$. Instead, $S_{T}^{\prime }$, written in the tilde parametrization,
contains parameters that can have positive, vanishing, or negative
dimensions, as well as factors $\Lambda ^{N_{-}-N+}$ and $\Lambda
^{-N_{-}-N+}$. However, we can show that the overall factor $\Lambda
^{2N_{+}}$ that multiplies $S_{T}^{\prime }/\tilde{\kappa}^{2}$ in formula (%
\ref{sltil}) allows us to turn $\Lambda ^{2N_{+}}S_{T}^{\prime }$ into a
functional that contains only parameters of positive (and arbitrarily large)
dimensions, at least within the truncation T1.

We begin with the functional $S_{K}$. By formula (\ref{sko}) and the
properties recalled right below formula (\ref{link}), we have, in the tilde
parametrization 
\begin{equation}
S_{K}=-\frac{1}{\tilde{\kappa}^{2}}\sum_{g}\int R^{\prime \alpha }(\tilde{%
\kappa}\tilde{\Phi},\tilde{r},\tilde{r}_{+})(\tilde{\kappa}\tilde{K}_{\alpha
}),  \label{ska}
\end{equation}%
which are of the form we want, that is to say, the tilde version of (\ref%
{liable2}).

Next, consider the $K$-independent contributions to $\Lambda
^{2N_{+}}S_{T}^{\prime }/\tilde{\kappa}^{2}$ in formula (\ref{sltil}). They
have the form 
\begin{equation}
\lambda \frac{\Lambda ^{2N_{+}}}{\Lambda _{-}^{u}\tilde{\kappa}^{2}}\partial
^{p}\prod_{g}(\tilde{\kappa}\tilde{\Phi}_{g}^{\prime
})^{q_{g}}\prod_{m}(\Lambda ^{N_{-}-N_{+}}\tilde{\kappa}\tilde{\phi}%
_{m})^{q_{m}},  \label{struct3}
\end{equation}%
where $u$, $p$, $q_{g}$, $q_{m}$ are non-negative integers and $\lambda $ is
a $\Lambda $-independent product of parameters of non-negative dimensions.
The truncated action $S_{T}=\Lambda ^{2N_{+}}S_{T}^{\prime }/\tilde{\kappa}%
^{2}$ contains a finite number of matter fields $\phi _{m}$, because $%
[\Lambda ^{N_{-}-N_{+}}\tilde{\kappa}\tilde{\phi}_{m}]=[\kappa \phi _{m}]>0$%
, $[\tilde{\kappa}\tilde{\Phi}_{g}^{\prime }]=[\kappa \Phi _{g}^{\prime
}]\geqslant 0$, by assumption (II-$i$) of subsection \ref{s21}, and $%
u\leqslant T$, by prescription T1. Thus, there exists a $q_{\text{max}}$
such that $\sum_{m}q_{m}\leqslant q_{\text{max}}$. Then, if we choose $N_{+}$
and $N_{-}$ such that the condition 
\begin{equation}
2N_{+}>q_{\text{max}}(N_{+}-N_{-})+T+2t+2|[\kappa ]|  \label{req2}
\end{equation}%
holds, besides (\ref{req1}), the structure (\ref{struct3}) becomes 
\begin{equation}
\frac{\tilde{\lambda}}{\tilde{\kappa}^{2}}\partial ^{p}\prod_{g}(\tilde{%
\kappa}\tilde{\Phi}_{g}^{\prime })^{q_{g}}\prod_{m}(\tilde{\kappa}\tilde{\phi%
}_{m})^{q_{m}},  \label{rew}
\end{equation}%
where the constants $\tilde{\lambda}=\lambda \Lambda ^{d_{\lambda }}/\Lambda
_{-}^{u}$ , with $d_{\lambda }=2N_{+}-(N_{+}-N_{-})\sum_{m}q_{m}$, have
dimensions greater than $2t+2|[\kappa ]|$.

For future use, we observe that if $\omega $ denotes $\zeta $, $\varsigma $
or $\eta $, all terms of $S_{c}(\phi )$, and $S_{\text{ev}}$ that just
depend on $\phi _{g}^{\prime }$ have the $\kappa $ structure 
\begin{equation}
\omega F(\phi _{g}^{\prime },\kappa ,r,r_{+})=\frac{\omega }{\kappa ^{2}}%
F^{\prime }(\kappa \phi _{g}^{\prime },r,r_{+})=\frac{\tilde{\omega}}{\tilde{%
\kappa}^{2}}F^{\prime }(\tilde{\kappa}\tilde{\phi}_{g}^{\prime },\tilde{r},%
\tilde{r}_{+}),  \label{sctr}
\end{equation}%
where $\tilde{\omega}=\omega \Lambda ^{2N_{+}}$.

Collecting (\ref{ska}) and (\ref{rew}), we can define a truncated functional 
$S_{d\Lambda T}^{\prime \prime }$ that depends analytically on $\tilde{%
\lambda}$, such that 
\begin{equation}
S_{d\Lambda T}(\Phi ,K,\kappa )=\frac{1}{\tilde{\kappa}^{2}}S_{d\Lambda
T}^{\prime \prime }(\tilde{\kappa}\tilde{\Phi}_{g}^{\prime },\tilde{\kappa}%
\tilde{\phi}_{m},\tilde{\kappa}\tilde{K}_{g}^{\prime },\tilde{\kappa}\tilde{K%
}_{m},\tilde{\lambda}).  \label{sevo}
\end{equation}

It remains to study the $K$-dependent contributions to the first term on the
right-hand side of (\ref{sltil}). Actually, we have already studied those
contained in $S_{K}$, which are rearranged in formula (\ref{ska}). The
remaining ones are contained in $S_{\text{ev}T}$. Write 
\begin{equation}
S_{\text{ev}T}(\Phi ,K,\kappa )=\frac{\Lambda ^{2N_{+}}}{\tilde{\kappa}^{2}}%
S_{\text{ev}T}^{\prime }(\tilde{\kappa}\tilde{\Phi}_{g}^{\prime },\Lambda
^{N_{-}-N_{+}}\tilde{\kappa}\tilde{\phi}_{m},\kappa K_{g}^{\prime },\kappa
K_{m}).  \label{sevt}
\end{equation}

Using $[\kappa K_{\Phi }]\geqslant 1/2$, which was proved in subsection \ref%
{s21}, a condition like (\ref{req2}), with a possibly different $q_{\text{max%
}}$, is also sufficient to rewrite each contribution to $S_{\text{ev}T}$ in
the form%
\begin{equation}
\frac{\tilde{\varsigma}}{\tilde{\kappa}^{2}}\partial ^{p}\prod_{g}(\tilde{%
\kappa}\tilde{\Phi}_{g}^{\prime })^{q_{g}}\prod_{m}(\tilde{\kappa}\tilde{\phi%
}_{m})^{q_{m}}\prod_{K}(\kappa K)^{q_{K}}  \label{rewa}
\end{equation}%
where $\tilde{\varsigma}$ are new parameters of dimensions greater than $%
2t+2|[\kappa ]|$, which include the tilde versions of both $\varsigma $ and $%
\eta $. Finally, we can write 
\begin{equation}
S_{\text{ev}T}(\Phi ,K,\kappa )=\frac{1}{\tilde{\kappa}^{2}}S_{\text{ev}%
T}^{\prime \prime }(\tilde{\kappa}\tilde{\Phi}_{g}^{\prime },\tilde{\kappa}%
\tilde{\phi}_{m},\tilde{\varsigma}\kappa ^{p}K^{p},\tilde{\varsigma}),
\label{seva}
\end{equation}%
with $S_{\text{ev}T}=0$ at $\tilde{\varsigma}=0$. The argument $\tilde{%
\varsigma}\kappa ^{p}K^{p}$ of $S_{\text{ev}T}^{\prime \prime }$ is there to
remind us that all nontilde products of $\kappa K$ must be multiplied by
parameters $\tilde{\varsigma}$. From now on we assume that the $q_{\text{max}%
}$ of condition (\ref{req2}) is raised to a value that is good for both (\ref%
{rew}) and (\ref{rewa}).

The T1 truncated HD theory has the basic features of a super-renormalizable
theory, since its parameters have non-negative dimensions in units of mass,
and $\tilde{\kappa}$ has a strictly positive dimension. The proof of
super-renormalizability is completed in the next sections, where we show
that the divergences can be renormalized by redefining a few parameters. In
the tilde parametrization, the action $S_{\Lambda T}$ becomes 
\begin{equation}
\tilde{S}_{\Lambda T}=\frac{1}{\tilde{\kappa}^{2}}S_{d\Lambda T}^{\prime
\prime }(\tilde{\kappa}\tilde{\Phi},\tilde{\kappa}\tilde{K},\tilde{\lambda})+%
\frac{1}{\tilde{\kappa}^{2}}S_{\text{ev}T}^{\prime \prime }(\tilde{\kappa}%
\tilde{\Phi},\tilde{\varsigma}\kappa ^{p}K^{p},\tilde{\varsigma})+\frac{1}{%
\tilde{\kappa}^{2}}S_{\mathrm{HD}}^{\prime \prime }(\tilde{\kappa}\tilde{\Phi%
},\tilde{\lambda}_{+})  \label{stl}
\end{equation}%
and $\tilde{\kappa},\tilde{\lambda}_{+}$ are the only tilde parameters that
may have (non-negative) dimensions smaller than or equal to $2t+2|[\kappa ]|$%
. Only the first and third functionals on the right-hand side of (\ref{stl})
have the expected form, which is the tilde version of (\ref{liable2}). The
second functional cannot be written like the rest. This will force us to do
some extra effort. However, since the terms of $S_{\text{ev}T}^{\prime
\prime }$ are multiplied by parameters $\tilde{\varsigma}$, which have
sufficiently large dimensions, we will still be able to prove the properties
we need.

Finally, it is possible to choose $N_{+}$ and $N_{-}$ so that the HD theory
satisfies other properties that will be important for the arguments of the
next subsections. For example, it is sufficient to require 
\begin{equation}
N_{+}+N_{-}>2t-\min_{K}[\kappa K],\qquad N_{+}-N_{-}>2t-\min_{m}[\kappa \phi
_{m}]  \label{req4}
\end{equation}%
to make all products $\tilde{\kappa}\tilde{K}$ and $\tilde{\kappa}\tilde{\phi%
}_{m}$ have dimensions (equal to their weights) greater than $2t$.

Another condition allows us to have control on the dependences on the
antighosts $\bar{C}$ and the Lagrange multipliers $B$. Checking the action (%
\ref{tHD}), we see that $\bar{C}$ and $B$ appear inside the term $-\int BK_{%
\bar{C}}$ of $S_{K}$ (which cannot contribute to nontrivial diagrams), as
well as $(S_{K},\Psi _{\mathrm{HD}})$ and $S_{\text{ev}\Lambda T}$. The
gauge fermion $\Psi _{\mathrm{HD}}$ contains $\bar{C}$ and $B$ according to
the structure (\ref{antigh}), where now the integers $N_{i}$ and $%
N_{i}^{\prime }$ are replaced by $N_{+}$. Since we have suppressed the
parameters $\lambda _{+}$ of $S_{\mathrm{HD}}$ that have $[\tilde{\lambda}%
_{+}]>[\lambda _{+}]$, the terms of $\Psi _{\mathrm{HD}}$ with $0<k<N_{+}$
are absent. Working out $(S_{K},\Psi _{\mathrm{HD}}-\Psi )$ explicitly, it
is easy to prove that at least $N_{+}$ derivatives $\bar{\partial}$ act on
the antighosts $\bar{C}$ and $N_{+}$ derivatives $\bar{\partial}$ act on the
Lagrange multipliers $B$. By construction, the formally evanescent
higher-derivative terms $S_{\text{ev}\Lambda T}-S_{\text{ev}T}$ depend on $%
\bar{C}$ and $B$ in the same way, with derivatives $\bar{\partial}$ possibly
replaced by $\hat{\partial}^{2}/M$. In the end, the dependence on $\bar{C}$
and $B$ of the full higher-derivative sector $S_{\mathrm{HD}}$ of the action 
$S_{\Lambda T}$ has this structure.

When we switch to the tilde parametrization, the powers of $\Lambda $
disappear from the denominators. With the sole exception of $-\int BK_{\bar{C%
}}$, every term of $\tilde{S}_{\Lambda T}$ that depends on $\tilde{\kappa}%
\widetilde{\bar{C}}$ and/or $\tilde{\kappa}\tilde{B}$ is multiplied by a
parameter $\tilde{\lambda}$ or $\tilde{\varsigma}$, or has at least $N_{+}$
derivatives $\bar{\partial}\sim \hat{\partial}^{2}/M$ acting on each leg $%
\tilde{\kappa}\widetilde{\bar{C}}$ and $\tilde{\kappa}\tilde{B}$. It is easy
to check that $(\tilde{S}_{\Lambda T},\tilde{S}_{\Lambda T})$ has the same
structure. These observations will be useful later on, because the
parameters $\tilde{\lambda}$ or $\tilde{\varsigma}$, as well as the
derivatives $\bar{\partial}\sim \hat{\partial}^{2}/M$ acting on the external
legs $\tilde{\kappa}\widetilde{\bar{C}}$ and $\tilde{\kappa}\tilde{B}$,
lower the degrees of divergence of the diagrams, and allow us to prove that
certain types of counterterms and local contributions to anomalies are
absent.

For our purposes, it is sufficient to require that the $N_{+}$ derivatives $%
\bar{\partial}\sim \hat{\partial}^{2}/M$ that act on $\tilde{\kappa}%
\widetilde{\bar{C}}$ and $\tilde{\kappa}\tilde{B}$ inside $S_{\mathrm{HD}}$
have weights greater than $2t$, which means 
\begin{equation}
N_{+}>2t.  \label{req5}
\end{equation}

There is no difficulty to choose $N_{+}$ and $N_{-}$ such that requirements (%
\ref{req1}), (\ref{req2}), (\ref{req4}), and (\ref{req5}) are fulfilled at
the same time, no matter how large we want $t$ to be. In the next
subsections we show that, if we choose $t$ in a clever way, we can ensure
that the higher-derivative theory has no divergences and no local
contributions to anomalies beyond one loop, and that the one-loop
divergences, as well as the one-loop potential anomalies, are independent of
the sources, the matter fields $\phi _{m}$, the antighosts, and the Lagrange
multipliers. We begin by studying the structure of the counterterms.

\subsection{HD theory: structure of counterterms}

\label{s31}

Ignoring the factors $\tilde{\kappa}$ and $\kappa $ attached to the sources $%
\tilde{K}$ and $K$, which are external to the diagrams, each vertex of the
action (\ref{stl}) is multiplied by a power of $\tilde{\kappa}$ that is
equal to the number of its $\Phi $ legs minus 2. Then each loop carries an
extra factor $\tilde{\kappa}^{2}$, and the counterterms have the form 
\begin{equation}
(\tilde{\kappa}^{2})^{L-1}\tilde{\lambda}_{+}^{u}\tilde{\lambda}^{r}\tilde{%
\varsigma}^{s}\partial ^{p}\prod_{g}(\tilde{\kappa}\tilde{\Phi}_{g}^{\prime
})^{q_{g}}\prod_{m}(\tilde{\kappa}\tilde{\phi}_{m})^{q_{m}}\prod_{K}(\tilde{%
\kappa}\tilde{K})^{q_{K}}\prod_{K}(\kappa K)^{q_{K}^{\prime }}  \label{expra}
\end{equation}%
where $u,r,s,p,q_{g}$, $q_{m}$, $q_{K}$, and $q_{K}^{\prime }$ are
non-negative integers. Every factor has a non-negative dimension for $%
L\geqslant 1$, since $[\tilde{\kappa}\tilde{\Phi}]\geqslant \lbrack \kappa
\Phi ]\geqslant 0$ and $[\tilde{\kappa}\tilde{K}]>[\kappa K]\geqslant 1/2$.
Recalling that $[\tilde{\kappa}^{2}]>2t$, we see that, if we choose $t>d/2$,
the expressions (\ref{expra}) have dimensions greater than $d$ for every $%
L\geqslant 2$. Thus, no divergences may be present beyond one loop.
Moreover, at $L=1$ we must have $r=s=0$, because the dimensions of $\tilde{%
\lambda}$ and $\tilde{\varsigma}$ are also greater than $d$. Then, we also
have $q_{K}^{\prime }=0$, because the last product of (\ref{expra}) is
always accompanied by some parameters $\tilde{\varsigma}$. Finally, since by
(\ref{req4}) the dimensions of $\tilde{\kappa}\tilde{\phi}_{m}$ and $\tilde{%
\kappa}\tilde{K}$ are greater than $d$, the divergences of the
higher-derivative theory are just one loop and have the form 
\begin{equation}
\tilde{\Gamma}_{\Lambda T\hspace{0.01in}\text{div}}^{(1)}(\tilde{\kappa}%
\tilde{\Phi}_{g}^{\prime },\tilde{\lambda}_{+})=\Gamma _{\Lambda T\hspace{%
0.01in}\text{div}}^{(1)}(\kappa \Phi _{g}^{\prime },\lambda _{+}).
\label{noneva}
\end{equation}%
To write the last equality we have used the fact that the parameters $%
\lambda _{+}$ with $[\tilde{\lambda}_{+}]>[\lambda _{+}]$ have been switched
off.

We can also show that $\tilde{\Gamma}_{\Lambda T\hspace{0.01in}\text{div}%
}^{(1)}$ cannot depend on the antighosts and the Lagrange multipliers,
since, by the observations of the previous subsection and condition (\ref%
{req5}), a nontrivial Feynman diagram that has $\tilde{\kappa}\widetilde{%
\bar{C}}$ and/or $\tilde{\kappa}\tilde{B}$ among its external legs either is
multiplied by parameters $\tilde{\lambda}$ and $\tilde{\varsigma}$ or has
derivative operators of weights greater than $d$ acting on all external legs 
$\tilde{\kappa}\widetilde{\bar{C}}$ and $\tilde{\kappa}\tilde{B}$. Finally,
since $\tilde{\Gamma}_{\Lambda T\hspace{0.01in}\text{div}}^{(1)}$ has ghost
number zero, it cannot even depend on the ghosts, because we have already
excluded all fields and sources that have negative ghost numbers. In the
end, we have 
\begin{equation}
\tilde{\Gamma}_{\Lambda T\hspace{0.01in}\text{div}}^{(1)}=\tilde{\Gamma}%
_{\Lambda T\hspace{0.01in}\text{div}}^{(1)}(\tilde{\kappa}\tilde{\phi}%
_{g}^{\prime },\tilde{\lambda}_{+})=\Gamma _{\Lambda T\hspace{0.01in}\text{%
div}}^{(1)}(\kappa \phi _{g}^{\prime },\lambda _{+}).  \label{nonev}
\end{equation}%
We stress that $\Gamma _{\Lambda T\hspace{0.01in}\text{div}}^{(1)}$ is
independent of $\Lambda $. Moreover, it is independent of $\Lambda _{-}$,
which implies that it is fully contained in every truncation T2 such that $%
T\geqslant 2\sigma $. From now on we assume that $T$ is larger than $2\sigma 
$.

\subsection{HD theory: structure of anomalies}

\label{s32}

We call \textquotedblleft local contributions to (potential)
anomalies\textquotedblright\ the local terms originated by the
simplification between overall divergences and evanescences in Feynman
diagrams (see section \ref{s6} for details). The local contributions to
anomalies may still be divergent, or nonevanescent, or even evanescent. What
is important for us is that they inherit the basic properties of
divergences. Besides being local, they are polynomial in the parameters that
have positive dimensions. If the gauge anomalies do not vanish at one loop,
the anomaly functional $\mathcal{A}$ receives in general nonlocal
contributions at higher orders. If the gauge anomalies vanish up to and
including $n$ loops, $\mathcal{A}$ receives only local contributions at $n+1$
loops, up to evanescent corrections. In view of the applications of the next
sections, now we investigate the structure of the local contributions to the
gauge anomalies of the HD theory.

We must concentrate on $(\tilde{S}_{\Lambda T},\tilde{S}_{\Lambda T})$ and
the average $\langle (\tilde{S}_{\Lambda T},\tilde{S}_{\Lambda T})\rangle _{%
\tilde{S}_{\Lambda T}}$. Using (\ref{stl}) we find 
\begin{equation}
(\tilde{S}_{\Lambda T},\tilde{S}_{\Lambda T})=\frac{1}{\tilde{\kappa}^{2}}U(%
\tilde{\kappa}\tilde{\Phi},\tilde{\kappa}\tilde{K},\tilde{\lambda},\tilde{%
\lambda}_{+})+\frac{\Lambda ^{-2N_{+}}}{\tilde{\kappa}^{2}}V(\tilde{\kappa}%
\tilde{\Phi},\tilde{\kappa}\tilde{K},\tilde{\varsigma}\kappa ^{p}K^{p},%
\tilde{\lambda},\tilde{\lambda}_{+},\tilde{\varsigma},\Lambda ),  \label{fev}
\end{equation}%
where $U$ and $V$ are formally evanescent functionals, and $V=0$ at $\tilde{%
\varsigma}=0$. We have added the argument $\Lambda $ to $V$, to emphasize
that $V$ can contain positive powers of $\Lambda $, which are generated,
together with the overall factor $\Lambda ^{-2N_{+}}$, by the presence of
nontilde products $\kappa K$ inside $S_{\text{ev}T}^{\prime \prime }$. The
factor $\Lambda ^{-2N_{+}}$ in front of $V$ deserves some attention, because
it can be a source of trouble, from the point of view of power counting. We
can bypass this \ difficulty as follows. Denoting the $\Gamma $ functional
associated with the action $\tilde{S}_{\Lambda T}$ by $\tilde{\Gamma}%
_{\Lambda T}$, the anomaly functional is 
\begin{equation}
\mathcal{\tilde{A}}_{\Lambda T}=(\tilde{\Gamma}_{\Lambda T},\tilde{\Gamma}%
_{\Lambda T})=\langle (\tilde{S}_{\Lambda T},\tilde{S}_{\Lambda T})\rangle _{%
\tilde{S}_{\Lambda T}}=\frac{1}{\tilde{\kappa}^{2}}\left\langle
U\right\rangle _{\tilde{S}_{\Lambda T}}+\frac{\Lambda ^{-2N_{+}}}{\tilde{%
\kappa}^{2}}\left\langle V\right\rangle _{\tilde{S}_{\Lambda T}}.
\label{attilda}
\end{equation}%
It is easy to see that the averages have the following structures: 
\begin{eqnarray}
\frac{1}{\tilde{\kappa}^{2}}\left\langle U\right\rangle _{\tilde{S}_{\Lambda
T}} &=&\sum_{L=0}^{\infty }(\tilde{\kappa}^{2})^{L-1}\mathcal{U}_{L}(\tilde{%
\kappa}\tilde{\Phi},\tilde{\kappa}\tilde{K},\tilde{\varsigma}\kappa
^{p}K^{p},\tilde{\lambda},\tilde{\lambda}_{+},\tilde{\varsigma}),  \label{l1}
\\
\frac{\Lambda ^{-2N_{+}}}{\tilde{\kappa}^{2}}\left\langle V\right\rangle _{%
\tilde{S}_{\Lambda T}} &=&\sum_{L=0}^{\infty }\kappa ^{2}(\tilde{\kappa}%
^{2})^{L-2}\mathcal{V}_{L}(\tilde{\kappa}\tilde{\Phi},\tilde{\kappa}\tilde{K}%
,\tilde{\varsigma}\kappa ^{p}K^{p},\tilde{\lambda},\tilde{\lambda}_{+},%
\tilde{\varsigma},\Lambda ),  \label{l2}
\end{eqnarray}%
where $\mathcal{V}_{L}=0$ at $\tilde{\varsigma}=0$. Recall that $[\tilde{%
\kappa}^{2}]>2t$ and $[\kappa ^{2}\tilde{\varsigma}]>2t$. If we choose a $t$
such that $2t>d+1$ (instead of $2t>d$, which was the condition of the
previous subsection), then all local contributions to anomalies (which must
be integrals of local functions of weight $d+1$) vanish by weighted power
counting for $L\geqslant 2$. Indeed, the right-hand side of (\ref{l1})
contains at least one factor $\tilde{\kappa}^{2}$ times objects of
non-negative weights, while the right-hand side of (\ref{l2}) contains one
factor $\kappa ^{2}\tilde{\varsigma}$ times objects of non-negative weights.

Now we study the functionals $\mathcal{U}_{1}$ and $\mathcal{V}_{1}$. Since
they collect one-loop diagrams that contain insertions of formally
evanescent vertices, they are sums of local divergent evanescences, plus
local nonevanescent terms (which arise from simplified divergences), plus
possibly nonlocal evanescent terms. We concentrate our attention on the
nonevanescent contributions $\mathcal{U}_{1\text{nonev}}$ and $\mathcal{V}_{1%
\text{nonev}}$ to $\mathcal{U}_{1}$ and $\mathcal{V}_{1}$.

The nonevanescent part $\mathcal{U}_{1\text{nonev}}$ of $\mathcal{U}_{1}$ is
independent of $\tilde{\lambda}$, $\tilde{\varsigma}$, $\tilde{\kappa}\tilde{%
K}$, and $\tilde{\kappa}\tilde{\phi}_{m}$, because such objects have weights
greater than $d+1$. Moreover, $\mathcal{U}_{1\text{nonev}}$ is independent
of the antighosts and the Lagrange multipliers, because the choice $2t>d+1$
and the condition (\ref{req5}) ensure that every Feynman diagram that
contributes to $\mathcal{\tilde{A}}_{\Lambda T}$ and has external legs $%
\tilde{\kappa}\widetilde{\bar{C}}$ and/or $\tilde{\kappa}\tilde{B}$ is
either multiplied by parameters $\tilde{\lambda}$ and $\tilde{\varsigma}$ or
has derivative operators of weights greater than $d+1$ acting on each
external leg $\tilde{\kappa}\widetilde{\bar{C}}$ and $\tilde{\kappa}\tilde{B}
$. In this respect, it is important to recall that not only $\tilde{S}%
_{\Lambda T}$ but also $(\tilde{S}_{\Lambda T},\tilde{S}_{\Lambda T})$ has
the structure explained before formula (\ref{req5}). Since $\mathcal{U}_{1%
\text{nonev}}$ has ghost number one, and cannot contain any fields or
sources of negative ghost numbers, it must be proportional to the ghosts.
Precisely, 
\begin{equation}
\mathcal{U}_{1\text{nonev}}=\int (\tilde{\kappa}\tilde{C})^{I}\mathcal{A}%
_{I}(\tilde{\kappa}\tilde{\phi}_{g}^{\prime },\tilde{\lambda}_{+})=\int
(\kappa C)^{I}\mathcal{A}_{I}(\kappa \phi _{g}^{\prime },\lambda _{+}),
\label{f1}
\end{equation}%
where $\mathcal{A}_{I}$ are local functions of the fields $\phi _{g}^{\prime
}$.

The nonevanescent part $\mathcal{V}_{1\text{nonev}}$ of $\mathcal{V}_{1}$
actually vanishes. We know that it must be polynomial in $\tilde{\varsigma}$
and vanish for $\tilde{\varsigma}=0$. If we differentiate the one-loop
contributions to (\ref{attilda}) with respect to $\tilde{\varsigma}$, and
take their nonevanescent parts, we find 
\begin{equation}
\frac{\Lambda ^{-2N_{+}}}{2}\tilde{\varsigma}\frac{\partial \mathcal{V}_{1%
\text{nonev}}}{\partial \tilde{\varsigma}}=\left. \left( \tilde{S}_{\Lambda
T},\tilde{\varsigma}\frac{\partial \tilde{\Gamma}_{\Lambda T\hspace{0.01in}%
}^{(1)}}{\partial \tilde{\varsigma}}\right) \right\vert _{\text{nonev}%
}+\left. \left( \tilde{\Gamma}_{\Lambda T\hspace{0.01in}}^{(1)},\tilde{%
\varsigma}\frac{\partial \tilde{S}_{\Lambda T}}{\partial \tilde{\varsigma}}%
\right) \right\vert _{\text{nonev}},  \label{res}
\end{equation}%
where $\tilde{\Gamma}_{\Lambda T\hspace{0.01in}}^{(1)}$ is the one-loop
contribution to the $\Gamma $ functional $\tilde{\Gamma}_{\Lambda T}$. We
have used the fact that $\mathcal{U}_{1\text{nonev}}$ is independent\ of $%
\tilde{\varsigma}$. Now, $\tilde{\varsigma}\partial \tilde{S}_{\Lambda
T}/\partial \tilde{\varsigma}$ is formally evanescent, so the last term of (%
\ref{res}) vanishes. On the other hand, we have%
\begin{equation}
\tilde{\varsigma}\left. \frac{\partial \tilde{\Gamma}_{\Lambda T\hspace{%
0.01in}}}{\partial \tilde{\varsigma}}\right\vert _{\text{nonev}}^{\text{%
one-loop}}=\left\langle \tilde{\varsigma}\frac{\partial \tilde{S}_{\Lambda T}%
}{\partial \tilde{\varsigma}}\right\rangle _{\text{nonev}}^{\text{one-loop}}.
\label{colleg}
\end{equation}%
The average appearing on the right-hand side of this formula collects the
diagrams that contain one insertion of $\tilde{\varsigma}\partial \tilde{S}%
_{\Lambda T}/\partial \tilde{\varsigma}$. At one loop, the formally
evanescent vertices provided by this functional can give a nonevanescent
result only by simplifying some divergences. Therefore, expression (\ref%
{colleg}) is a local functional. It is equal to the integral of a local
function of dimension $d$ that has the structure (\ref{expra}), with $L=1$
and $s>0$. This means that it vanishes, since $[\tilde{\varsigma}]>d$.
Consequently, (\ref{res}) also vanishes, and so does $\mathcal{V}_{1\text{%
nonev}}$.

In the end, we take 
\begin{equation}
t>\frac{d+1}{2},  \label{nuke}
\end{equation}%
because with this choice ($a$)\ the truncated HD theory is
super-renormalizable, ($b$) there are no divergences and no local
contributions to anomalies beyond one loop, ($c$)\ the one-loop divergences
have the form (\ref{nonev}), and ($d$) the one-loop nonevanescent
contributions to anomalies have the form (\ref{f1}).

We have not discussed the divergent evanescences contained in $\mathcal{U}%
_{1}$ and $\mathcal{V}_{1}$. The reason is that we do not need to, because
as soon as we renormalize the one-loop divergences of the $\Gamma $
functional $\tilde{\Gamma}_{\Lambda T}$, the anomaly functional $\mathcal{%
\tilde{A}}_{\Lambda T}=(\tilde{\Gamma}_{\Lambda T},\tilde{\Gamma}_{\Lambda
T})$ is automatically one-loop convergent.

\subsection{The CDHD limit}

\label{s33}

In the CDHD limit, it is important to avoid conflicts between the
higher-derivative terms contained in the action $S_{\Lambda T}$ and the
powerlike divergences. In particular, if $\Gamma _{nRT}$ denotes the $\Gamma 
$ functional CDHD renormalized up to and including $n$ loops, when we take
the $(n+1)$-loop $\Lambda $-divergent part of expressions such as $(\Gamma
_{nRT},\Gamma _{nRT})$, we have to be sure that $(S_{\mathrm{HD}},\Gamma
_{nRT\hspace{0.01in}\text{div}}^{(n+1)})$ vanishes for $\Lambda \rightarrow
\infty $, where $\Gamma _{nRT\hspace{0.01in}\text{div}}^{(n+1)}$ denotes the 
$(n+1)$-loop divergent part of $\Gamma _{nRT}$. It is impossible to satisfy
this requirement without a truncation, because the powerlike divergences $%
\sim \Lambda ^{k}$ of $\Gamma _{nR\hspace{0.01in}\text{div}}^{(n+1)}$ can
have $k$ arbitrarily large and beat the powers $\Lambda ^{-2N+}$ and $%
\Lambda ^{-2N-}$ that appear in $S_{\Lambda }$. This is the main reason why
we cannot provide a subtraction scheme where the Adler-Bardeen theorem is
manifest to all orders.

Given a truncation, on the other hand, it is possible to fulfill a
satisfactory requirement by choosing higher-derivative regularizing terms $%
S_{\mathrm{HD}}$ that lie \textit{well beyond the truncation} and
subtracting just the contributions to $\Gamma _{nRT\hspace{0.01in}\text{div}%
}^{(n+1)}$ that lie within the truncation. We recall that the truncation T2
of subsection \ref{s21} prescribes that we ignore the $L$-loop contributions
that are $o(1/\Lambda _{-}^{T-2L\sigma })$. We anticipate that, to
provide a scheme where the Adler-Bardeen theorem is almost manifest within
the truncation, we need to satisfy 
\begin{equation}
\lim_{\Lambda \rightarrow \infty }(S_{\mathrm{HD}},\Gamma _{nRT\hspace{0.01in%
}\text{div}}^{(n+1)})=o(1/\Lambda _{-}^{T-2(n+1)\sigma }).  \label{gial}
\end{equation}%
By this formula we mean that the limit exists and vanishes up to corrections 
$o(1/\Lambda _{-}^{T-2(n+1)\sigma })$ (but such corrections may not have a
regular limit for $\Lambda \rightarrow \infty $).

To find a condition that ensures (\ref{gial}), we first observe that the
powerlike divergences of $\Gamma _{nRT\hspace{0.01in}\text{div}}^{(n+1)}$
have the form 
\begin{equation}
\ln ^{q^{\prime }}\Lambda \hspace{0.01in}\frac{\Lambda ^{q}}{\Lambda
_{-}^{q_{-}}}\delta _{+}(\kappa ^{2})^{n}\partial ^{p}(\kappa \Phi
)^{n_{\Phi }}(\kappa K)^{n_{K}},  \label{powdiv}
\end{equation}%
where $q>0$, $q^{\prime },q_{\_}\geqslant 0$, and $\delta _{+}$ is a product
of parameters of non-negative dimensions. We can concentrate on the
contributions (\ref{powdiv}) that have $q_{-}\leqslant T-2(n+1)\sigma $,
because the ones with $q_{-}>T-2(n+1)\sigma $ satisfy (\ref{gial}) in an
obvious way. We know that $[\kappa \Phi ]\geqslant 0$ and $[\kappa
K]\geqslant 1/2$. Then, distinguishing the cases $[\kappa ]\geqslant 0$ and $%
[\kappa ]<0$, we can easily check that 
\begin{equation}
q\leqslant T+d-2\sigma .  \label{qt}
\end{equation}%
In perturbatively unitary, power-counting renormalizable theories with $T=0$
we obviously have $q\leqslant d$.

To ensure that $(S_{\mathrm{HD}},\Gamma _{nRT\hspace{0.01in}\text{div}%
}^{(n+1})$ vanishes for $\Lambda \rightarrow \infty $ within the truncation,
it is sufficient to require $S_{\mathrm{HD}}=\mathcal{O}(1/\Lambda
^{T+d-2\sigma +1})$. In particular, we must have 
\begin{equation}
2N_{+}>2N_{-}>T+d-2\sigma .  \label{nuq}
\end{equation}%
Moreover, the HD\ regularized theory cannot contain higher-derivative terms
of orders $\mathcal{O}(1/\Lambda ^{k})$ with $k\leqslant T+d-2\sigma $.
However, this is an automatic consequence of another choice we have already
made, when we switched off the parameters $\lambda _{+}$ of $S_{\mathrm{HD}}$
such that $[\tilde{\lambda}_{+}]>[\lambda _{+}]$. Thus, in our framework
condition (\ref{nuq}) is sufficient to ensure (\ref{gial}).

Given any truncation $T$, it is always possible to satisfy all the
conditions on $N_{+}$ and $N_{-}$ mentioned so far, at the same time. They
are (\ref{req1}), (\ref{req2}), (\ref{req4}), (\ref{req5}), (\ref{nuke}),
and (\ref{nuq}).

\section{Renormalization of the HD theory}

\label{s4}

\setcounter{equation}{0}

\label{subase}

In this section and the next two, we study the truncated higher-derivative
theory with action $\tilde{S}_{\Lambda T}$, which is defined by keeping $%
\Lambda $ fixed and regularized by means of the CD technique. We mostly use
the tilde parametrization, but sometimes need to switch to the nontilde one.
The first task is to work out the renormalization of this theory. Then we
must study its one-loop anomalies, and finally prove that it satisfies the
manifest Adler-Bardeen theorem.

The anomaly functional (\ref{anom}) of the higher-derivative theory is (\ref%
{attilda}), in the tilde parametrization. Its one-loop contribution $%
\mathcal{\tilde{A}}_{\Lambda T}^{(1)}$ is 
\begin{equation}
\mathcal{\tilde{A}}_{\Lambda T}^{(1)}=2(\tilde{S}_{\Lambda T},\tilde{\Gamma}%
_{\Lambda T}^{(1)})=\left. \langle (\tilde{S}_{\Lambda T},\tilde{S}_{\Lambda
T})\rangle _{\tilde{S}_{\Lambda T}}\right\vert _{\text{one-loop}}.
\label{ana2}
\end{equation}%
We know that $(\tilde{S}_{\Lambda T},\tilde{S}_{\Lambda T})=\mathcal{\hat{O}}%
(\varepsilon )$. The right-hand side of (\ref{ana2}) collects one-loop
Feynman diagrams that contain insertions of formally evanescent vertices.
The formal evanescences can either remain as such or generate factors of $%
\varepsilon $. In the former case, they give local divergent evanescences,
plus evanescences. In the latter case, a factor $\varepsilon $ can simplify
a local divergent part and give local nonevanescent contributions, in
addition to evanescences. Therefore, we can write 
\begin{equation}
\mathcal{\tilde{A}}_{\Lambda T}^{(1)}=\mathcal{\tilde{A}}_{\Lambda T\hspace{%
0.01in}\text{nev}}^{(1)}+\mathcal{\tilde{A}}_{\Lambda T\hspace{0.01in}\text{%
divev}}^{(1)}+\mathcal{\tilde{A}}_{\Lambda T\hspace{0.01in}\text{ev}}^{(1)},
\label{adivev}
\end{equation}%
where $\mathcal{\tilde{A}}_{\Lambda T\hspace{0.01in}\text{nev}}^{(1)}$ is
local, convergent, and nonevanescent, $\mathcal{\tilde{A}}_{\Lambda T\hspace{%
0.01in}\text{divev}}^{(1)}$ is local and divergent evanescent and $\mathcal{%
\tilde{A}}_{\Lambda T\hspace{0.01in}\text{ev}}^{(1)}$ is evanescent and
possibly nonlocal. The analysis of subsection \ref{s32} and formula (\ref{f1}%
)\ tell us that 
\begin{equation}
\mathcal{\tilde{A}}_{\Lambda T\hspace{0.01in}\text{nev}}^{(1)}=\int (\tilde{%
\kappa}\tilde{C})^{I}\mathcal{A}_{I}(\tilde{\kappa}\tilde{\phi}_{g}^{\prime
},\tilde{\lambda}_{+})=\int (\kappa C)^{I}\mathcal{A}_{I}(\kappa \phi
_{g}^{\prime },\lambda _{+}).  \label{algol}
\end{equation}%
Clearly, $\mathcal{\tilde{A}}_{\Lambda T\hspace{0.01in}\text{nev}}^{(1)}$ is
independent of $\Lambda _{-}$ and $\Lambda $. In particular, it is fully
contained in any truncation that has $T\geqslant 2\sigma $.

Taking the divergent part of equation (\ref{ana2}), we find 
\begin{equation}
(\tilde{S}_{\Lambda T},\tilde{\Gamma}_{\Lambda T\hspace{0.01in}\text{div}%
}^{(1)})=\frac{1}{2}\mathcal{\tilde{A}}_{\Lambda T\hspace{0.01in}\text{divev}%
}^{(1)}.  \label{noveva}
\end{equation}%
Formula (\ref{nonev}) tells us that $\tilde{\Gamma}_{\Lambda T\hspace{0.01in}%
\text{div}}^{(1)}$ is just a functional of $\tilde{\kappa}\tilde{\phi}%
_{g}^{\prime }$, fully contained within any truncation T2 with $T\geqslant
2\sigma $. In particular, its antiparentheses with $\tilde{S}_{\Lambda T}$
are only sensitive to $\tilde{S}_{K}$ and the $K$-dependent contributions to 
$\tilde{S}_{\text{ev}T}$. Moreover, we can further decompose $\tilde{\Gamma}%
_{\Lambda T\hspace{0.01in}\text{div}}^{(1)}$ as the sum of a nonevanescent
divergent part $\tilde{\Gamma}_{\Lambda T\hspace{0.01in}\text{nev\hspace{%
0.01in}div}}^{(1)}$ and a divergent evanescence $\tilde{\Gamma}_{\Lambda T%
\hspace{0.01in}\text{divev}}^{(1)}$. Taking the nonevanescent divergent part
of (\ref{noveva}),\ we obtain 
\begin{equation}
(\tilde{S}_{K},\tilde{\Gamma}_{\Lambda T\hspace{0.01in}\text{nev\hspace{%
0.01in}div}}^{(1)})=0,  \label{stg}
\end{equation}%
which just states that $\tilde{\Gamma}_{\Lambda T\hspace{0.01in}\text{nev%
\hspace{0.01in}div}}^{(1)}$ is gauge invariant.

Since $\tilde{\Gamma}_{\Lambda T\hspace{0.01in}\text{div}}^{(1)}$ is $%
\Lambda _{-}$ independent, the arguments that lead to formula (\ref{custom})
ensure that $\tilde{\Gamma}_{\Lambda T\hspace{0.01in}\text{nev\hspace{0.01in}%
div}}^{(1)}$ is a linear combination of the invariants $\mathcal{G}_{i}$
contained in the T1 truncated classical action $S_{cT}(\phi )$ with $%
T=2\sigma $ [check formula (\ref{scf})]. Since we are assuming $T\geqslant
2\sigma $, we can remove $\tilde{\Gamma}_{\Lambda T\hspace{0.01in}\text{nev%
\hspace{0.01in}div}}^{(1)}$ by redefining a few parameters $\lambda _{i}$ of 
$S_{cT}$. The rest, which is $\tilde{\Gamma}_{\Lambda T\hspace{0.01in}\text{%
divev}}^{(1)}$, can be subtracted by redefining the parameters $\varsigma $
and $\eta $ of $S_{\text{ev}T}$.

In the case of the standard model coupled to quantum gravity, $\Gamma
_{\Lambda T\hspace{0.01in}\text{nev\hspace{0.01in}div}}^{(1)}$ is a linear
combination of terms of dimensions smaller than or equal to four, such as%
\begin{eqnarray*}
\Gamma _{\Lambda T\hspace{0.01in}\text{nev\hspace{0.01in}div}}^{(1)} &=&\int 
\sqrt{|g|}\left( c_{1}+c_{2}R+c_{3}R^{2}+c_{4}R_{\bar{\mu}\bar{\nu}}R^{\bar{%
\mu}\bar{\nu}}+c_{5}\kappa ^{2}F_{\bar{\mu}\bar{\nu}}^{a}F^{a\bar{\mu}\bar{%
\nu}}+c_{6}\kappa ^{2}F_{\bar{\mu}\bar{\nu}}^{a}\mathcal{D}^{2}F^{a\bar{\mu}%
\bar{\nu}}\right. \\
&&\qquad \qquad \qquad +\left. c_{7}\kappa ^{2}RF_{\bar{\mu}\bar{\nu}%
}^{a}F^{a\bar{\mu}\bar{\nu}}+c_{8}\kappa ^{4}F_{\bar{\mu}\bar{\nu}}^{a}F^{a%
\bar{\mu}\bar{\nu}}F_{\bar{\rho}\bar{\sigma}}^{b}F^{b\bar{\rho}\bar{\sigma}%
}+\cdots \right)
\end{eqnarray*}%
where the coefficients $c_{i}$ are products of parameters of non-negative
dimensions. This list also contains invariants that in principle can be
subtracted by means of field redefinitions, rather than redefinitions of
parameters. Among those invariants, we mention $\int \sqrt{|g|}R^{2}$ and $%
\int \sqrt{|g|}R_{\bar{\mu}\bar{\nu}}R^{\bar{\mu}\bar{\nu}}$. However, if we
use the Einstein equations, which read%
\begin{equation*}
R_{\bar{\mu}\bar{\nu}}-\frac{1}{2}Rg_{\bar{\mu}\bar{\nu}}-\Lambda _{\text{c}%
}g_{\bar{\mu}\bar{\nu}}=\kappa ^{2}T_{\bar{\mu}\bar{\nu}},
\end{equation*}%
where the energy-momentum tensor $T_{\bar{\mu}\bar{\nu}}$ can contain purely
gravitational contributions due to the higher-derivative corrections, we do
not really remove the invariants in question, but rather convert them into
other invariants, such as $\int \sqrt{|g|}\kappa ^{4}T_{\bar{\mu}\bar{\nu}%
}T^{\bar{\mu}\bar{\nu}}$, which may depend on the matter fields $\phi _{m}$
and spoil the nice structure of the HD theory. For this reason, it is not
convenient to use canonical transformations to remove $\Gamma _{\Lambda T%
\hspace{0.01in}\text{nev\hspace{0.01in}div}}^{(1)}$, or parts of it. As
anticipated in section \ref{s2}, all the invariants of $\Gamma _{\Lambda T%
\hspace{0.01in}\text{nev\hspace{0.01in}div}}^{(1)}$ are included in the
basis $\{\mathcal{G}_{i}(\phi )\}$, so we can completely remove $\Gamma
_{\Lambda T\hspace{0.01in}\text{nev\hspace{0.01in}div}}^{(1)}$ by redefining
the parameters $\lambda _{i}$. We recall that it is possible to get rid of
the redundant invariants at the very end (after subtracting the $\Lambda $
divergences and proving the almost manifest Adler-bardeen theorem), by means
of a procedure like the one described in ref. \cite{ABward}, which consists
of making a canonical transformation, re-renormalize the theory, and
re-fine-tune the finite local counterterms to recover the cancellation of
gauge anomalies.

In the end, to renormalize the HD theory we just need to redefine some
parameters $\lambda _{i}$, $\varsigma $ and $\eta $ of $S_{cT}$, and $S_{%
\text{ev}T}$, which multiply terms of the form (\ref{sctr}). The
renormalized action, which we denote by $\hat{S}_{\Lambda T}$, is obtained
by making the replacements 
\begin{equation}
\tilde{\lambda}_{i}\rightarrow \tilde{\lambda}_{i}+\frac{f_{i}}{\varepsilon }%
\tilde{\kappa}^{2},\text{\qquad }\tilde{\varsigma}\rightarrow \tilde{%
\varsigma}+\frac{f_{\varsigma }}{\varepsilon }\tilde{\kappa}^{2},\text{%
\qquad }\tilde{\eta}\rightarrow \tilde{\eta}+\frac{f_{\eta }}{\varepsilon }%
\tilde{\kappa}^{2},  \label{rena0}
\end{equation}%
inside $S_{\Lambda T}$, where $f_{i}$, $f_{\varsigma }$, and $f_{\eta }$ are
calculable factors that may depend on the parameters $\tilde{\lambda}_{+}$
appearing in (\ref{nonev}). Switching to the nontilde parametrization, the
redefinitions (\ref{rena0}) are equivalent to 
\begin{equation}
\lambda _{i}\rightarrow \lambda _{i}+\frac{f_{i}}{\varepsilon }\kappa ^{2},%
\text{\qquad }\varsigma \rightarrow \varsigma +\frac{f_{\varsigma }}{%
\varepsilon }\kappa ^{2}\text{\qquad }\eta \rightarrow \eta +\frac{f_{\eta }%
}{\varepsilon }\kappa ^{2}.  \label{rena}
\end{equation}

Since $S_{\Lambda T}$ is linear in $\lambda _{i}$, $\varsigma $, and $\eta $%
, we have 
\begin{equation}
\hat{S}_{\Lambda T}=\tilde{S}_{\Lambda T}-\tilde{\Gamma}_{\Lambda T\hspace{%
0.01in}\text{div}}^{(1)}.  \label{hatta}
\end{equation}%
Using (\ref{noveva}) and $(\tilde{\Gamma}_{\Lambda T\hspace{0.01in}\text{div}%
}^{(1)},\tilde{\Gamma}_{\Lambda T\hspace{0.01in}\text{div}}^{(1)})=0$ (which
holds because $\tilde{\Gamma}_{\Lambda T\hspace{0.01in}\text{div}}^{(1)}$ is 
$K$ independent), we find 
\begin{equation}
(\hat{S}_{\Lambda T},\hat{S}_{\Lambda T})=(\tilde{S}_{\Lambda T},\tilde{S}%
_{\Lambda T})-\mathcal{\tilde{A}}_{\Lambda T\hspace{0.01in}\text{divev}%
}^{(1)}.  \label{hattadiv}
\end{equation}

The generating functional $\hat{\Gamma}_{\Lambda T}$ defined by $\hat{S}%
_{\Lambda T}$ is convergent to all orders within the truncation, because it
is convergent at one loop and the tilde structure of $\tilde{\Gamma}%
_{\Lambda T\hspace{0.01in}\text{div}}^{(1)}$ has the expected form, that is
to say, the tilde version of (\ref{liable2}). Then, the counterterms keep
the form (\ref{expra}), which forbids divergences beyond one loop. Finally, $%
\hat{\Gamma}_{\Lambda T}$ and the anomaly functional $\mathcal{\hat{A}}%
_{\Lambda T}=(\hat{\Gamma}_{\Lambda T},\hat{\Gamma}_{\Lambda T})$ are
obtained by making the replacements (\ref{rena0}) inside $\tilde{\Gamma}%
_{\Lambda T}$ and $\mathcal{\tilde{A}}_{\Lambda T}=(\tilde{\Gamma}_{\Lambda
T},\tilde{\Gamma}_{\Lambda T})$, respectively. Clearly, $\mathcal{\hat{A}}%
_{\Lambda T}$ is convergent, because $\hat{\Gamma}_{\Lambda T}$ is
convergent.

\section{One-loop anomalies}

\label{s5}

\setcounter{equation}{0}

In this section we study the one-loop anomalies and relate those of the
basic theory, which are trivial by assumption (IV) of subsection \ref{key},
to those of the HD theory, which turn out to also be trivial.

We begin with the relation between the one-loop contributions $\mathcal{\hat{%
A}}_{\Lambda T}^{(1)}$ and $\mathcal{\tilde{A}}_{\Lambda T\hspace{0.01in}%
}^{(1)}$ to $\mathcal{\hat{A}}_{\Lambda T}$ and $\mathcal{\tilde{A}}%
_{\Lambda T}$. Observe that 
\begin{equation*}
\mathcal{\hat{A}}_{\Lambda T}=(\hat{\Gamma}_{\Lambda T},\hat{\Gamma}%
_{\Lambda T})=\langle (\hat{S}_{\Lambda T},\hat{S}_{\Lambda T})\rangle _{%
\hat{S}_{\Lambda T}}=\langle (\hat{S}_{\Lambda T},\hat{S}_{\Lambda
T})\rangle _{\tilde{S}_{\Lambda T}-\tilde{\Gamma}_{\Lambda T\hspace{0.01in}%
\text{div}}^{(1)}}=\langle (\hat{S}_{\Lambda T},\hat{S}_{\Lambda T})\rangle
_{\tilde{S}_{\Lambda T}}+\mathcal{O}(\hbar ^{2}).
\end{equation*}%
The last equality is proved as follows. The functional $\mathcal{\hat{A}}%
_{\Lambda T}$ collects the one-particle irreducible diagrams that contain
one insertion of a vertex coming from $(\hat{S}_{\Lambda T},\hat{S}_{\Lambda
T})$. If we also use $\mathcal{O}(\hbar )$ vertices provided by $\tilde{%
\Gamma}_{\Lambda T\hspace{0.01in}\text{div}}^{(1)}$, we must close at least
one loop, to connect them with the vertex of $(\hat{S}_{\Lambda T},\hat{S}%
_{\Lambda T})$. This can only give $\mathcal{O}(\hbar ^{2})$ corrections.

Using (\ref{hattadiv}), we have 
\begin{equation*}
\mathcal{\hat{A}}_{\Lambda T}=\langle (\tilde{S}_{\Lambda T},\tilde{S}%
_{\Lambda T})\rangle _{\tilde{S}_{\Lambda T}}-\mathcal{\tilde{A}}_{\Lambda T%
\hspace{0.01in}\text{divev}}^{(1)}+\mathcal{O}(\hbar ^{2})=\mathcal{\tilde{A}%
}_{\Lambda T}-\mathcal{\tilde{A}}_{\Lambda T\hspace{0.01in}\text{divev}%
}^{(1)}+\mathcal{O}(\hbar ^{2}),
\end{equation*}%
and thus (\ref{adivev}) gives 
\begin{equation}
\mathcal{\hat{A}}_{\Lambda T}^{(1)}=\mathcal{\tilde{A}}_{\Lambda T\hspace{%
0.01in}\text{nev}}^{(1)}+\mathcal{\tilde{A}}_{\Lambda T\hspace{0.01in}\text{%
ev}}^{(1)}.  \label{adivevhat}
\end{equation}%
The divergent evanescences $\mathcal{\tilde{A}}_{\Lambda T\hspace{0.01in}%
\text{divev}}^{(1)}$ had to disappear from $\mathcal{\hat{A}}_{\Lambda
T}^{(1)}$, because $\mathcal{\hat{A}}_{\Lambda T}$ is convergent.

Since the structure of $\tilde{\Gamma}_{\Lambda T\hspace{0.01in}\text{div}%
}^{(1)}$ is the one of formula (\ref{nonev}), we can straightforwardly
extend the analysis of subsection \ref{s32} to the renormalized action $\hat{%
S}_{\Lambda T}$. The anomaly functional is still the sum of contributions of
the forms (\ref{l1}) and (\ref{l2}). Therefore, all local contributions to
anomalies vanish from two loops onwards.

Anomalies satisfy the Wess-Zumino consistency conditions \cite{wesszumino},
which, in the Batalin-Vilkovisky formalism, are consequences of a well-known
property of the antiparentheses, stating that every functional $X$ satisfies
the identity $(X,(X,X))=0$. Taking $X=\hat{\Gamma}_{\Lambda T}$, we obtain 
\begin{equation}
(\hat{\Gamma}_{\Lambda T},\mathcal{\hat{A}}_{\Lambda T})=0.  \label{acca}
\end{equation}%
At one loop we have 
\begin{equation}
(\tilde{S}_{\Lambda T},\mathcal{\hat{A}}_{\Lambda T}^{(1)})=-(\hat{\Gamma}%
_{\Lambda T}^{(1)},(\tilde{S}_{\Lambda T},\tilde{S}_{\Lambda T})).
\label{rhs}
\end{equation}%
Since the antiparentheses of an evanescent functional, such as $(\tilde{S}%
_{\Lambda },\tilde{S}_{\Lambda })$, with a convergent functional, such as $%
\hat{\Gamma}_{\Lambda T}^{(1)}$, are evanescent, we have 
\begin{equation*}
(\tilde{S}_{\Lambda T},\mathcal{\hat{A}}_{\Lambda T}^{(1)})=\mathcal{O}%
(\varepsilon ).
\end{equation*}%
Using (\ref{adivevhat}) we also find 
\begin{equation}
(\tilde{S}_{\Lambda T},\mathcal{\tilde{A}}_{\Lambda T\hspace{0.01in}\text{nev%
}}^{(1)})=\mathcal{O}(\varepsilon ).  \label{ali}
\end{equation}%
By formula (\ref{algol}), $\mathcal{\tilde{A}}_{\Lambda T\hspace{0.01in}%
\text{nev}}^{(1)}$ is independent of the sources $K$. Then, only the $K$%
-dependent terms of $\tilde{S}_{\Lambda T}$, which are contained in $\tilde{S%
}_{K}$ and $\tilde{S}_{\text{ev}T}$, can contribute to the left-hand side of
(\ref{ali}). Taking the nonevanescent part of both sides, we find 
\begin{equation}
(\tilde{S}_{K},\mathcal{\tilde{A}}_{\Lambda T\hspace{0.01in}\text{nev}%
}^{(1)})=0.  \label{aqua}
\end{equation}

\subsection*{Relation between the anomalies of the HD theory and those of
the basic theory}

Now we relate the potential one-loop anomalies $\mathcal{\tilde{A}}_{\Lambda
T\hspace{0.01in}\text{nev}}^{(1)}$ of the HD theory to the potential
one-loop anomalies $\mathcal{A}_{\hspace{0.01in}\text{b}}^{(1)}$ of the
basic theory, which are trivial by assumption (IV) of subsection \ref{key}.
We recall that the action $S_{d\text{b}}$ of the basic theory can be
retrieved by taking the formal limit $\Lambda _{-}\rightarrow \infty $ of $%
S_{dT}$. In the same limit, the CD regularized action $S_{T}$ is equal to
the basic action $S_{d\text{b}}$ plus the evanescent terms $S_{\text{ev}T}$
(calculated at $\Lambda _{-}=\infty $). The CDHD regularized action is still
obtained by adding $S_{\mathrm{HD}}$ (which is $\Lambda _{-}$ independent),
or by taking the formal limit $\Lambda _{-}\rightarrow \infty $ of $%
S_{\Lambda T}$.

Once the formal limit $\Lambda _{-}\rightarrow \infty $ is taken, the
one-loop CDHD divergences must be subtracted just as they come, rather than
by redefining parameters (since the basic action misses the parameters of
the subset $s_{-}$). For example, the one-loop divergences $\tilde{\Gamma}%
_{\Lambda T\hspace{0.01in}\text{div}}^{(1)}$ of the HD theory can still be
subtracted by formula (\ref{hatta}), which, however, cannot be seen as
implied by the redefinitions (\ref{rena0}) or (\ref{rena}). In this section
we understand that $\Lambda _{-}=\infty $ everywhere, so the final theory is
the one associated with the basic action. Since $\tilde{\Gamma}_{\Lambda T%
\hspace{0.01in}\text{div}}^{(1)}$ and $\mathcal{\tilde{A}}_{\Lambda T\hspace{%
0.01in}\text{nev}}^{(1)}$ do not depend on $\Lambda _{-}$, we do not lose
any relevant information.

The last expression of formula (\ref{algol}) tells us that $\mathcal{\tilde{A%
}}_{\Lambda T\hspace{0.01in}\text{nev}}^{(1)}$ is $\Lambda $ independent in
the nontilde parametrization, where we denote it by $\mathcal{A}_{\Lambda T%
\hspace{0.01in}\text{nev}}^{(1)}$. Now we show that actually $\mathcal{A}%
_{\Lambda T\hspace{0.01in}\text{nev}}^{(1)}$ is equivalent to the one-loop
anomaly $\mathcal{A}_{\hspace{0.01in}\text{b}}^{(1)}$ of the basic theory.

To prove this fact, we need to study the $\Lambda $-divergent parts and take
the CDHD\ limit at one loop. In this subsection we denote the terms that are 
$\Lambda $ divergent in the CDHD\ limit as \textquotedblleft
Ddiv\textquotedblright , to distinguish them from the poles in $\varepsilon $%
. Recall that the $\Lambda $ divergences can be nonevanescent or formally
evanescent, from the point of view of the dimensional regularization, but
not analytically evanescent. They are the terms $\varepsilon ^{0}\Lambda $
and $\hat{\delta}\varepsilon ^{0}\Lambda $ of the list (\ref{assu}).

Consider $\mathcal{\hat{A}}_{\Lambda T}=(\hat{\Gamma}_{\Lambda T},\hat{\Gamma%
}_{\Lambda T})$ and take the one-loop CDHD-divergent part of this equation.
Using (\ref{adivevhat}) and recalling that $\mathcal{A}_{\Lambda T\hspace{%
0.01in}\text{nev}}^{(1)}$ is $\Lambda $ independent, we get 
\begin{equation}
\frac{1}{2}\left. \mathcal{A}_{\Lambda T\hspace{0.01in}\text{ev\hspace{0.01in%
}}}^{(1)}\right\vert _{\text{Ddiv}}=\left. (S_{\Lambda T},\hat{\Gamma}%
_{\Lambda T}^{(1)})\right\vert _{\text{Ddiv}}=\left. (S_{\Lambda T},\hat{%
\Gamma}_{\Lambda T\hspace{0.01in}\text{Ddiv}}^{(1)})\right\vert _{\text{Ddiv}%
}=(S_{T},\hat{\Gamma}_{\Lambda T\hspace{0.01in}\text{Ddiv}}^{(1)})+\left.
(S_{\mathrm{HD}},\hat{\Gamma}_{\Lambda T\hspace{0.01in}\text{Ddiv}%
}^{(1)})\right\vert _{\text{Ddiv}},  \label{bingo}
\end{equation}%
where $\hat{\Gamma}_{\Lambda T\hspace{0.01in}\text{Ddiv}}^{(1)}$ is the
one-loop CDHD-divergent part of $\hat{\Gamma}_{\Lambda T}$. Note that
although $\left. \mathcal{A}_{\Lambda T\hspace{0.01in}\text{ev\hspace{0.01in}%
}}^{(1)}\right\vert _{\text{Ddiv}}$ is evanescent from the point of view of
the CD regularization, it can be nontrivial, because it can contain the
terms $\hat{\delta}\varepsilon ^{0}\Lambda $ of the list (\ref{assu}).

The one-loop powerlike divergences at $\Lambda _{-}=\infty $ have the form 
\begin{equation*}
\Lambda ^{q}\hspace{0.01in}\hspace{0.01in}\delta _{+}\partial ^{p}(\kappa
\Phi )^{n_{\Phi }}(\kappa K)^{n_{K}},
\end{equation*}%
where $q>0$, and $\delta _{+}$ is a product of parameters of non-negative
dimensions. Recalling that $[\kappa \Phi ]\geqslant 0$ and $[\kappa
K]\geqslant 1/2$, the exponent $q$ is smaller than or equal to $d$. Since $%
T\geqslant 2\sigma $ and $S_{\mathrm{HD}}=\mathcal{O}(1/\Lambda
^{T+d-2\sigma +1})$, by inequality (\ref{nuq}), the antiparentheses $(S_{%
\mathrm{HD}},\hat{\Gamma}_{\Lambda T\hspace{0.01in}\text{Ddiv}}^{(1)})$,
specialized to the basic theory, tend to zero in the CDHD limit. Thus, (\ref%
{bingo}) gives%
\begin{equation}
\frac{1}{2}\left. \mathcal{A}_{\Lambda T\hspace{0.01in}\text{ev\hspace{0.01in%
}}}^{(1)}\right\vert _{\text{Ddiv}}=(S_{T},\hat{\Gamma}_{\Lambda T\hspace{%
0.01in}\text{Ddiv}}^{(1)}).  \label{bingo2}
\end{equation}

The one-loop CDHD-renormalized action $\hat{S}_{fT\hspace{0.01in}}$ of the
final theory associated with the basic action reads 
\begin{equation}
\hat{S}_{f\hspace{0.01in}T}=\hat{S}_{\Lambda T}-\hat{\Gamma}_{\Lambda T%
\hspace{0.01in}\text{Ddiv}}^{(1)}-\hat{\Gamma}_{\Lambda T\hspace{0.01in}%
\text{fin}}^{(1)}+\mathcal{O}(\hbar ^{2}),  \label{subtra}
\end{equation}%
where $\hat{\Gamma}_{\Lambda T\hspace{0.01in}\text{fin}}^{(1)}$ denote
arbitrary local counterterms that are finite and nonevanescent in the CDHD\
limit [i.e. terms of the type $\varepsilon ^{0}\Lambda ^{0}$ of the list (%
\ref{survo})]. For the purposes of this section, the generic subtraction (%
\ref{subtra}) is enough. In section \ref{s7} we will be more precise about
the removal of divergences (at $\Lambda _{-}<\infty $), as well as the
finite local counterterms $\hat{\Gamma}_{\Lambda T\hspace{0.01in}\text{fin}%
}^{(1)}$ and the higher-order corrections $\mathcal{O}(\hbar ^{2})$. The
anomaly is then 
\begin{equation*}
\mathcal{A}_{fT}=\langle (\hat{S}_{f\hspace{0.01in}T},\hat{S}_{fT\hspace{%
0.01in}})\rangle _{\hat{S}_{fT\hspace{0.01in}}},
\end{equation*}%
and its one-loop nonevanescent part $\mathcal{A}_{\text{b}}^{(1)}$ is the
quantity we want. Denoting the sum $\hat{\Gamma}_{\Lambda T\hspace{0.01in}%
\text{Ddiv}}^{(1)}+\hat{\Gamma}_{\Lambda T\hspace{0.01in}\text{fin}}^{(1)}$
by $\Delta \hat{\Gamma}_{\Lambda T\hspace{0.01in}}^{(1)}$ and using (\ref%
{adivevhat}), we find 
\begin{eqnarray}
&&\mathcal{A}_{fT}=\langle (\hat{S}_{\Lambda T}-\Delta \hat{\Gamma}_{\Lambda
T\hspace{0.01in}}^{(1)},\hat{S}_{\Lambda T}-\Delta \hat{\Gamma}_{\Lambda T%
\hspace{0.01in}}^{(1)})\rangle _{\hat{S}_{\Lambda T}-\Delta \hat{\Gamma}%
_{\Lambda T\hspace{0.01in}}^{(1)}}+\mathcal{O}(\hbar ^{2})=\mathcal{\hat{A}}%
_{\Lambda T}-2(S_{\Lambda T},\Delta \hat{\Gamma}_{\Lambda T\hspace{0.01in}%
}^{(1)})+\mathcal{O}(\hbar ^{2})  \notag \\
&&\,=(S_{\Lambda T},S_{\Lambda T})+\mathcal{A}_{\Lambda T\hspace{0.01in}%
\text{nev}}^{(1)}+\mathcal{A}_{\Lambda T\hspace{0.01in}\text{ev}%
}^{(1)}-2(S_{T},\Delta \hat{\Gamma}_{\Lambda T\hspace{0.01in}}^{(1)})-2(S_{%
\mathrm{HD}},\Delta \hat{\Gamma}_{\Lambda T\hspace{0.01in}}^{(1)})+\mathcal{O%
}(\hbar ^{2}).  \label{7}
\end{eqnarray}%
In these manipulations we have used the formula 
\begin{equation*}
\mathcal{\hat{A}}_{\Lambda T}=\langle (\hat{S}_{\Lambda T},\hat{S}_{\Lambda
T})\rangle _{\hat{S}_{\Lambda T}}=\langle (\hat{S}_{\Lambda T},\hat{S}%
_{\Lambda T})\rangle _{\hat{S}_{\Lambda T}-\Delta \hat{\Gamma}_{\Lambda T%
\hspace{0.01in}}^{(1)}}+\mathcal{O}(\hbar ^{2}),
\end{equation*}%
which holds because at one loop the vertices of $\Delta \hat{\Gamma}%
_{\Lambda T\hspace{0.01in}}^{(1)}$, which are already $\mathcal{O}(\hbar )$,
cannot contribute to one-particle irreducible diagrams that contain one
insertion of $(\hat{S}_{\Lambda T},\hat{S}_{\Lambda T})$.

At one loop, using (\ref{bingo2}), we obtain 
\begin{equation}
\mathcal{A}_{fT}^{(1)}=\mathcal{A}_{\Lambda T\hspace{0.01in}\text{nev}%
}^{(1)}+\mathcal{A}_{\Lambda T\hspace{0.01in}\text{ev}}^{(1)}-\left. 
\mathcal{A}_{\Lambda T\hspace{0.01in}\text{ev\hspace{0.01in}}%
}^{(1)}\right\vert _{\text{Ddiv}}-2(S_{T},\hat{\Gamma}_{\Lambda T\hspace{%
0.01in}\text{fin}}^{(1)})-2(S_{\mathrm{HD}},\Delta \hat{\Gamma}_{\Lambda T%
\hspace{0.01in}}^{(1)}).  \label{8}
\end{equation}

Now we take the CDHD limit. Since $\Delta \hat{\Gamma}_{\Lambda T\hspace{%
0.01in}}^{(1)}$ is $\Lambda $ independent, the antiparentheses $(S_{\mathrm{%
HD}},\Delta \hat{\Gamma}_{\Lambda T\hspace{0.01in}}^{(1)})$ vanish when $%
\Lambda \rightarrow \infty $. Moreover, $\mathcal{A}_{\Lambda T\hspace{0.01in%
}\text{nev}}^{(1)}$ is independent of $\Lambda $. On the other hand, $%
\mathcal{A}_{\Lambda T\hspace{0.01in}\text{ev}}^{(1)}-\left. \mathcal{A}%
_{\Lambda T\hspace{0.01in}\text{ev\hspace{0.01in}}}^{(1)}\right\vert _{\text{%
Ddiv}}$ vanishes in the CDHD limit, because the terms $\hat{\delta}%
\varepsilon ^{0}\Lambda $ are subtracted away in the difference. Since $%
S_{T}-S_{d\text{b}}=\mathcal{O}(\varepsilon )$ at $\Lambda _{-}=\infty $, we
can replace $(S_{T},\hat{\Gamma}_{\Lambda T\hspace{0.01in}\text{fin}}^{(1)})$
by $(S_{d\text{b}},\hat{\Gamma}_{\Lambda T\hspace{0.01in}\text{fin}}^{(1)})$%
. Finally, using formula (\ref{algol}) we get 
\begin{equation}
\mathcal{A}_{\text{b}}^{(1)}=\mathcal{A}_{\Lambda T\hspace{0.01in}\text{nev}%
}^{(1)}-2(S_{d\text{b}},\hat{\Gamma}_{\Lambda T\hspace{0.01in}\text{fin}%
}^{(1)})=\int (\kappa C)^{I}\mathcal{A}_{I}(\kappa \phi _{g}^{\prime
},\lambda _{+})-2(S_{d\text{b}},\hat{\Gamma}_{\Lambda T\hspace{0.01in}\text{%
fin}}^{(1)}).  \label{line2}
\end{equation}%
In particular, by formula (\ref{aqua}) and $(S_{d\text{b}},S_{d\text{b}})=0$%
, the one-loop anomaly functional $\mathcal{A}_{\text{b}}^{(1)}$ of the
basic theory solves the condition%
\begin{equation}
(S_{d\text{b}},\mathcal{A}_{\text{b}}^{(1)})=0.  \label{sdbab}
\end{equation}

At this point, we are ready to use assumption (IV) of subsection \ref{key},
which tells us that there exists a local functional $\mathcal{X}(\Phi ,K)$
such that $\mathcal{A}_{\hspace{0.01in}\text{b}}^{(1)}=(S_{d\text{b}},%
\mathcal{X})$. Using this piece of information and (\ref{line2}), we obtain 
\begin{equation}
\mathcal{A}_{\Lambda T\hspace{0.01in}\text{nev}}^{(1)}=\int (\kappa C)^{I}%
\mathcal{A}_{I}(\kappa \phi _{g}^{\prime },\lambda _{+})=(S_{d\text{b}},%
\mathcal{X}^{\prime })  \label{ju}
\end{equation}%
for $\mathcal{X}^{\prime }=\mathcal{X}+2\hat{\Gamma}_{\Lambda T\hspace{0.01in%
}\text{fin}}^{(1)}$.

We know that the functional $\mathcal{A}_{\Lambda T\hspace{0.01in}\text{nev}%
}^{(1)}$ satisfies both (\ref{aqua}) and (\ref{ju}). To subtract it in a way
that preserves the structure of the HD\ theory, we need to know that, in
addition, we can find a $K$-independent $\mathcal{X}^{\prime }$. This is
ensured by assumption (V)\ of subsection \ref{key}, which tells us that
there exists a local functional of vanishing ghost number $\chi (\kappa \Phi
,\lambda _{+})$, equal to the integral of a local function of dimension $d$,
such that 
\begin{equation}
\mathcal{A}_{\Lambda T\hspace{0.01in}\text{nev}}^{(1)}=(S_{K},\chi ).
\label{anomcanc}
\end{equation}%
Since $\mathcal{A}_{\Lambda T\hspace{0.01in}\text{nev}}^{(1)}$ is $\phi _{m}$
independent, we can assume that $\chi $ is also $\phi _{m}$ independent.
Indeed, recall that the transformations $R_{g}(\Phi )$ of the fields $\Phi
_{g}^{\prime }$ are independent of $\phi _{m}$ and the transformations $%
R_{m}(\Phi )$ of the fields $\phi _{m}$ are proportional to $\phi _{m}$.
Write $\chi (\kappa \Phi )=\chi _{0}(\kappa \Phi _{g}^{\prime })+\chi _{m}$,
where $\chi _{m}=0$ at $\phi _{m}=0$. Then, $(S_{K},\chi )=(S_{K},\chi _{0})$%
, as we can see by calculating these expressions at $\phi _{m}=0$. From now
on we drop $\chi _{m}$ and just write $\chi =\chi (\kappa \Phi _{g}^{\prime
},\lambda _{+})$.

Clearly, assumption (IV$^{\prime }$) of subsection \ref{key} is sufficient
to justify (\ref{anomcanc}), with $\chi =\chi (\kappa \Phi _{g}^{\prime
},\lambda _{+})$, in alternative to assumptions (IV) and (V).

Since $\chi $ is one loop, its $\kappa $ structure agrees with the $L=1$
sector of formula (\ref{liable2}).

\subsection*{Cancellation of anomalies in the HD theory}

Now we go back to the HD\ theory. We can cancel its potential anomalies by
redefining the action. Indeed, if we take 
\begin{equation}
\breve{S}_{\Lambda T}=\hat{S}_{\Lambda T}-\frac{1}{2}\chi =S_{\Lambda
T}-\Gamma _{\Lambda T\hspace{0.01in}\text{div}}^{(1)}-\frac{1}{2}\chi
\label{suba}
\end{equation}%
as the new action, we find 
\begin{equation}
\mathcal{\breve{A}}_{\Lambda T}=\langle (\breve{S}_{\Lambda T},\breve{S}%
_{\Lambda T})\rangle _{\breve{S}_{\Lambda T}}=\mathcal{\hat{A}}_{\Lambda
T}-(S_{\Lambda T},\chi )+\mathcal{O}(\hbar ^{2}).  \label{appa}
\end{equation}%
Since $\chi $ is $K$ independent, only the $K$-dependent sector of $%
S_{\Lambda T}$, which is made of $S_{K}$ and $S_{\text{ev}T}$, can
contribute to $(S_{\Lambda T},\chi )$. Taking the one-loop nonevanescent
part of (\ref{appa}), and using (\ref{adivevhat}) and (\ref{anomcanc}), we
get\ 
\begin{equation}
\mathcal{\breve{A}}_{\Lambda T\hspace{0.01in}\text{nev}}^{(1)}=\mathcal{A}%
_{\Lambda T\hspace{0.01in}\text{nev}}^{(1)}-(S_{K},\chi )=0.  \label{nnp}
\end{equation}

The new $\Gamma $ functional $\breve{\Gamma}_{\Lambda T}$ defined by the
action $\breve{S}_{\Lambda T}$ of formula (\ref{suba}) is still convergent
to all orders. Indeed, it is convergent at one loop and, once we switch to
the tilde parametrization, the functional $\chi $ is written as a functional 
$\tilde{\chi}(\tilde{\kappa}\tilde{\Phi}_{g}^{\prime },\tilde{\lambda}_{+})$%
. This fact, together with formulas (\ref{stl}) and (\ref{nonev}), ensures
that the counterterms keep the form (\ref{expra}), which forbids divergences
beyond one loop. The anomaly functional $\mathcal{\breve{A}}_{\Lambda T}=(%
\breve{\Gamma}_{\Lambda T},\breve{\Gamma}_{\Lambda T})$ is also convergent
to all orders. Since its one-loop contribution $\mathcal{\breve{A}}_{\Lambda
T}^{(1)}$ has no divergent part and, by formula (\ref{nnp}), no
nonevanescent part, it is just evanescent: $\mathcal{\breve{A}}_{\Lambda
T}^{(1)}=$ $\mathcal{O}(\varepsilon )$. Including the tree-level
contribution $(S_{\Lambda T},S_{\Lambda T})$, which is also $\mathcal{O}%
(\varepsilon )$, we can write 
\begin{equation}
\mathcal{\breve{A}}_{\Lambda T}=\mathcal{O}(\varepsilon )+\mathcal{O}(\hbar
^{2}).  \label{becau}
\end{equation}

The next step is to prove the anomaly cancellation to all orders in the
higher-derivative theory, which we do in the next section. After that, we
complete the CDHD limit by renormalizing the $\Lambda $ divergences.

\section{Manifest Adler-Bardeen theorem in the HD theory}

\label{s6}

\setcounter{equation}{0}

In this section we prove that, from two loops onwards, the gauge anomalies
manifestly vanish in the HD theory. We have to study the diagrams with two
or more loops, with one insertion of 
\begin{equation}
\mathcal{E}_{T}\equiv (\breve{S}_{\Lambda T},\breve{S}_{\Lambda
T})=(S_{\Lambda T},S_{\Lambda T})-\mathcal{A}_{\Lambda T\hspace{0.01in}\text{%
nev}}^{(1)}-\mathcal{A}_{\Lambda T\hspace{0.01in}\text{divev}}^{(1)}-(S_{%
\text{ev}T},\chi ),  \label{e}
\end{equation}%
calculated with the action (\ref{suba}). \ To derive the right-hand side of (%
\ref{e}), we have used the fact that $\Gamma _{\Lambda T\hspace{0.01in}\text{%
div}}^{(1)}$ and $\chi $ are $K$ independent, then applied formula (\ref%
{noveva}) and replaced $(S_{K},\chi )$ with $\mathcal{A}_{\Lambda T\hspace{%
0.01in}\text{nev}}^{(1)}$. The action $\breve{S}_{\Lambda T}$ has the
structure (\ref{stl}) plus one-loop corrections of the form $F(\tilde{\kappa}%
\tilde{\Phi}_{g}^{\prime },\tilde{\lambda}_{+})$. Therefore, its
counterterms have the structure (\ref{expra}). On the other hand, $\mathcal{E%
}_{T}$ has the structure (\ref{fev}) plus (possibly nonevanescent and
divergent-evanescent) one-loop corrections that have the same form times $%
\tilde{\kappa}^{2}$, such that $V$ still vanishes at $\tilde{\varsigma}=0$.
This fact implies that $\mathcal{\breve{A}}_{\Lambda T}=\left\langle 
\mathcal{E}_{T}\right\rangle $ is still the sum of contributions that have
the structures (\ref{l1}) and (\ref{l2}), with $\mathcal{V}_{L}=0$ at $%
\tilde{\varsigma}=0$.

The functional $\mathcal{E}_{T}$ is made of the tree-level local evanescent
functional $(S_{\Lambda T},S_{\Lambda T})$, plus one-loop local corrections.
Formula (\ref{becau}) tells us that such corrections make the average $%
\left\langle \mathcal{E}_{T}\right\rangle $ evanescent at one loop. Then,
the theory of evanescent operators \cite{collins,ABrenoYMLR} tells us that
the two-loop nonevanescent part of $\left\langle \mathcal{E}%
_{T}\right\rangle $ is local. Briefly, the reason is as follows. Writing $%
\hat{\partial}^{\mu }=\hat{\eta}^{\mu \nu }\partial _{\nu }$ and $\hat{p}%
^{\mu }=\hat{\eta}^{\mu \nu }p_{\nu }$ everywhere inside $(S_{\Lambda
T},S_{\Lambda T})$, we can express each vertex of $(S_{\Lambda T},S_{\Lambda
T})$ in a factorized form $\mathcal{T}_{k}\hat{\delta}_{k}$, where $\hat{%
\delta}_{k}$ denotes a formally evanescent part, made of tensors $\eta ^{%
\hat{\mu}\hat{\nu}}$ and other structures that stay outside of the diagrams,
while $\mathcal{T}_{k}$ is a nonevanescent local functional and collects the
momenta. The average $\langle \mathcal{T}_{k}\hat{\delta}_{k}\rangle $ is
the sum of the one-particle irreducible diagrams $G$ that contain one
insertion of $\mathcal{T}_{k}\hat{\delta}_{k}$. Leaving $\hat{\delta}_{k}$
outside the diagrams, consider the average $\langle \mathcal{T}_{k}\rangle $%
, and let $\mathcal{T}_{k\text{div}}^{(1)}$ denote its one-loop divergent
part. Using (\ref{ana2}) and (\ref{adivev}), we find 
\begin{equation*}
\sum_{k}\mathcal{T}_{k\text{div}}^{(1)}\hat{\delta}_{k}=\mathcal{A}_{\Lambda
T\hspace{0.01in}\text{nev}}^{(1)}+\mathcal{A}_{\Lambda T\hspace{0.01in}\text{%
divev}}^{(1)}+L_{\text{ev}}^{(1)},
\end{equation*}%
where $L_{\text{ev}}^{(1)}$ are unspecified local evanescences. The theorem
on the locality of counterterms ensures that the divergent part of $\langle 
\mathcal{T}_{k}-\mathcal{T}_{k\text{div}}^{(1)}\rangle $ is local at two
loops. Accordingly, the nonevanescent and divergent parts of 
\begin{equation*}
\left\langle \mathcal{E}_{T}\right\rangle =\left\langle \sum_{k}\mathcal{(T}%
_{k}-\mathcal{T}_{k\text{div}}^{(1)})\hat{\delta}_{k}+L_{\text{ev}%
}^{(1)}-(S_{\text{ev}T},\chi )\right\rangle
\end{equation*}%
are also local at two loops. In subsection \ref{s32} we proved that the
local functionals that have the structures (\ref{l1}) and (\ref{l2}) vanish
from two loops onwards, by simple power counting. Therefore, $\left\langle 
\mathcal{E}_{T}\right\rangle $ is evanescent at two loops, which means that
formula (\ref{becau}) can be improved by one order and turned into 
\begin{equation*}
\mathcal{\breve{A}}_{\Lambda T}=\mathcal{O}(\varepsilon )+\mathcal{O}(\hbar
^{3}).
\end{equation*}

The argument can be iterated to all orders, because if an evanescent
operator $\mathsf{E}$ is renormalized, and equipped with finite local
subtractions such that its average $\left\langle \mathsf{E}\right\rangle $
is evanescent up to and including $\ell $ loops, then the $\mathcal{O}(\hbar
^{\ell +1})$ nonevanescent and divergent parts $\left\langle \mathsf{E}%
\right\rangle _{\text{nonev}}^{(\ell +1)}$ and $\left\langle \mathsf{E}%
\right\rangle _{\text{div}}^{(\ell +1)}$ of $\left\langle \mathsf{E}%
\right\rangle $ must be local. In the case we are considering here, which is 
$\mathsf{E}=\mathcal{E}_{T}$, $\left\langle \mathsf{E}\right\rangle _{\text{%
nonev}}^{(\ell +1)}$ and $\left\langle \mathsf{E}\right\rangle _{\text{div}%
}^{(\ell +1)}$ must also have the structures (\ref{l1}) and (\ref{l2}), but
then they vanish.

We infer that the anomaly functional $\mathcal{\breve{A}}_{\Lambda T}$ is
evanescent to all orders, that is to say, 
\begin{equation}
\mathcal{\breve{A}}_{\Lambda T}=(\breve{\Gamma}_{\Lambda T},\breve{\Gamma}%
_{\Lambda T})=\mathcal{O}(\varepsilon ),  \label{evaeva}
\end{equation}%
which proves the manifest Adler-Bardeen theorem for the HD theory $%
S_{\Lambda T}$. Therefore, the HD theory is free of gauge anomalies to all
orders in the limit $D\rightarrow d$.

This concludes the proof that the HD theory is super-renormalizable and
anomaly free to all orders. We stress again that only the truncation T1 of
the action $S_{\Lambda }$ is necessary, and the result (\ref{evaeva}) holds
to all orders in $\hbar $ and for arbitrarily large powers of $1/\Lambda
_{-} $. The truncation T2 of subsection \ref{s21} is important for the
second part of the proof, which is worked out in the next section.

\section{Almost manifest Adler-Bardeen theorem in the final theory}

\label{s7}

\setcounter{equation}{0}

We are finally ready to prove the cancellation of gauge anomalies to all
orders in the final theory. The task consists of studying the $\Lambda $
dependence of the HD theory, for $\Lambda $ large, subtract the $\Lambda $
divergences, and complete the CDHD limit, according to the rules explained
in subsection \ref{s23}. The subtraction of the $\Lambda $ divergences is
done inductively and preserves the master equation up to $\mathcal{O}%
(\varepsilon )$ terms that vanish in the CDHD limit.

Before beginning the proof, let us recall that our approach uses two
regularizations, the\ chiral dimensional one, with regularizing parameter $%
\varepsilon $, and the higher-derivative one, with energy scale $\Lambda $.
So far, we have taken care of the renormalization and the cancellation of
anomalies to all orders at the CD\ level. Now we consider the $\Lambda $
divergences. As far as those are concerned, once we have adjusted the orders 
$\hbar ^{n}$, $k\leqslant n$, we can concentrate on the order $\hbar ^{n+1}$
and neglect higher-order corrections, as is done in most common
renormalization procedures. However, at each step of the subtraction of the $%
\Lambda $ divergences, we must preserve the properties gained so far with
respect to the CD renormalization, and those must hold to all orders in $%
\hbar $, like equation (\ref{evaeva}).

Because of the truncation T2, we say that an action $S_{k}$ is CDHD
renormalized up to and including $k$ loops, when the $\ell $-loop
contributions to its $\Gamma $ functional $\Gamma _{k}$ are CDHD convergent
up to $o(1/\Lambda _{-}^{T-2\ell \sigma })$, for $0\leqslant \ell \leqslant
k $.

We work inductively in the number $n$ of loops. We assume that for every $%
k\leqslant n<\bar{\ell}$, where $\bar{\ell}$ is given by (\ref{lmax}), there
exists an action $S_{kT}=$ $S_{\Lambda T}+\mathcal{O}(\hbar )$, obtained
from $S_{\Lambda T}$ by means of $\varepsilon $-convergent, possibly $%
\Lambda $-divergent canonical transformations and redefinitions of
parameters, with the following properties: we can $\varepsilon $-renormalize 
$S_{kT}$ at $\Lambda $ fixed, to all orders in $\hbar $, and fine-tune its
finite local counterterms, so that the so-renormalized action $S_{kRT}$ is
also CDHD renormalized up to and including $k$ loops, and the renormalized $%
\Gamma $ functional $\Gamma _{kRT}$ associated with $S_{kRT}$ is free of
gauge anomalies to all orders in $\hbar $ at $\Lambda $ fixed, i.e.%
\begin{equation}
(\Gamma _{kRT},\Gamma _{kRT})=\langle (S_{kRT},S_{kRT})\rangle _{S_{kRT}}=%
\mathcal{O}(\varepsilon ),\qquad k\leqslant n.  \label{assumo}
\end{equation}%
At $n=0$ we take $S_{0T}=S_{\Lambda T}$, so $S_{0RT}=\breve{S}_{\Lambda T}$.
Clearly, $\Gamma _{0RT}$ coincides with $\breve{\Gamma}_{\Lambda T}$ and
satisfies (\ref{evaeva}).

Note that, by assumption, $\Gamma _{kRT}$ has a regular limit for $%
\varepsilon \rightarrow 0$ at $\Lambda $ fixed, and not just within the
truncation T2, but also beyond. More precisely, $\Gamma _{kRT}$ is a sum of $%
\ell $-loop contributions of the form (\ref{survo}) up to $o(1/\Lambda
_{-}^{T-2\ell \sigma })$ for $0\leqslant \ell \leqslant k$ (because it is
CDHD convergent in that sector), and a sum of terms (\ref{survi}) everywhere
else. Instead, $(\Gamma _{kRT},\Gamma _{kRT})$ is a sum of $\ell $-loop
contributions (\ref{survo}) except $\varepsilon ^{0}\Lambda ^{0}$ and $%
\varepsilon ^{0}/\Lambda $ up to $o(1/\Lambda _{-}^{T-2\ell \sigma })$ for $%
0\leqslant \ell \leqslant k$, plus terms (\ref{survi}) except $\varepsilon
^{0}$ everywhere else. Note that assumption (\ref{assumo}) also holds beyond
the truncation T2 [where the \textquotedblleft $\mathcal{O}(\varepsilon )$%
\textquotedblright\ may contain terms $\hat{\delta}\varepsilon ^{0}\Lambda $%
].

The theorem on the locality of counterterms ensures that the $(n+1)$-loop
CDHD\ divergent part $\Gamma _{nRT\hspace{0.01in}\text{div}}^{(n+1)}$ of $%
\Gamma _{nRT}$ is a local functional, up to $o(1/\Lambda _{-}^{T-2n\sigma })$%
. Since $\Gamma _{nRT}$ has a regular limit for $\varepsilon \rightarrow 0$
at $\Lambda $ fixed, $\Gamma _{nRT\hspace{0.01in}\text{div}}^{(n+1)}$
contains only divergences in $\Lambda $, but not in $\varepsilon $.
Precisely, we can write 
\begin{equation}
\Gamma _{nRT\hspace{0.01in}\text{div}}^{(n+1)}=\Gamma _{nRT\hspace{0.01in}%
\text{div\hspace{0.01in}nev}}^{(n+1)}+\Gamma _{nRT\hspace{0.01in}\text{div%
\hspace{0.01in}fev}}^{(n+1)}+o(1/\Lambda _{-}^{T-2(n+1)\sigma }),
\label{gdiv}
\end{equation}%
where $\Gamma _{nRT\hspace{0.01in}\text{div\hspace{0.01in}nev}}^{(n+1)}$ and 
$\Gamma _{nRT\hspace{0.01in}\text{div\hspace{0.01in}fev}}^{(n+1)}$ collect
the terms $\varepsilon ^{0}\Lambda $ and $\hat{\delta}\varepsilon
^{0}\Lambda $ of the list (\ref{assu}), respectively.

Now, we take the $(n+1)$-loop CDHD-divergent non-$\varepsilon $-evanescent
part of equation (\ref{assumo}) for $k=n$, within the truncation, which
means the terms of types $\varepsilon ^{0}\Lambda $ of the list (\ref{assu}%
), up to $o(1/\Lambda _{-}^{T-2(n+1)\sigma })$. Expand $\Gamma _{nRT}$ in
powers of $\hbar $, by writing it as $\sum_{k=0}^{\infty }\hbar ^{k}\Gamma
_{nRT}^{(k)}$. Observe that the contributions $(\Gamma _{nRT}^{(k)},\Gamma
_{nRT}^{(n+1-k)})$ with $0<k<n+1$ can be dropped, because they are
convergent in the CDHD limit, up to $o(1/\Lambda _{-}^{T-2(n+1)\sigma })$.
We remain with $2(\Gamma _{nRT}^{(0)},\Gamma _{nRT}^{(n+1)})=2(S_{\Lambda
T},\Gamma _{nRT}^{(n+1)})$. Taking the $\Lambda $ divergent part of this
expression, and recalling that, by formula (\ref{gial}), $(S_{\mathrm{HD}%
},\Gamma _{nRT\hspace{0.01in}\text{div}}^{(n+1)})$ tends to zero for $%
\Lambda \rightarrow \infty $ within the truncation, we get $2(S_{T},\Gamma
_{nRT\hspace{0.01in}\text{div}}^{(n+1)})+o(1/\Lambda _{-}^{T-2(n+1)\sigma })$%
. Taking the non-$\varepsilon $-evanescent part and recalling that $S_{T}$
is equal to $S_{dT}+S_{\text{ev}T}$, where $S_{dT}$ is non-$\varepsilon $%
-evanescent, the left-hand side of (\ref{assumo}) at $k=n$ gives $%
2(S_{dT},\Gamma _{nRT\hspace{0.01in}\text{div\hspace{0.01in}nev}%
}^{(n+1)})+o(1/\Lambda _{-}^{T-2(n+1)\sigma })$. Noting that the
CDHD-divergent part of the right-hand side is just made of terms $\hat{\delta%
}\varepsilon ^{0}\Lambda $, within the truncation, we obtain 
\begin{equation}
(S_{dT},\Gamma _{nRT\hspace{0.01in}\text{div\hspace{0.01in}nev}%
}^{(n+1)})=o(1/\Lambda _{-}^{T-2(n+1)\sigma }).  \label{questa}
\end{equation}

\subsection{Solution of the cohomological problem}

\label{s71}

We work out the solution of the cohomological problem (\ref{questa}) by
applying the assumption (III) of subsection \ref{key}. Let us imagine that,
instead of working with the classical action $S_{c}$, we work with its
extension $\check{S}_{c}$, which includes the invariants $\mathcal{\check{G}}%
_{i}$ that break the nonanomalous accidental symmetries belonging to the
group $G_{\text{nas}}$. Similarly, we extend $S_{d}$ to $\check{S}_{d}$, $S_{%
\text{ev}}$ to $\check{S}_{\text{ev}}$, and $S=S_{d}+S_{\text{ev}}$ to $%
\check{S}=\check{S}_{d}+\check{S}_{\text{ev}}$. Every extended functional
reduces to the nonextended one when we set $\check{\lambda}=\check{\eta}=0$,
where $\check{\lambda}$ and $\check{\eta}$ are the extra parameters
contained in $\check{S}_{c}$ and $\check{S}_{\text{ev}}$, respectively.
There is no need to extend the higher-derivative sector $S_{\mathrm{HD}}$.

If we repeat the operations that lead to (\ref{questa}), we obtain an
extended, nonevanescent local functional $\check{\Gamma}_{nRT\hspace{0.01in}%
\text{div\hspace{0.01in}nev}}^{(n+1)}$ that satisfies $(\check{S}_{dT},%
\check{\Gamma}_{nRT\hspace{0.01in}\text{div\hspace{0.01in}nev}%
}^{(n+1)})=o(1/\Lambda _{-}^{T-2(n+1)\sigma })$. Taking the limit $\Lambda
_{-}\rightarrow \infty $ of this equation and recalling that $T\geqslant
2(n+1)\sigma $ (because $n<\bar{\ell}$), we get%
\begin{equation*}
(\check{S}_{d\text{b}},\check{V}_{0})=0,
\end{equation*}%
where $\check{V}_{0}$ denotes the $\Lambda _{-}\rightarrow \infty $ limit of 
$\check{\Gamma}_{nRT\hspace{0.01in}\text{div\hspace{0.01in}nev}}^{(n+1)}$.
Assumption (III) tells us that the action $\check{S}_{d\text{b}}$ satisfies
the extended Kluberg-Stern--Zuber assumption, and the group $G_{\text{nas}}$
is compact. Thus, there exist constants $a_{i0}$ and $b_{i0}$, which depend
on the parameters of $\check{V}_{0}$, and a local functional $\check{Y}_{0}$
such that%
\begin{equation*}
\check{V}_{0}=\sum_{i}a_{i0}\mathcal{G}_{i}+\sum_{i}b_{i0}\mathcal{\check{G}}%
_{i}+(\check{S}_{d\text{b}},\check{Y}_{0}).
\end{equation*}%
Recall that in subsection \ref{s21} we showed that only integer and
semi-integer powers of $1/\Lambda _{-}$ can appear. Define 
\begin{equation*}
\check{X}_{1}=\Lambda _{-}^{1/2}\left[ \check{\Gamma}_{nRT\hspace{0.01in}%
\text{div\hspace{0.01in}nev}}^{(n+1)}-\sum_{i}a_{i0}\mathcal{G}%
_{i}-\sum_{i}b_{i0}\mathcal{\check{G}}_{i}-(\check{S}_{dT},\check{Y}_{0})%
\right] .
\end{equation*}%
The local functional $\check{X}_{1}$ is analytic in $1/\Lambda _{-}^{1/2}$,
because $\check{\Gamma}_{nRT\hspace{0.01in}\text{div\hspace{0.01in}nev}%
}^{(n+1)}=\check{V}_{0}+\mathcal{O}(1/\Lambda _{-}^{1/2})$ and $\check{S}%
_{dT}=\check{S}_{d\text{b}}+\mathcal{O}(1/\Lambda _{-}^{1/2})$. Moreover,
since $(\check{S}_{dT},\mathcal{G}_{i})=(\check{S}_{dT},\mathcal{\check{G}}%
_{i})=(\check{S}_{dT},\check{S}_{dT})=0$, $\check{X}_{1}$ satisfies $(\check{%
S}_{dT},\check{X}_{1})=o(1/\Lambda _{-}^{T-2(n+1)\sigma -1/2})$. Then we
repeat the argument just given with $\check{\Gamma}_{nRT\hspace{0.01in}\text{%
div\hspace{0.01in}nev}}^{(n+1)}$ replaced by $\check{X}_{1}$, and continue
like this till we can. For $0\leqslant m\leqslant 2T-4(n+1)\sigma +1$, we
find constants $a_{i\hspace{0.01in}m-1}$ and $b_{i\hspace{0.01in}m-1}$,
depending on the parameters, and local functionals $\check{Y}_{m-1}$ such
that the combinations 
\begin{equation*}
\check{X}_{m}=\Lambda _{-}^{1/2}\left[ \check{X}_{m-1}-\sum_{i}a_{i\hspace{%
0.01in}m-1}\mathcal{G}_{i}-\sum_{i}b_{i\hspace{0.01in}m-1}\mathcal{\check{G}}%
_{i}-(\check{S}_{dT},\check{Y}_{m-1})\right]
\end{equation*}%
are analytic in $1/\Lambda _{-}^{1/2}$ and satisfy $(\check{S}_{dT},\check{X}%
_{m})=o(1/\Lambda _{-}^{T-2(n+1)\sigma -m/2})$, with $\check{X}_{0}=\check{%
\Gamma}_{nRT\hspace{0.01in}\text{div\hspace{0.01in}nev}}^{(n+1)}$. In the
end, there exist constants $\Delta \lambda _{ni}^{\prime }$, $\Delta \check{%
\lambda}_{ni}^{\prime }$ depending on the parameters, and local functionals $%
\check{\chi}_{nT\hspace{0.01in}}$, 
\begin{equation}
\Delta \lambda _{ni}^{\prime }=\sum_{m=0}^{2T-4(n+1)\sigma }\frac{a_{im}}{%
\Lambda _{-}^{m/2}},\qquad \Delta \check{\lambda}_{ni}^{\prime
}=\sum_{m=0}^{2T-4(n+1)\sigma }\frac{b_{im}}{\Lambda _{-}^{m/2}},\qquad 
\check{\chi}_{nT\hspace{0.01in}}=\sum_{m=0}^{2T-4(n+1)\sigma }\frac{\check{Y}%
_{m}}{\Lambda _{-}^{m/2}},  \label{aby}
\end{equation}%
such that 
\begin{equation}
\check{\Gamma}_{nRT\hspace{0.01in}\text{div\hspace{0.01in}nev}%
}^{(n+1)}=\sum_{i}\Delta \lambda _{ni}^{\prime }\mathcal{G}%
_{i}+\sum_{i}\Delta \check{\lambda}_{ni}^{\prime }\mathcal{\check{G}}_{i}+(%
\check{S}_{dT},\check{\chi}_{nT\hspace{0.01in}})+o(1/\Lambda
_{-}^{T-2(n+1)\sigma }).  \label{ro1}
\end{equation}%
Clearly, $\Delta \lambda _{ni}^{\prime }$, $\Delta \check{\lambda}%
_{ni}^{\prime }$, and $\check{\chi}_{nT\hspace{0.01in}}$ are of order $\hbar
^{n+1}$. If we set $\check{\lambda}=\check{\eta}=0$ in equation (\ref{ro1}),
we obtain 
\begin{equation}
\Gamma _{nRT\hspace{0.01in}\text{div\hspace{0.01in}nev}}^{(n+1)}=\sum_{i}%
\bar{\Delta}\lambda _{ni}\mathcal{G}_{i}+\sum_{i}\Delta \check{\lambda}_{ni}%
\mathcal{\check{G}}_{i}+(S_{dT},\bar{\chi}_{\hspace{0.01in}nT})+o(1/\Lambda
_{-}^{T-2(n+1)\sigma }),  \label{inn}
\end{equation}%
where $\bar{\Delta}\lambda _{ni}$, $\Delta \check{\lambda}_{ni}$, and $\bar{%
\chi}_{\hspace{0.01in}nT}$ are equal to $\Delta \lambda _{ni}^{\prime }$, $%
\Delta \check{\lambda}_{ni}^{\prime }$, and $\check{\chi}_{nT\hspace{0.01in}%
} $ at $\check{\lambda}=\check{\eta}=0$. However, $\Gamma _{nRT\hspace{0.01in%
}\text{div\hspace{0.01in}nev}}^{(n+1)}$ is invariant under the nonanomalous
accidental symmetries that belong to the group $G_{\text{nas}}$, while the
functionals $\mathcal{\check{G}}_{i}$ are not. Since $G_{\text{nas}}$ is
assumed to be compact, we can average on it. When we do that, the invariants 
$\mathcal{\check{G}}_{i}$ disappear (or turn into linear combinations of $%
\mathcal{G}_{i}$) and $\bar{\chi}_{\hspace{0.01in}nT}$ turns into some $\chi
_{\hspace{0.01in}nT}$. We finally obtain 
\begin{equation}
\Gamma _{nRT\hspace{0.01in}\text{div\hspace{0.01in}nev}}^{(n+1)}=\sum_{i}%
\Delta \lambda _{ni}\mathcal{G}_{i}+(S_{dT},\chi _{\hspace{0.01in}%
nT})+o(1/\Lambda _{-}^{T-2(n+1)\sigma }),  \label{both}
\end{equation}%
for possibly new constants $\Delta \lambda _{ni}$ of order $\hbar ^{n+1}$
that depend on the parameters.

The arguments of this subsection, which lead from formula (\ref{questa}) to
formula (\ref{both}), are purely algebraic and can be applied in more
general contexts. For example, taking $T\rightarrow \infty $, formula (\ref%
{ro1}) proves that the action $S_{d}$ is also cohomologically complete.
Instead, formula (\ref{both}) at $T=\infty $ proves that $S_{d}$ satisfies
what we can call the \textit{physical} Kluberg-Stern--Zuber assumption,
which states that if a nonevanescent local functional $\Gamma _{\text{div}}$
solves $(S_{d},\Gamma _{\text{div}})=0$ and is generated by renormalization
as a local divergent part of the $\Gamma $ functional, then there exists
constants $a_{i}$ and a local functional $Y$ of ghost number $-1$ such that 
\begin{equation}
\Gamma _{\text{div}}=\sum_{i}a_{i}\mathcal{G}_{i}+(S_{d},Y).  \label{phyz}
\end{equation}%
Indeed, we can always lift the discussion to the extended theory $\check{S}%
_{d}$, which gives an extended functional $\check{\Gamma}_{\text{div}}$ that
solves $(\check{S}_{d},\check{\Gamma}_{\text{div}})=0$. Then $\check{\Gamma}%
_{\text{div}}$ can be expanded like the right-hand side of (\ref{ro1}) at $%
T=\infty $. When we go back down to $S_{d}$, we find (\ref{phyz}).

\subsection{Subtraction of divergences}

Now we work out the operations that subtract the divergences $\Gamma _{nRT%
\hspace{0.01in}\text{div}}^{(n+1)}$ within the truncation. We recall from
subsection \ref{s21} that the truncated classical action $S_{cT}$ contains
enough independent parameters $\lambda _{i}$ to subtract the divergences
proportional to $\mathcal{G}_{i}$ of (\ref{both}) by means of $\lambda _{i}$
redefinitions, within the truncation T2. If we make the canonical
transformation generated by 
\begin{equation}
F_{n}(\Phi ,K^{\prime })=\int \Phi ^{\alpha }K_{\alpha }^{\prime }-\chi _{%
\hspace{0.01in}nT}(\Phi ,K^{\prime })  \label{op1}
\end{equation}%
and the redefinitions 
\begin{equation}
\lambda _{i}\rightarrow \lambda _{i}-\Delta \lambda _{ni}  \label{op2}
\end{equation}%
on $S_{dT}$, we get 
\begin{equation}
S_{dT}\rightarrow S_{dT}-\sum_{i}\Delta \lambda _{ni}\mathcal{G}%
_{i}-(S_{dT},\chi _{\hspace{0.01in}nT})+\mathcal{O}(\hbar ^{n+2}).
\label{arr}
\end{equation}%
Observe that the operations (\ref{op1}) and (\ref{op2}) are independent of $%
\varepsilon $ and divergent in $\Lambda $, because\ so is $\Gamma _{nRT%
\hspace{0.01in}\text{div\hspace{0.01in}nev}}^{(n+1)}$.

Formula (\ref{arr}) is equivalent to 
\begin{equation*}
S_{dT}\rightarrow S_{dT}-\Gamma _{nRT\hspace{0.01in}\text{div\hspace{0.01in}%
nev}}^{(n+1)}+\mathcal{O}(\hbar ^{n+2})+\mathcal{O}(\hbar ^{n+1})o(1/\Lambda
_{-}^{T-2(n+1)\sigma }),
\end{equation*}%
which shows that we can fully subtract the $\varepsilon $-nonevanescent $%
\Lambda $ divergences $\Gamma _{nRT\hspace{0.01in}\text{div\hspace{0.01in}nev%
}}^{(n+1)}$, by making the operations (\ref{op1}) and (\ref{op2}) on $S_{dT}$%
, up to $\mathcal{O}(\hbar ^{n+1})o(1/\Lambda _{-}^{T-2(n+1)\sigma })$.

However, the truncated classical action we have been using is not $S_{dT}$,
nor $S_{T}=S_{dT}+S_{\text{ev}T}$, but $S_{nT}$, whose classical limit is $%
S_{\Lambda T}$, so we must inquire what happens when we make the operations (%
\ref{op1}) and (\ref{op2})\ on $S_{\Lambda T}$.

Let us begin from $S_{T}=S_{dT}+S_{\text{ev}T}$. Since the operations (\ref%
{op1}) and (\ref{op2}) are independent of $\varepsilon $ and divergent in $%
\Lambda $, when we apply them to $S_{\text{ev}T}$ we generate new formally $%
\varepsilon $-evanescent, $\Lambda $-divergent terms of order $\hbar ^{n+1}$%
, which change $\Gamma _{nRT\hspace{0.01in}\text{div\hspace{0.01in}fev}%
}^{(n+1)}$ [check formula (\ref{gdiv})] into some new $\Gamma _{nRT\hspace{%
0.01in}\text{div\hspace{0.01in}fev}}^{\prime \hspace{0.01in}(n+1)}$, plus $%
\mathcal{O}(\hbar ^{n+2})$. The divergences $\Gamma _{nRT\hspace{0.01in}%
\text{div\hspace{0.01in}fev}}^{\prime \hspace{0.01in}(n+1)}$ are not
constrained by gauge invariance, but just locality, weighted power counting
and the nonanomalous global symmetries of the theory. In subsection \ref{s21}
we remarked that, within the truncation T2, that is to say, up to $%
o(1/\Lambda _{-}^{T-2(n+1)\sigma })$, they can be subtracted by redefining
the parameters $\varsigma $ and $\eta $ of $S_{\text{ev}T}$.

Let $\mathcal{R}_{n}$ denote the set of operations made by the canonical
transformation (\ref{op1}), the $\lambda $ redefinitions (\ref{op2}), and
the $\varsigma $ and $\eta $ redefinitions that subtract $\Gamma _{nRT%
\hspace{0.01in}\text{div\hspace{0.01in}fev}}^{\prime \hspace{0.01in}(n+1)}$.
We have 
\begin{equation}
\mathcal{R}_{n}S_{T}=S_{T}-\Gamma _{nRT\hspace{0.01in}\text{div}}^{(n+1)}+%
\mathcal{O}(\hbar ^{n+2})+\mathcal{O}(\hbar ^{n+1})o(1/\Lambda
_{-}^{T-2(n+1)\sigma }).  \label{bal}
\end{equation}

It remains to check what happens when the operations $\mathcal{R}_{n}$ act
on $S_{\mathrm{HD}}=S_{\Lambda T}-S_{T}$. Note that $\mathcal{R}_{n}$ are
equal to the identity plus $\mathcal{O}(\hbar ^{n+1})$, and they are
independent of $\varepsilon $ and divergent in $\Lambda $. Moreover, by
formula (\ref{aby}) and the arguments of subsection \ref{s33}, they do not
involve powers of $\Lambda $ greater than $T+d-2\sigma $, at the order $%
\mathcal{O}(\hbar ^{n+1})$. Recalling that the difference $S_{\mathrm{HD}}$
is $\mathcal{O}(1/\Lambda ^{T+d-2\sigma +1})$, we have that $(\mathcal{R}%
_{n}-1)S_{\mathrm{HD}}$ vanishes in the CDHD limit to the order $\mathcal{O}%
(\hbar ^{n+1})$.

Define 
\begin{equation}
S_{n+1\hspace{0.01in}T}=\mathcal{R}_{n}S_{nT}=\mathcal{R}_{n}\circ \cdots
\circ \mathcal{R}_{0}S_{\Lambda T}\equiv \mathcal{U}_{n}S_{\Lambda T}.
\label{inert}
\end{equation}%
Using (\ref{bal}), we find 
\begin{eqnarray}
S_{n+1\hspace{0.01in}T} &=&S_{nT}+(\mathcal{R}_{n}-1)S_{\Lambda T}+\mathcal{O%
}(\hbar ^{n+2})  \notag \\
&=&S_{nT}+(\mathcal{R}_{n}-1)S_{T}+(\mathcal{R}_{n}-1)S_{\mathrm{HD}}+%
\mathcal{O}(\hbar ^{n+2})  \label{tells} \\
&=&S_{nT}-\Gamma _{nRT\hspace{0.01in}\text{div}}^{(n+1)}+(\mathcal{R}%
_{n}-1)S_{\mathrm{HD}}+\mathcal{O}(\hbar ^{n+2})+\mathcal{O}(\hbar
^{n+1})o(1/\Lambda _{-}^{T-2(n+1)\sigma }).  \notag
\end{eqnarray}%
Thus, the operations $\mathcal{R}_{n}$ do renormalize the $\Lambda $
divergences to the order $n+1$, as we want.

The operations $\mathcal{U}_{n}=\mathcal{R}_{n}\circ \cdots \circ \mathcal{R}%
_{0}$ are combinations of local canonical transformations and redefinitions
of parameters. They act on the action $S_{\Lambda T}$, and, from the point
of view of the HD theory, where $\Lambda $ is fixed, they are convergent. In
general, a canonical transformation may destroy the nice properties of the
HD theory, such as its manifest super-renormalizability, its structure in
the tilde parametrization, and the manifest cancellation of its gauge
anomalies. To overcome these problems, we must re-renormalize the $%
\varepsilon $ divergences and recancel the gauge anomalies after making the
operations $\mathcal{U}_{n}$. We can achieve these goals with the help of
the theorem proved in ref. \cite{ABward}.

\subsection{Renormalization and almost manifest Adler-Bardeen theorem}

Now we must renormalize $S_{n+1\hspace{0.01in}T}$ at $\Lambda $ fixed. We
use the theorem proved in ref. \cite{ABward}, which ensures that if we make
a convergent local canonical transformation [equal to the identity
transformation plus $\mathcal{O(}\theta )$, where $\theta $ is some
expansion parameter] on the action $S$ of a theory that is free of gauge
anomalies, it is possible to re-renormalize the divergences of the
transformed theory and re-fine-tune its finite local counterterms,
continuously in $\theta $, so as to preserve the cancellation of gauge
anomalies to all orders. Clearly, we can achieve the same goal if we combine
canonical transformations and redefinitions of parameters, as long as they
are both convergent.

Before proceeding, let us recapitulate the situation. The HD theory has the
action $S_{\Lambda T}$, which is super-renormalizable and has a particularly
nice structure, once we use the tilde parametrization. Its renormalized
action is the action $\breve{S}_{\Lambda T}$ of formula (\ref{suba}), which
contains both the counterterms $\Gamma _{\Lambda T\hspace{0.01in}\text{div}%
}^{(1)}$ that subtract the $\varepsilon $-divergences at $\Lambda $ fixed,
and the finite local counterterms $-\chi /2$ that subtract the trivial
anomalous terms. Formula (\ref{evaeva}) ensures that $\breve{\Gamma}%
_{\Lambda T}$ is free of gauge anomalies to all orders.

Now we need to make the operations $\mathcal{U}_{n}$ on the action $%
S_{\Lambda T}$. From the point of view of the HD theory, where $\Lambda $ is
fixed, those operations are completely convergent, because they are
convergent in $\varepsilon $ (although possibly divergent in $\Lambda $).
However, the canonical transformations can ruin the manifest
super-renormalizability of $S_{\Lambda T}$, as well as the nice structure
exhibited by $S_{\Lambda T}$ in the tilde parametrization. Because of this,
the arguments that allowed us to prove the cancellation of gauge anomalies
in the HD theory cannot be used after the transformations. Nevertheless, we
expect that the super-renormalizability of $S_{\Lambda T}$ and the
cancellation of its gauge anomalies survive in some nonmanifest form.

What happens is that, after the operations $\mathcal{U}_{n}$, the (nonlinear
part of the) canonical transformation generates new poles in $\varepsilon $,
and not just at one loop, but at each order of the perturbative expansion.
Then, the first thing to do is re-renormalize the transformed HD theory at $%
\Lambda $ fixed, to remove the new divergences. Moreover, the cancellation
of gauge anomalies, which is in general ruined by the operations $\mathcal{U}%
_{n}$, can be enforced again by re-fine-tuning all sorts of finite local
counterterms. The theorem proved in ref. \cite{ABward} ensures that this
goal can indeed be achieved, to all orders in $\hbar $ and $1/\Lambda _{-}$.
In these arguments, the truncation T2 plays no role.

We know that each $\mathcal{R}_{n}$ is equal to the identity plus $\mathcal{O%
}(\hbar ^{n+1})$, and so is the canonical transformation (\ref{op1}). If we
replace the factor $\hbar ^{n+1}$ by a parameter $\theta _{n}$, we can
define operations $\mathcal{R}_{n}(\theta _{n})$ that are equal to the
identity plus $\mathcal{O(}\theta _{n})$. Then we also have operations $%
\mathcal{U}_{n}(\theta _{1},\cdots ,\theta _{n})$, which we sometimes denote
for brevity by $\mathcal{U}_{n}(\theta )$. Clearly, $\mathcal{U}%
_{n-1}(\theta _{1},\cdots ,\theta _{n-1})=\mathcal{U}_{n}(\theta _{1},\cdots
,\theta _{n-1},0)$. For a while, we work on the actions $\bar{S}_{k+1\hspace{%
0.01in}T}\equiv \mathcal{U}_{k}(\theta )S_{\Lambda T}$ at $\Lambda $ fixed,
for $0\leqslant k\leqslant n$. Applying the results of ref. \cite{ABward} to
the operations $\mathcal{U}_{k}(\theta )$, we know that we can $\varepsilon $%
-renormalize the actions $\bar{S}_{k+1\hspace{0.01in}T}$ at $\Lambda $ fixed
and fine-tune the finite local counterterms, continuously in $\theta $, so
as to preserve the cancellation of gauge anomalies for arbitrary values of
each $\theta $. Call the so-renormalized actions $\bar{S}_{k+1\hspace{0.01in}%
RT}$ and their $\Gamma $ functionals $\bar{\Gamma}_{k+1\hspace{0.01in}RT}$.
We have%
\begin{equation}
(\bar{\Gamma}_{k+1\hspace{0.01in}RT},\bar{\Gamma}_{k+1\hspace{0.01in}RT})=%
\mathcal{O}(\varepsilon ),\qquad k\leqslant n.  \label{gk}
\end{equation}

Observe that 
\begin{equation*}
\bar{S}_{k+1\hspace{0.01in}RT}=\bar{S}_{k+1\hspace{0.01in}T}+\breve{S}%
_{\Lambda T}-S_{\Lambda T}+\mathcal{O}(\hbar )\mathcal{O}(\theta ),\qquad
k\leqslant n.
\end{equation*}%
Indeed, $\breve{S}_{\Lambda T}-S_{\Lambda T}$ are the counterterms that $%
\varepsilon $-renormalize the theory and cancel the gauge anomalies at $%
\theta =0$. Every other counterterm must be both $\mathcal{O}(\hbar )$ and $%
\mathcal{O}(\theta )$. Thus, 
\begin{equation}
\bar{S}_{k+1\hspace{0.01in}RT}-\bar{S}_{kRT}=\bar{S}_{k+1\hspace{0.01in}T}-%
\bar{S}_{kT}+\mathcal{O}(\hbar )\mathcal{O}(\theta _{k}),\qquad k\leqslant n.
\label{uno}
\end{equation}%
We have replaced $\mathcal{O}(\hbar )\mathcal{O}(\theta )$ with $\mathcal{O}%
(\hbar )\mathcal{O}(\theta _{k})$ in this formula, because at $\theta _{k}=0$
we have $\bar{S}_{k+1\hspace{0.01in}RT}=\bar{S}_{kRT}$ and $\bar{S}_{k+1%
\hspace{0.01in}T}=\bar{S}_{kT}$.

By formula (\ref{inert}), when we replace $\theta _{i}$ with $\hbar ^{i+1}$, 
$i=1,\ldots ,k$, inside $\bar{S}_{k+1\hspace{0.01in}T}$, we obtain the
actions $S_{k+1\hspace{0.01in}T}$, $k\leqslant n$. When we replace $\theta
_{i}$ with $\hbar ^{i+1}$ inside $\bar{S}_{k+1\hspace{0.01in}RT}$, we obtain
the renormalized actions $S_{k+1\hspace{0.01in}RT}$. The actions $S_{k+1%
\hspace{0.01in}T}$ and $S_{k+1\hspace{0.01in}RT}$ with $k<n$ are those that
are assumed to satisfy the inductive hypotheses mentioned at the beginning
of this section. We must show that the actions 
\begin{equation}
S_{n+1\hspace{0.01in}T}=\left. \bar{S}_{n+1\hspace{0.01in}T}\right\vert
_{\theta _{i}=\hbar ^{i+1}},\qquad S_{n+1\hspace{0.01in}RT}=\left. \bar{S}%
_{n+1\hspace{0.01in}RT}\right\vert _{\theta _{i}=\hbar ^{i+1}},  \label{due}
\end{equation}%
satisfy analogous properties, that is to say: ($a$) $S_{n+1\hspace{0.01in}%
RT} $ is $\varepsilon $-renormalized to all orders in $\hbar $ at $\Lambda $
fixed; ($b$) it is CDHD\ renormalized up to and including $n+1$ loops; and ($%
c$) the $\Gamma $ functional $\Gamma _{n+1\hspace{0.01in}RT}$ associated
with $S_{n+1\hspace{0.01in}RT}$ is free of gauge anomalies to all orders in $%
\hbar $ at $\Lambda $ fixed.

The action $S_{n+1\hspace{0.01in}RT}$ defined by formula (\ref{due}) is $%
\varepsilon $-renormalized to all orders at $\Lambda $ fixed, because so is
the action $\bar{S}_{n+1\hspace{0.01in}RT}$, by construction. To show that $%
S_{n+1\hspace{0.01in}RT}$ is properly CDHD renormalized, we use, in the
order, (\ref{due}), (\ref{uno}), and (\ref{tells}). We obtain%
\begin{eqnarray}
S_{n+1\hspace{0.01in}RT}-S_{nRT} &=&\left. \bar{S}_{n+1\hspace{0.01in}%
RT}\right\vert _{\theta _{i}=\hbar ^{i+1}}-\left. \bar{S}_{n\hspace{0.01in}%
RT}\right\vert _{\theta _{i}=\hbar ^{i+1}}=\left. \bar{S}_{n+1\hspace{0.01in}%
T}\right\vert _{\theta _{i}=\hbar ^{i+1}}-\left. \bar{S}_{n\hspace{0.01in}%
T}\right\vert _{\theta _{i}=\hbar ^{i+1}}+\mathcal{O}(\hbar ^{n+2})  \notag
\\
&=&S_{n+1\hspace{0.01in}T}-S_{n\hspace{0.01in}T}+\mathcal{O}(\hbar ^{n+2}) 
\notag \\
&=&-\Gamma _{nRT\hspace{0.01in}\text{div}}^{(n+1)}+(\mathcal{R}_{n}-1)S_{%
\mathrm{HD}}+\mathcal{O}(\hbar ^{n+2})+\mathcal{O}(\hbar ^{n+1})o(1/\Lambda
_{-}^{T-2(n+1)\sigma }).  \label{prov}
\end{eqnarray}%
By the inductive assumption, the action $S_{nRT}$ is CDHD renormalized up to
and including $n$ loops, which means that the $\ell $-loop contributions to $%
\Gamma _{nRT}$ are CDHD convergent up to $o(1/\Lambda _{-}^{T-2\ell \sigma
}) $, for $0\leqslant \ell \leqslant n$. Moreover, $\Gamma _{n+1\hspace{%
0.01in}RT}$ and $\Gamma _{nRT}$ coincide up to $\mathcal{O}(\hbar ^{n+1})$,
as well as $S_{n+1\hspace{0.01in}RT}$ and $S_{nRT}$. Now, $\Gamma _{n+1%
\hspace{0.01in}RT}=\Gamma _{nRT}+S_{n+1\hspace{0.01in}RT}-S_{nRT}+\mathcal{O}%
(\hbar ^{n+2}) $, and $(\mathcal{R}_{n}-1)S_{\mathrm{HD}}$ vanishes in the
CDHD limit, up to $\mathcal{O}(\hbar ^{n+2})$. Thus, formula (\ref{prov})\
proves that the $\ell $-loop contributions to $\Gamma _{n+1\hspace{0.01in}%
RT} $ are CDHD convergent up to $o(1/\Lambda _{-}^{T-2\ell \sigma })$, for $%
0\leqslant \ell \leqslant n+1$, which means that $\Gamma _{n+1\hspace{0.01in}%
RT}$ is CDHD renormalized up to and including $n+1$ loops.

The last thing to do is show that $\Gamma _{n+1\hspace{0.01in}RT}$ is free
of gauge anomalies. This result follows from formula (\ref{gk}) for $k=n$.
Indeed, by (\ref{due}), when we replace $\theta _{i}$ with $\hbar ^{i+1}$, $%
i=1,\ldots ,n$, the functional $\bar{\Gamma}_{n+1\hspace{0.01in}RT}$ turns
into $\Gamma _{n+1\hspace{0.01in}RT}$. We finally obtain%
\begin{equation}
(\Gamma _{n+1\hspace{0.01in}RT},\Gamma _{n+1\hspace{0.01in}RT})=\mathcal{O}%
(\varepsilon ),  \label{lma}
\end{equation}%
which means that we have successfully promoted the inductive hypotheses to $%
n+1$ loops.

Iterating the argument, we can make it work till it makes sense, which means
for $n=0,\ldots ,\bar{\ell}-1$, where $\bar{\ell}$ is given by formula (\ref%
{lmax}) for $[\kappa ]<0$ and $\infty $ for $[\kappa ]\geqslant 0$. Finally,
we obtain%
\begin{equation}
\mathcal{A}_{RT}\equiv (\Gamma _{RT},\Gamma _{RT})=\mathcal{O}(\varepsilon ),
\label{AMAB}
\end{equation}%
where $\Gamma _{RT}=\Gamma _{\bar{\ell}RT}$. Observe that the right-hand
side of (\ref{AMAB}) tends to zero everywhere at $\Lambda $ fixed. However,
only within the truncation T2 is $\Gamma _{\bar{\ell}RT}$ convergent in the
CDHD limit. Thus, the $\ell $-loop contributions to the right-hand side
vanish in the CDHD\ limit up to $o(1/\Lambda _{-}^{T-2\ell \sigma })$, for $%
0\leqslant \ell \leqslant \bar{\ell}$. In other words, $\Gamma _{RT}$ is
free of gauge anomalies within the truncation T2. This proves the almost
manifest Adler-Bardeen theorem.

\subsection{Adler-Bardeen theorem}

The result just achieved is also sufficient to prove the Adler-Bardeen
theorem, i.e. statement \ref{bardo} of the introduction. So far, we have
suppressed the $o(1/\Lambda _{-}^{T})$ terms of the action $S$ and its HD
regularized extension $S_{\Lambda }$, according to the prescription T1 of
subsection \ref{s21}. Now we restore those terms, all of which fall outside
the truncation T2. Clearly, the results we have obtained still hold within
the truncation T2. The CD, HD, and CDHD regularizations are still well
defined, because the divergences not cured by the HD technique are cured by
the dimensional one. Note that, however, the HD\ theory $S_{\Lambda }$ is
not super-renormalizable, but nonrenormalizable.

Consider the contributions to the gauge anomalies that lie outside the
truncation $T$, and classify them according to the number of loops and the
power of $1/\Lambda _{-}$. Let $\mathcal{A}_{>T}$ denote any finite class of
them. Clearly, the terms of $\mathcal{A}_{>T}$ lie inside some other
truncation $T^{\prime }>T$, as long as $T^{\prime }$ is sufficiently large.
Now, different truncations just define different subtraction schemes (by
means of different higher-derivative theories and different CDHD
regularizations), and different subtraction schemes differ by finite local
counterterms. Let $s_{T}$ and $s_{T}^{\prime }$ denote the schemes defined
by the truncations $T$ and $T^{\prime }$, respectively. We can assume that
they give exactly the same results (which means that $\Gamma _{RT}$ and $%
\Gamma _{RT^{\prime }}$ coincide) within the truncation $T$, up to
corrections $\mathcal{E}_{\text{CDHD}}$ that vanish in the CDHD limit. We
prove this fact by proceeding inductively. Assume that%
\begin{equation}
\Gamma _{RT^{\prime }}=\Gamma _{RT}+\mathcal{O}(\hbar ^{n+1})+\sum_{k=0}^{n}%
\mathcal{O}(\hbar ^{k})o(1/\Lambda _{-}^{T-2k\sigma })+\mathcal{E}_{\text{%
CDHD}}  \label{assuma}
\end{equation}%
till some order $n<\bar{\ell}$. The assumption is certainly true for $n=0$.
Then, the CDHD nonevanescent $(n+1)$-loop contributions to $\Gamma _{RT}$
and $\Gamma _{RT^{\prime }}$ differ by finite local terms $\Delta S_{n+1}$,
up to $o(1/\Lambda _{-}^{T-2(n+1)\sigma })$, which means%
\begin{equation}
\Gamma _{RT^{\prime }}=\Gamma _{RT}+\Delta S_{n+1}+\mathcal{O}(\hbar
^{n+2})+\sum_{k=0}^{n+1}\mathcal{O}(\hbar ^{k})o(1/\Lambda _{-}^{T-2k\sigma
})+\mathcal{E}_{\text{CDHD}}.  \label{grp}
\end{equation}%
Both $\Gamma _{RT}$ and $\Gamma _{RT^{\prime }}$ satisfy the almost manifest
Adler-Bardeen theorem, that is to say, formula (\ref{AMAB}) and its $%
T^{\prime }$ version. The right-hand sides of (\ref{AMAB}) and its $%
T^{\prime }$ version vanish in the CDHD limit, within the respective
truncations, because $\Gamma _{RT}$ and $\Gamma _{RT^{\prime }}$ are
convergent there. Thus, 
\begin{eqnarray}
\mathcal{A}_{RT} &=&(\Gamma _{RT},\Gamma _{RT})=\mathcal{E}_{\text{CDHD}}+%
\mathcal{O}(\hbar ^{\bar{\ell}+1})+\sum_{k=0}^{\bar{\ell}}\mathcal{O}(\hbar
^{k})o(1/\Lambda _{-}^{T-2k\sigma }),  \notag \\
\mathcal{A}_{RT^{\prime }} &=&(\Gamma _{RT^{\prime }},\Gamma _{RT^{\prime
}})=\mathcal{E}_{\text{CDHD}}+\mathcal{O}(\hbar ^{\bar{\ell}^{\prime
}+1})+\sum_{k=0}^{\bar{\ell}^{\prime }}\mathcal{O}(\hbar ^{k})o(1/\Lambda
_{-}^{T^{\prime }-2k\sigma }).  \label{anjo}
\end{eqnarray}%
Using (\ref{grp}) inside these equations, and taking the CDHD convergent $%
(n+1)$-loop contributions to the difference, we obtain%
\begin{equation*}
(S_{dT},\Delta S_{n+1})=o(1/\Lambda _{-}^{T-2(n+1)\sigma }),
\end{equation*}%
which is a cohomological problem analogous to (\ref{questa}). It can be
solved in the same way, and the solution is the analogue of (\ref{both}),
i.e.%
\begin{equation*}
\Delta S_{n+1}=\sum_{i}\Delta \tilde{\lambda}_{ni}\mathcal{G}%
_{i}+(S_{dT},\Delta \tilde{\chi}_{\hspace{0.01in}nT})+o(1/\Lambda
_{-}^{T-2(n+1)\sigma }),
\end{equation*}%
where $\Delta \tilde{\lambda}_{ni}$ are convergent constants and $\Delta 
\tilde{\chi}_{\hspace{0.01in}nT}$ is a convergent local functional. At this
point, we can attach $\Delta \tilde{\lambda}_{ni}\mathcal{\ }$and $\Delta 
\tilde{\chi}_{\hspace{0.01in}nT}$ to the constants $\Delta \lambda
_{niT^{\prime }}$ and the functional $\chi _{\hspace{0.01in}nT^{\prime }}$
that subtract the $(n+1)$-loop divergences belonging to the truncation $%
T^{\prime }$, given by the $T^{\prime }$ version of formula (\ref{both}).
After that, we can go through the $T^{\prime }$ versions of the arguments
that lead from formula (\ref{both}) to formula (\ref{lma}) with no
difficulty. So doing, we promote assumption (\ref{assuma}) to the order $n+1$
and iterate the procedure till we get%
\begin{equation*}
\Gamma _{RT^{\prime }}=\Gamma _{RT}+\mathcal{O}(\hbar ^{\bar{\ell}%
+1})+\sum_{k=0}^{\bar{\ell}}\mathcal{O}(\hbar ^{k})o(1/\Lambda
_{-}^{T-2k\sigma })+\mathcal{E}_{\text{CDHD}}.
\end{equation*}%
Once this is done, the subtraction schemes $s_{T}$ and $s_{T}^{\prime }$
give the same results within the truncation $T$, up to $\mathcal{E}_{\text{%
CDHD}}$.

Now we compare $s_{T}$ and $s_{T}^{\prime }$ in between the truncations $T$
and $T^{\prime }$. First, we extend the subtraction scheme $s_{T}$ in a
generic way beyond the truncation $T$ and within the truncation $T^{\prime }$%
, and renormalize the action $S_{\Lambda T}$ accordingly. Then, we adapt the
extended scheme order by order to make it give the same results as the
scheme $s_{T^{\prime }}$ within the truncation $T^{\prime }$, up to $%
\mathcal{E}_{\text{CDHD}}$. Let $s_{n,TT^{\prime }}$ denote the extended
scheme adapted up to and including $n<\bar{\ell}^{\prime }$ loops.
Precisely, we assume that $s_{n,TT^{\prime }}$ gives 
\begin{equation}
\Gamma _{RT^{\prime }}=\Gamma _{RT}+\mathcal{O}_{n+1}+\sum_{k=0}^{n}\mathcal{%
O}(\hbar ^{k})o(1/\Lambda _{-}^{T^{\prime }-2k\sigma })+\mathcal{E}_{\text{%
CDHD}},  \label{promo}
\end{equation}%
where%
\begin{eqnarray*}
\mathcal{O}_{n+1} &=&\mathcal{O}(\hbar ^{n+1})\qquad \text{for }n\geqslant 
\bar{\ell}, \\
\mathcal{O}_{n+1} &=&\mathcal{O}(\hbar ^{\bar{\ell}+1})+\sum_{k=n+1}^{\bar{%
\ell}}\mathcal{O}(\hbar ^{k})o(1/\Lambda _{-}^{T-2k\sigma })\qquad \text{for 
}n<\bar{\ell}.
\end{eqnarray*}%
Again, this assumption is satisfied at $n=0$. Then, within the truncation $%
T^{\prime }$ the $(n+1)$-loop contributions to $\Gamma _{RT^{\prime }}$ and $%
\Gamma _{RT}$ differ by finite local terms, which we call $\Delta
S_{n+1,T^{\prime }}$, up to $\mathcal{E}_{\text{CDHD}}$:%
\begin{equation*}
\Gamma _{RT^{\prime }}=\Gamma _{RT}+\Delta S_{n+1,T^{\prime }}+\mathcal{O}%
_{n+2}+\sum_{k=0}^{n+1}\mathcal{O}(\hbar ^{k})o(1/\Lambda _{-}^{T^{\prime
}-2k\sigma })+\mathcal{E}_{\text{CDHD}}.
\end{equation*}%
Note that for $n<\bar{\ell}$, $\Delta S_{n+1,T^{\prime }}=\mathcal{O}(\hbar
^{n+1})o(1/\Lambda _{-}^{T-2(n+1)\sigma })$. Now, replacing the renormalized
action $S_{RT}$ that defines $\Gamma _{RT}$ with $S_{RT}-\Delta
S_{n+1,T^{\prime }}$, we cancel out $\Delta S_{n+1,T^{\prime }}$ and promote
the inductive assumption (\ref{promo}) from order $n$ to order $n+1$.
Iterating the procedure, we arrive at formula (\ref{promo}) with $n=\bar{\ell%
}^{\prime }$. In the end, $\Gamma _{RT^{\prime }}$ coincides with $\Gamma
_{RT}$ within the truncation $T^{\prime }$, up to $\mathcal{E}_{\text{CDHD}}$%
. Finally, formula (\ref{anjo}) ensures that $\Gamma _{RT}$ is free of gauge
anomalies within the truncation $T^{\prime }$.

In other words, it is possible to modify the scheme $s_{T}$ by fine-tuning
the finite local counterterms so as to cancel the potentially anomalous
contributions that belong to the class $\mathcal{A}_{>T}$. Since this
conclusion applies to every class $\mathcal{A}_{>T}$, theorem \ref{bardo}
follows.

\section{Standard model coupled to quantum gravity}

\label{s9}

\setcounter{equation}{0}

In this section we prove that the standard model coupled to quantum gravity
satisfies the assumptions of the proof. In particular, although it does not
satisfy the Kluberg-Stern--Zuber assumption (\ref{coho}), it satisfies
assumption (III) of section \ref{key}, since its basic action $S_{d\text{b}}$
is cohomologically complete, and the group $G_{\text{nas}}$ is compact. We
also comment on the physical meaning of that assumption. We also show that
the standard model coupled to quantum gravity satisfies assumptions (IV)\
and (V)\ of subsection \ref{key}, which concern the one-loop gauge anomalies.

We start by considering the class of four-dimensional Einstein--Yang-Mills
theories that have classical actions of the form 
\begin{equation}
S_{c\text{EYM}}=\int \sqrt{|g|}\left[ -\frac{1}{2\kappa ^{2}}(R+2\Lambda _{%
\text{c}})-\frac{1}{4}F_{\mu \nu }^{a}F^{a\mu \nu }+\mathcal{L}_{\varphi
}(\varphi ,D\varphi )+\mathcal{L}_{\psi }(\psi ,D\psi )+\mathcal{L}_{\varphi
\psi }(\varphi ,\psi )\right] ,  \label{seym}
\end{equation}%
where $F_{\mu \nu }^{a}$ are the field strengths of the Abelian and
non-Abelian Yang-Mills gauge fields, while $\mathcal{L}_{\varphi }$, $%
\mathcal{L}_{\psi }$, and $\mathcal{L}_{\varphi \psi }$ are the matter
Lagrangians, which depend on the scalar fields $\varphi $, the fermions $%
\psi $, and their covariant derivatives $D\varphi $, $D\psi $, as specified
by their arguments. Moreover, $\mathcal{L}_{\varphi }$ is at most quadratic
in $D\varphi $, and $\mathcal{L}_{\psi }$ is at most linear in $D\psi $. The
actions $\bar{S}_{d\text{EYM}}$ and $S_{d\text{EYM}}$ of formulas (\ref{sk})
and (\ref{sid}), built by taking $S_{c\text{EYM}}$ as the classical action $%
S_{c}$, are known to satisfy the Kluberg-Stern--Zuber assumption (\ref{coho}%
) in two cases: when the Yang-Mills gauge group is semisimple and when there
are no accidental symmetries \cite{coho2}. When the Yang-Mills gauge group
contains $U(1)$ factors and $S_{c\text{EYM}}$ is invariant under accidental
symmetries, there exist extra local solutions $X$ of $(S_{d\text{EYM}},X)=0$
that cannot be written in the form $(S_{d\text{EYM}},Y)$ with $Y$ a local
functional \cite{coho2}. We denote them by $\mathcal{G}_{I}^{\text{new}}$.
They depend on the sources $K$, the $U(1)$ gauge fields and the Noether
currents associated with the accidental symmetries.

Consider first the standard model in flat space. We denote its basic action $%
S_{d\text{b}}$ by $S_{d\text{SM}}$. Clearly, $S_{d\text{SM}}$ has the form (%
\ref{seym}) (with gravity switched off), but does not satisfy the
Kluberg-Stern--Zuber assumption (\ref{coho}), because the Yang-Mills gauge
group $SU(3)\times SU(2)\times U(1)_{Y}$ is not semisimple and $S_{d\text{SM}%
}$ has accidental symmetries. One accidental symmetry is the conservation of
the baryon number $B$. If the right-handed neutrinos are present and have
Majorana masses, there are no other accidental symmetries. If the
right-handed neutrinos are present, but do not have Majorana masses, there
is an additional accidental symmetry, which is the conservation of the
lepton number $L$. If the right-handed neutrinos are absent, the lepton
numbers $L_{e}$, $L_{\mu }$ and $L_{\tau }$ of each family are also
conserved. The group of accidental symmetries is $U(1)^{I_{\text{max}}}$,
where $I_{\text{max}}=1$, $2$, or $4$, depending on the case.

The extra solutions $X$ to the condition $(S_{d\text{SM}},X)=0$ can be built
as follows. It is well-known that the hypercharges of the matter fields are
not uniquely fixed by the symmetries of the standard model Lagrangian. If we
deform the standard model action $S_{d\text{SM}}$ by giving arbitrary
hypercharges to the matter fields, and later impose $U(1)_{Y}$ invariance,
then one, two, or four arbitrary charges $q_{I}$ ($I=1,\ldots ,I_{\text{max}%
} $) survive (depending on the group of accidental symmetries), besides the
overall $U(1)_{Y}$ charge. Call the deformed action $S_{d\text{SM}q}(\Phi
,K,q_{I})$. Clearly, $S_{d\text{SM}q}$ satisfies the master equation 
\begin{equation}
(S_{d\text{SM}q},S_{d\text{SM}q})=0  \label{smg}
\end{equation}%
in arbitrary $D$ dimensions and for arbitrary values of the charges $q_{I}$.
If we differentiate (\ref{smg}) with respect to each $q_{I}$, and then set
the $q_{J}$ to zero, we get 
\begin{equation*}
(S_{d\text{SM}},\mathcal{G}_{I\text{SM}}^{\text{new}})=0,\qquad \mathcal{G}%
_{I\text{SM}}^{\text{new}}\equiv \left. \frac{\partial S_{d\text{SM}q}}{%
\partial q_{I}}\right\vert _{q=0}.
\end{equation*}%
The local functionals $\mathcal{G}_{I\text{SM}}^{\text{new}}(\Phi ,K)$
depend explicitly on the sources $K$, because the charges $q_{I}$ appear in
the functional $S_{K}$ of formula (\ref{skexpl}). It can be shown \cite%
{coho2} that $\mathcal{G}_{I\text{SM}}^{\text{new}}$ cannot be written in
the form $(S_{d\text{SM}},Y)$ for a local $Y$. This is why the
Kluberg-Stern--Zuber requirement is not satisfied by the standard model.

The argument just given in flat space can be repeated for the standard model
coupled to quantum gravity, with obvious modifications. Let us denote its
basic action $S_{d\text{b}}$ by $S_{d\text{SMG}}$. It is built on the
classical action $S_{c\text{SMG}}$ of formula (\ref{basi}), which has the
form (\ref{seym}). If we deform it into $S_{d\text{SMG}q}(\Phi ,K,q_{I})$
and differentiate with respect to $q_{I}$, we find extra solutions $X$ of $%
(S_{d\text{SMG}},X)=0$ that cannot be written in the form $(S_{d\text{SMG}%
},Y)$ for a local $Y$. We denote them by $\mathcal{G}_{I\text{SMG}}^{\text{%
new}}(\Phi ,K)$.

In principle, the invariants $\mathcal{G}_{I\text{SM}}^{\text{new}}$, or $%
\mathcal{G}_{I\text{SMG}}^{\text{new}}$, could be generated as counterterms
by renormalization, because they satisfy $(S_{d\text{SM}},\mathcal{G}_{I%
\text{SM}}^{\text{new}})=0$, or $(S_{d\text{SMG}},\mathcal{G}_{I\text{SMG}}^{%
\text{new}})=0$. If this happened, however, we would have a big problem:
some hypercharges would be allowed to run independently from one another and
violate the conditions for the cancellation of gauge anomalies at one loop,
required by assumption (IV). Indeed, it is easy to check that, in general,
the deformation $S_{d\text{SM}q}$ (and therefore also $S_{d\text{SMG}q}$) is
not compatible with the one-loop cancellation of the gauge anomalies \cite%
{chargequant}.

In fact, in subsection \ref{s71} it was shown that, if assumption (III) of
subsection \ref{key} holds, the extra invariants $\mathcal{G}_{I}^{\text{new}%
}$, such as $\mathcal{G}_{I\text{SM}}^{\text{new}}$ or $\mathcal{G}_{I\text{%
SMG}}^{\text{new}}$, \textit{are not} generated by renormalization. Indeed,
they do not appear on the right-hand side of formula (\ref{both}), which
just contains the invariants $\mathcal{G}_{i}(\phi )$. Thus, the meaning of
cohomological completeness is to ensure that renormalization has this key
property.

To show that the standard model coupled to quantum gravity satisfies
assumption (III), we lift the discussion to the extended theory $\check{S}%
_{d}$ of section \ref{s2} and denote its basic action by $\check{S}_{d\text{%
SMG}}$. It is easy to show that $\check{S}_{d\text{SMG}}$ has no accidental
symmetries, because it contains both the four fermion vertices and the
vertex $(LH)^{2}$ that break $B$, $L_{e}$, $L_{\mu }$, and $L_{\tau }$.
Indeed, in the parametrization (\ref{para}) such vertices are not multiplied
by parameters $\zeta $ belonging to the subsets $s_{-}$: the coefficients $%
\zeta $ of the four fermion vertices are dimensionless, while the
coefficient $\zeta $ of $(LH)^{2}$ has dimension one, as shown by formula (%
\ref{4f}). The functionals $\mathcal{G}_{I}^{\text{new}}$ do not satisfy $(%
\check{S}_{d\text{SMG}},\mathcal{G}_{I}^{\text{new}})=0$, and the theory
with action $\check{S}_{d\text{SMG}}$ cannot generate them as counterterms.
By the results of ref. \cite{coho2}, the action $\check{S}_{d\text{SMG}}$,
which has the form (\ref{seym}), satisfies the extended Kluberg-Stern--Zuber
assumption (\ref{coho2}); i.e. $S_{d\text{SMG}}$ is cohomologically
complete. The group $G_{\text{nas}}$ of nonanomalous accidental symmetries
of the action $S_{d\text{SMG}}$ is certainly compact, so assumption (III)
holds.

Let us now move to assumption (IV). Formula (\ref{sdbab}) tells us that the
one-loop anomaly functional $\mathcal{A}_{\text{b}}^{(1)}$ associated with
the basic action $S_{d\text{SMG}}$ of the standard model coupled to quantum
gravity solves the equation $(S_{d\text{SMG}},\mathcal{A}_{\text{b}%
}^{(1)})=0 $. The most general solution to this condition reads%
\begin{equation}
\mathcal{A}_{\text{b}}^{(1)}=\mathcal{A}_{\text{nt}}+(S_{d\text{SMG}}%
\mathcal{X}),  \label{adb}
\end{equation}%
and is the sum of nontrivial terms $\mathcal{A}_{\text{nt}}$ plus trivial
terms $(S_{d\text{SMG}},\mathcal{X})$, where $\mathcal{X}$ is a local
functional of ghost number zero. The nontrivial terms have been classified
in ref. \cite{coho2}. They are ($i$)\ Bardeen terms \cite{bardeen}%
\begin{equation*}
\int \mathrm{d}^{D}x\hspace{0.01in}\varepsilon ^{\mu \nu \rho \sigma }%
\mathrm{Tr}\left[ \partial _{\mu }C\left( A_{\nu }\partial _{\rho }A_{\sigma
}+\frac{g}{2}A_{\nu }A_{\rho }A_{\sigma }\right) \right]
\end{equation*}%
for non-Abelian Yang-Mills symmetries, where $C=C^{\dot{a}}T^{\dot{a}}$, $%
A_{\mu }=A_{\mu }^{\dot{a}}T^{\dot{a}}$, while $C^{\dot{a}}$, $A_{\mu }^{%
\dot{a}}$ are the non-Abelian Yang-Mills ghosts and gauge fields,
respectively, and the index $\dot{a}$ runs on each simple subalgebra of the
Yang-Mills Lie algebra; ($ii$)\ terms of the Bardeen type%
\begin{eqnarray*}
&&\int \mathrm{d}^{D}x\mathcal{\hspace{0.01in}}\varepsilon ^{\mu \nu \rho
\sigma }C_{V}(\partial _{\mu }V_{\nu })(\partial _{\rho }V_{\sigma }),\qquad
\int \mathrm{d}^{D}x\hspace{0.01in}\varepsilon ^{\mu \nu \rho \sigma }C^{%
\dot{a}}(\partial _{\mu }V_{\nu })(\partial _{\rho }A_{\sigma }^{\dot{a}%
}),\qquad \int \mathrm{d}^{D}x\hspace{0.01in}\varepsilon ^{\mu \nu \rho
\sigma }C_{V}F_{\mu \nu }^{\dot{a}}F_{\rho \sigma }^{\dot{a}}, \\
&&\int \mathrm{d}^{D}x\hspace{0.01in}\varepsilon ^{\mu \nu \rho \sigma
}C_{V}R_{\mu \nu }^{\bar{a}\bar{b}}R_{\rho \sigma }^{\bar{a}\bar{b}},\qquad
\int \mathrm{d}^{D}x\hspace{0.01in}\varepsilon ^{\mu \nu \rho \sigma
}C_{V}R_{\mu \nu }^{\bar{a}\bar{b}}R_{\rho \sigma }^{\bar{c}\bar{d}%
}\varepsilon _{\bar{a}\bar{b}\bar{c}\bar{d}},
\end{eqnarray*}%
involving $U(1)$ gauge fields $V_{\mu }$ and/or $U(1)$ ghosts $C_{V}$; ($iii$%
) terms of the form $\int C_{V}\mathcal{L}$, where $\mathcal{L}$ is a
Lagrangian density that depends only on the fields, is not a total
derivative, and satisfies $(S_{K},\int \mathcal{L})=0$; ($iv$) $K$-dependent
extra terms $\mathcal{A}_{I\text{SMG}}^{\text{new}}$ of ghost number one,
analogous to the extra terms $\mathcal{G}_{I\text{SMG}}^{\text{new}}$ of
ghost number zero discussed above. The terms of class ($iv$) are absent
unless the gauge group contains $U(1)$ factors and the theory has accidental
symmetries. We recall that there are no Lorentz anomalies in four dimensions.

To study the anomalies $\mathcal{A}_{\text{nt}}$ of equation (\ref{adb}) we
can switch to the framework we prefer. A change of framework affects the
finite local counterterms contained in the functional $\hat{\Gamma}_{\Lambda
T\hspace{0.01in}\text{fin}}^{(1)}$ of formula (\ref{subtra}). As far as $%
\mathcal{A}_{\text{b}}^{(1)}$ is concerned, formula (\ref{line2}) ensures
that it only affects the functional $\mathcal{X}$ of (\ref{adb}).

Consider first the terms $\mathcal{A}_{\text{nt}}$ that belong to the
classes ($i$) and ($ii$). The most economic framework to study them is the
standard dimensional regularization. For definiteness, we use a basis where
all the fermionic fields are left handed, and we denote them by $\psi _{L}$.
Associate a right-handed partner $\psi _{R}$ with each $\psi _{L}$ and
extend the action $S_{d\text{SMG}}$ by adding the correction 
\begin{equation*}
S_{LR}(\Phi )=\int \bar{\psi}_{R}i\tilde{\gamma}^{\mu }\partial _{\mu }\psi
_{L}+\int \bar{\psi}_{L}i\tilde{\gamma}^{\mu }\partial _{\mu }\psi _{R}+\int 
\bar{\psi}_{R}i\tilde{\gamma}^{\mu }\partial _{\mu }\psi _{R}
\end{equation*}%
to it, where the flat-space vielbein is used and $\tilde{\gamma}^{\mu }$
denote the standard $\gamma $ matrices in $D$ dimensions, which satisfy $\{%
\tilde{\gamma}^{\mu },\tilde{\gamma}^{\nu }\}=2\eta ^{\mu \nu }$. Let $S_{%
\text{ext}}(\Phi ,K)=S_{d\text{SMG}}(\Phi ,K)+S_{LR}(\Phi )$ denote the
extended action. Expanding around flat space as usual, the total kinetic
terms of $\psi _{L}$ and $\psi _{R}$ are $\int i\bar{\psi}\tilde{\gamma}%
^{\mu }\partial _{\mu }\psi $, where $\psi =\psi _{L}+\psi _{R}$. Since $%
\psi _{R}$ appears just in $S_{LR}$, no nontrivial one-particle irreducible
diagrams with $\psi _{R}$ external legs can be built, so the partners $\psi
_{R}$ decouple at $\varepsilon =0$. Moreover, $S_{d\text{SMG}}$ is gauge
invariant, while $S_{LR}$ is not, which means that $(S_{\text{ext}},S_{\text{%
ext}})$ is cubic in the fields $\Phi $. More precisely, $(S_{\text{ext}},S_{%
\text{ext}})$ is bilinear in the fermions and linear in the ghosts. The
anomaly functional is $\mathcal{A}=\langle (S_{\text{ext}},S_{\text{ext}%
})\rangle $. The nontrivial terms $\mathcal{A}_{\text{nt}}$ of classes ($i$)
and ($ii$) do not contain fermions, so they can only arise from the one-loop
polygon diagrams that have $(S_{\text{ext}},S_{\text{ext}})$ and gauge
currents (including the energy-momentum tensor) at their vertices, and
fermions circulating inside. It is well known \cite{peskin} that the
contributions of such diagrams vanish at $\varepsilon =0$ in the standard
model coupled to quantum gravity.

Next, consider the terms $\mathcal{A}_{\text{nt}}$ of class ($iii$). They
are anomalies of the global $U(1)_{Y}$ symmetry. To prove that they are
absent, it is sufficient to choose a regularization technique that is
globally $U(1)_{Y}$ invariant. Again, the standard dimensional
regularization has this property, while the CD technique does not [because
of the terms (\ref{w2}), which are of the Majorana type]. Finally, formula (%
\ref{line2}) ensures that the terms of class ($iv$) are not generated,
because they depend on the sources $K$.

This proves that $\mathcal{A}_{\text{nt}}=0$; i.e. the basic action $S_{d%
\text{SMG}}$ of the standard model coupled to quantum gravity satisfies
assumption (IV). We also note that the arguments of subsection \ref{s71}
imply that the action $S_{d}$ of the standard model coupled to quantum
gravity, which is equal to $S_{d\text{SMG}}$ plus corrections multiplied by
powers of $1/\Lambda _{-}$, is also cohomologically complete and satisfies
the physical Kluberg-Stern--Zuber conjecture (\ref{phyz}).

The absence of the terms of class ($iv$) is a general fact, not tied to the
particular model we are considering. It can also be proved by lifting the
discussion to $\check{S}_{d\text{SMG}}$, where all accidental symmetries are
broken. The one-loop anomaly functional $\mathcal{\check{A}}_{\text{b}%
}^{(1)} $ of the theory with action $\check{S}_{d\text{SMG}}$ satisfies $(%
\check{S}_{d\text{SMG}},\mathcal{\check{A}}_{\text{b}}^{(1)})=0$ and can be
decomposed as $\mathcal{\check{A}}_{\text{b}}^{(1)}=\mathcal{\check{A}}_{%
\text{nt}}+(\check{S}_{d\text{SMG}},\mathcal{\check{X}})$, where the
nontrivial anomalous terms $\mathcal{\check{A}}_{\text{nt}}$ can only belong
to the classes ($i$-$iii$), and $\mathcal{\check{X}}$ is a local functional
of $\Phi $ and $K$. The functional $\mathcal{A}_{\text{b}}^{(1)}$ can be
retrieved from $\mathcal{\check{A}}_{\text{b}}^{(1)}$ by switching off the
coefficients $\check{\lambda}$ and $\check{\eta}$ of the terms that break
the nonanomalous accidental symmetries. This operation gives a result of the
form (\ref{adb}), where $\mathcal{A}_{\text{nt}}$ and $\mathcal{X}$ are
equal to $\mathcal{\check{A}}_{\text{nt}}$ and $\mathcal{\check{X}}$ at $%
\check{\lambda}=\check{\eta}=0$, respectively. If, in addition, we average
on the group $G_{\text{nas}}$, we can assume that $\mathcal{X}$ is invariant
under $G_{\text{nas}}$. It follows that $\mathcal{A}_{\text{nt}}$ is a
linear combination of terms belonging to the classes ($i$-$iii$).

It remains to study assumption (V) of subsection \ref{key}. If\ a functional 
$\mathcal{F}(\kappa \Phi )$ of ghost number one can be written in the form $%
(S_{d\text{b}},\mathcal{X})$, it clearly satisfies $(S_{d\text{b}},\mathcal{F%
})=0$. Then it also satisfies $(S_{K},\mathcal{F})=0$, since $\mathcal{F}$
is $K$ independent. We want to show that $\mathcal{F}$ can be written as $%
(S_{K},\chi )$, where $\chi (\kappa \Phi )$ is a local functional of the
fields $\Phi $.

The most general solution of the problem $(S_{K},\mathcal{F})=0$, when the
gauge symmetries are diffeomorphisms, local Lorentz symmetry and Abelian and
non-Abelian Yang-Mills symmetries, is worked out in ref. \cite{brandt}. The
functional $\mathcal{F}$ is the sum of nontrivial terms $\mathfrak{A}_{\text{%
nt}}$ belonging to the classes ($i$-$iii$) listed above, plus trivial terms
of the correct form $(S_{K},\chi (\kappa \Phi ))$. Combining this fact with $%
\mathcal{F}=(S_{d\text{b}},\mathcal{X})$, we obtain 
\begin{equation*}
\mathcal{F}=(S_{d\text{b}},\mathcal{X})=\mathfrak{A}_{\text{nt}}+(S_{K},\chi
).
\end{equation*}%
Turning this equation around, we also get $\mathfrak{A}_{\text{nt}}=(S_{d%
\text{b}},\mathcal{X}^{\prime \prime })$, with $\mathcal{X}^{\prime \prime }=%
\mathcal{X}-\chi $. In other words, the functional $\mathfrak{A}_{\text{nt}}$
is trivial in the $S_{d\text{b}}$ cohomology and nontrivial in the $S_{K}$
cohomology. The results of ref. \cite{coho2} ensure that in four-dimensional
Einstein--Yang-Mills theories that have an action of the form (\ref{seym}),
this is impossible, unless $\mathfrak{A}_{\text{nt}}$ vanishes. Thus, the
standard model coupled to quantum gravity satisfies assumption (IV). 

We stress again that assumptions (IV)\ and (V)\ are just needed to prove
that the one-loop anomalies (\ref{algol}) of the HD\ theory are trivial in
the $S_{K}$ cohomology, which means that they have the form (\ref{anomcanc}%
). The same result is more quickly implied by assumption (IV$^{\prime }$) of
subsection \ref{key}. In several practical cases, it may be simpler to prove
assumption (IV$^{\prime }$), rather than assumptions (IV)\ and (V).

We conclude that the standard model coupled to quantum gravity satisfies all
the assumptions made in this paper. Therefore, it is free of gauge anomalies
to all orders in perturbation theory. In a generic framework, the
Adler-Bardeen theorem \ref{bardo} of the introduction tells us that the
cancellation of gauge anomalies is nonmanifest, and can be enforced by
fine-tuning finite local counterterms order by order. If we use the
framework elaborated in this paper, theorem \ref{bardo3} tells us that the
cancellation is manifest within any given truncation and nonmanifest outside.

The arguments of this section apply with simple modifications to most
standard model extensions, irrespectively of their gauge groups and
accidental symmetries. When the other assumptions are met, it is sufficient
to check that the gauge anomalies are trivial at one loop to infer that they
can be canceled to all orders. It is also clear how to generalize the
analysis of this section to theories living in spacetime dimensions
different than four.

\section{Conclusions}

\label{s10}

\setcounter{equation}{0}

In this paper we proved the Adler-Bardeen theorem for the cancellation of
gauge anomalies in nonrenormalizable theories, which is the statement that
there exists a subtraction scheme where the gauge anomalies cancel to all
orders, when they are trivial at one loop. We assumed that the gauge
symmetries are diffeomorphisms, local Lorentz symmetry and Yang-Mills
symmetries, and that the local functionals of vanishing ghost number satisfy
a variant of the Kluberg-Stern--Zuber conjecture. In our approach, the
cancellation is \textquotedblleft almost manifest\textquotedblright , which
means that, given a truncation of the theory, once the gauge anomalies are
canceled at one loop, they manifestly vanish from two loops onwards within
the truncation, while outside the truncation their cancellation can be
achieved by fine-tuning finite local counterterms. The truncation can
contain arbitrarily many terms.

Although some arguments of the proof are technically involved, the key ideas
are actually intuitive. The hardest part of the job is building the right
framework. We used a regularization technique that combines a modified
version of the dimensional regularization with a suitable higher-derivative
gauge invariant regularization. This trick allows us to isolate the sources
of potential anomalies, which are just one loop, from the nonanomalous
sector of the theory. When the HD energy scale $\Lambda $ is kept fixed, we
have a super-renormalizable theory that satisfies the manifest Adler-Bardeen
theorem to all orders in $\hbar $ by simple power counting arguments. When $%
\Lambda $ is taken to infinity, the $\Lambda $ divergences are subtracted by
means of canonical transformations and redefinitions of parameters. At each
step, the HD theory must be re-renormalized at $\Lambda $ fixed, to subtract
the newly generated divergences in $\varepsilon $. While doing so, it is
possible to enforce the cancellation of gauge anomalies again by fine-tuning
finite local counterterms.

The standard model coupled to quantum gravity satisfies the assumptions we
have made, so it is free of gauge anomalies to all orders. The theorem we
have proved also applies to most extensions of the standard model, coupled
to quantum gravity or not, and to a variety of other theories, including
higher-derivative and Lorentz violating theories, in arbitrary dimensions.

Among the prospects for the future, we mention the generalization of the
proof to supergravity. The complexity of local supersymmetry makes this task
quite challenging, especially in the presence of scalar multiplets and when
it is not known how to achieve closure off shell.

\end{document}